\documentclass[]{esub2acm}
\usepackage[]{graphicx,color,multicol,rotating}
\begin{document}

\title{Rectangular Full Packed Format for Cholesky's Algorithm:
Factorization, Solution and Inversion}

\author{Fred G. Gustavson \\
          IBM T.J. Watson Research Center
\and Jerzy Wa{\'{s}}niewski \\
         Technical University of Denmark
\and Jack J. Dongarra \\ 
          University of Tennessee, Oak Ridge National Laboratory and
          University of Manchester
\and Julien Langou \\
          University of Colorado Denver
}

\begin{abstract}

We describe a new data format for storing triangular, symmetric, and Hermitian
matrices called RFPF (Rectangular Full Packed Format). The standard two
dimensional arrays of Fortran and C (also known as full format) that are used
to represent triangular and symmetric matrices waste nearly half of the storage
space but provide high performance via the use of Level~3 BLAS.  Standard
packed format arrays fully utilize storage (array space) but provide low
performance as there is no Level~3 packed BLAS. We combine the good features
of packed and full storage using RFPF to obtain high performance via using
Level~3 BLAS as RFPF is a standard full format representation. Also, RFPF
requires exactly the same minimal storage as packed format. Each LAPACK full
and/or packed triangular, symmetric, and Hermitian routine becomes a single new
RFPF routine based on eight possible data layouts of RFPF.  This new RFPF
routine usually consists of two calls to the corresponding LAPACK full format
routine and two calls to Level~3 BLAS routines. This means {\it no} new
software is required. As examples, we present LAPACK routines for Cholesky
factorization, Cholesky solution and Cholesky inverse computation in RFPF to
illustrate this new work and to describe its performance on several commonly
used computer platforms. Performance of LAPACK full routines using RFPF versus
LAPACK full routines using standard format for both serial and SMP parallel
processing is about the same while using half the storage. Performance gains
are roughly one to a factor of 43 for serial and one to a factor of 97 for SMP
parallel times faster using vendor LAPACK full routines with RFPF than with
using vendor and/or reference packed routines.

\end{abstract}




\markboth{Fred G. Gustavson, Jerzy Wa{\'{s}}niewski, Jack J. Dongarra, and Julien Langou}
{Rectangular Full Packed Data Format (Cholesky Algorithm)}

\begin{bottomstuff}
Authors' addresses:
F.G. Gustavson, IBM T.J. Watson Research Center, Yorktown Heights, 
NY-10598, USA, email: \hbox{fg2@us.ibm.com};
J. Wa{\'{s}}niewski, Department of Informatics and Mathematical Modelling, 
Technical University of Denmark, 
Richard Petersens Plads, Building 321, DK-2800 Kongens Lyngby, 
Denmark, email: \hbox{jw@imm.dtu.dk};
J.J. Dongarra,  Electrical Engineering and Computer Science Department, 
University of Tennessee, 1122 Volunteer Blvd, 
Knoxville, TN 37996-3450, USA, email: \hbox{dongarra@eecs.utk.edu};
Julien Langou, Department of Mathematical and Statistical Sciences, 
University of Colorado Denver,
1250, 14th Street -- Room 646, Denver, Colorado 80202, USA,
email: \hbox{julien.langou@ucdenver.edu}.
\end{bottomstuff}

\maketitle

\section{Introduction}
A very important class of linear algebra problems deals with a
coefficient matrix $A$ that is symmetric and positive definite~\cite{HPCdongduffsorvorst98,demmel97,GV98,Trefethen97}.
Because of symmetry it is only necessary to store either the
upper or lower triangular part of the matrix $A$.
\begin{figure}[h]
\caption{\label{lafull}The {\bf full} format array layout of an order N symmetric matrix
required by LAPACK. LAPACK requires ${\tt LDA} \ge N$. Here we set {\tt LDA}=$N$=7.}
$$
    \begin{array}{c} \mbox{Lower triangular case} \\ \\
    \left( \begin{array}{rrrrrrrrrrrrrrrrrrrrrrrr}
      1 \\
      2 &  9 \\
      3 & 10 & 17 \\
      4 & 11 & 18 & 25 \\
      5 & 12 & 19 & 26 & 33 \\
      6 & 13 & 20 & 27 & 34 & 41 \\
      7 & 14 & 21 & 28 & 35 & 42 & 49 \\
     \end{array} \right) \end{array} \; \; \; \;
    \begin{array}{c} \mbox{Upper triangular case} \\ \\
    \left( \begin{array}{rrrrrrrrrrrrrrrrrrrrrrrr}
      1 &  8 & 15 & 22 & 29 & 36 & 43 \\
        &  9 & 16 & 23 & 30 & 37 & 44 \\
        &    & 17 & 24 & 31 & 38 & 45 \\
        &    &    & 25 & 32 & 39 & 46 \\
        &    &    &    & 33 & 40 & 47 \\
        &    &    &    &    & 41 & 48 \\
        &    &    &    &    &    & 49 \\
     \end{array} \right) \end{array}
$$
\end{figure}
\begin{figure}[h]
\caption{\label{lapacked}The {\bf packed} format array layout of an order 7 symmetric matrix
required by LAPACK.}
$$
    \begin{array}{c} \mbox{Lower triangular case} \\ \\
    \left( \begin{array}{rrrrrrrrrrrrrrrrrrrrrrrr}
      1 \\
      2 &  8 \\
      3 &  9 & 14 \\
      4 & 10 & 15 & 19 \\
      5 & 11 & 16 & 20 & 23 \\
      6 & 12 & 17 & 21 & 24 & 26 \\
      7 & 13 & 18 & 22 & 25 & 27 & 28 \\
     \end{array} \right) \end{array} \; \; \; \;
    \begin{array}{c} \mbox{Upper triangular case} \\ \\
    \left( \begin{array}{rrrrrrrrrrrrrrrrrrrrrrrr}
      1 &  2 &  4 &  7 & 11 & 16 & 22 \\
        &  3 &  5 &  8 & 12 & 17 & 23 \\
        &    &  6 &  9 & 13 & 18 & 24 \\
        &    &    & 10 & 14 & 19 & 25 \\
        &    &    &    & 15 & 20 & 26 \\
        &    &    &    &    & 21 & 27 \\
        &    &    &    &    &    & 28 \\
     \end{array} \right) \end{array}
$$
\end{figure}
\subsection{LAPACK full and packed storage formats}
The LAPACK library~\cite{laug}
offers two different kinds of subroutines to solve the same problem:
POTRF\footnote{\label{blasexp}Four names SPOTRF, DPOTRF, CPOTRF
and ZPOTRF are used in LAPACK for real symmetric and complex
Hermitian matrices~\cite{laug}, where the first
character indicates the
precision and arithmetic versions: S -- single precision, D --
double precision, C -- complex and Z -- double complex.
LAPACK95 uses one name LA\_POTRF for all
versions~\cite{laug95}. In this paper, POTRF and/or PPTRF express, 
any precision, any arithmetic and any language version of the PO
and/or PP matrix factorization algorithms.}
and PPTRF both factorize symmetric, positive
definite matrices by means of the Cholesky algorithm. A major difference in
these two routines is the way they access the array holding the 
triangular matrix (see
Figures \ref{lafull} and \ref{lapacked}). 

In the POTRF case, the matrix is stored in one of the lower left or upper right
triangles of a full square matrix (\cite[pages~139 and 140]{laug}
and~\cite[page~64]{ibmessl97})\footnote{In Fortran column major, in C row
major.}, the other triangle is wasted (see Figure~\ref{lafull}). Because of the
uniform storage scheme, blocked LAPACK and Level~3 BLAS
subroutines~\cite{blas3,blas3alg} can be employed, resulting in a fast
solution.

In the PPTRF case, the matrix is stored in \emph{packed}
storage (\cite[pages~140 and 141]{laug}, \cite{AGZ94}
and \cite[pages~74 and 75]{ibmessl97}),
which means that the columns of the lower or upper triangle
are stored consecutively in a one dimensional array (see
Figure~\ref{lapacked}). Now the
triangular matrix occupies the strictly necessary storage space
but the nonuniform storage scheme means that use of full storage BLAS
is impossible and only the Level~2 BLAS packed subroutines~\cite{blas1, blas2} can be
employed, resulting in a slow solution.

To summarize: LAPACK offers a choice between
high performance and wasting half of the memory space (POTRF)
versus low performance with optimal memory space (PPTRF).

\subsection{Packed Minimal Storage Data Formats related to RFPF}

Recently many new data formats for matrices have been introduced for improving the
performance of Dense Linear Algebra (DLA) algorithms.
The survey article~\cite{EGKJ04} gives an excellent overview.

{\em Recursive Packed Format (RPF)}~\cite{AGW01,espoo:jerzy:02}:
A new compact way to store a triangular, symmetric or Hermitian matrix called Recursive
Packed Format is described in~\cite{AGW01} as are novel ways to transform
RPF to and from standard packed format.  New algorithms, called Recursive
Packed Cholesky (RPC)~\cite{AGW01,espoo:jerzy:02} that operate on the RPF
format are presented. RPF format operates almost entirely by calling
Level~3 BLAS GEMM~\cite{blas3,blas3alg} but requires variants of algorithms
TRSM and SYRK~\cite{blas3,blas3alg} that are designed t  work on RPF. The authors call
these algorithms RPTRSM and RPSYRK~\cite{AGW01} and find that they do most of
their FLOPS by calling GEMM~\cite{blas3,blas3alg}. It follows that almost all
of execution time of the RPC algorithm is done in calls to GEMM.\\
\noindent
There are three advantages of this storage scheme compared to traditional packed
and full storage. First, the RPF storage format
uses the minimum amount of storage required for symmetric, triangular, or
Hermitian matrices. Second, the RPC algorithm is a Level~3 implementation of
Cholesky factorization. Finally, RPF requires no block size tuning parameter.
A disadvantage of the RPC algorithm was that it had a high recursive calling overhead.
The paper~\cite{GJ99} removed this overhead and added other novel features
to the RPC algorithm.
 
{\em Square Block Packed Format (SBPF)}~\cite{SBPF}:
SBPF is described in Section 4 of~\cite{SBPF}. A strong point of SBPF is that
it requires minimum block storage and all its blocks are contiguous and of
equal size. If one uses SBPF with kernel routines then data copying is mostly
eliminated during Cholesky factorization. 

{\em Block Packed Hybrid Format (BPHF)}~\cite{AGGRW05,GRW07}:
We consider an efficient implementation of the Cholesky solution of symmetric
positive-definite full linear systems of equations using packed storage. We
take the same starting point as that of LINPACK~\cite{DBMS79} and
LAPACK~\cite{laug}, with the upper (or lower) triangular part of the matrix
being stored by columns. Following LINPACK~\cite{DBMS79} and
LAPACK~\cite{laug}, we overwrite the given matrix by its Cholesky factor. The
paper~\cite{AGGRW05} uses the BPHF where blocks of the matrix are held
contiguously. The paper compares BPHF versus conventional full format storage,
packed format and the RPF for the algorithms.
BPF is a variant of SBPF in which the diagonal blocks are stored in
packed format and so its storage requirement is equal to that of packed
storage.

We mention that for packed matrices SBPF and BPHF have become the format of choice for multicore
processors when one stores the blocks in register block format~\cite{para06ggs}.
Recently, there have been many papers published on new algorithms for multicore processors.
This literature is extensive. So, we only mention two projects, PLASMA~\cite{BLKD} and FLAME~\cite{CQ2V}, and refer
the interested reader to the literature for additional references. 

In regard to other references on new data structures, the survey article~\cite{EGKJ04} gives an excellent overview.
However, since 2005 at least two new data formats for Cholesky type factorizations have emerged,~\cite{H06} 
and the subject matter of this paper, RFPF~\cite{para06fgjw}. In the next subsection we
highlight the main features of RFPF. 

\subsection{A novel way of representing triangular, symmetric, and Hermitian matrices in LAPACK}

LAPACK has two types of subroutines for triangular, symmetric, and Hermitian
matrices called packed and full format routines.  LAPACK has about 300 these
kind of subroutines. So, in either format, a variety of problems can be solved
by these LAPACK subroutines.  From a user point of view, RFPF can replace both
these LAPACK data formats.  Furthermore, and this is important, using RFPF does
not require any new LAPACK subroutines to be written. Using RFPF in LAPACK only
requires the use of already existing LAPACK and BLAS routines.  RFPF strongly
relies on the existence of the BLAS and LAPACK routines for full storage
format.

\subsection{Overview of the Paper}\label{overviev}

First we introduce the RFPF in general, see Section~\ref{rfpformat}.  Secondly
we show how to use RFPF on symmetric and Hermitian positive definite matrices;
e.g., for the factorization (Section~\ref{fact}), solution (Section~\ref{sol}),
and inversion (Section~\ref{inv}) of these matrices. Section~\ref{algorithms}
describes LAPACK subroutines for the Cholesky factorization, Cholesky solution, and Cholesky inversion of
symmetric and Hermitian positive definite matrices using RFPF.
Section~\ref{sec:stability} indicates that the stability results of using RFPF
is unaffected by this format choice as RFPF uses existing LAPACK algorithms
which are already known to be stable. Section~\ref{performance} describes a
variety of performance results on commonly used platforms both for serial and
parallel SMP execution.  These results show that performance of LAPACK full
routines using RFPF versus LAPACK full routines using standard format for both
serial and SMP parallel processing is about the same while using half the
storage. Also, performance gains are roughly one to a factor of 43 for serial
and one to a factor of 97 for SMP parallel times faster using vendor LAPACK
full routines with RFPF than with using vendor and/or reference packed routines.
Section~\ref{sec:lapack} explains how some new RFPF routines have been
integrated in LAPACK.
LAPACK software for Cholesky algorithm (factorization, solution and inversion)
using RFPF has been released with LAPACK-3.2 on November 2008.
Section~\ref{sec:summary}  gives a short summary and
brief conclusions.

\section{Description of Rectangular Full Packed Format}\label{rfpformat}

We describe Rectangular Full Packed Format (RFPF).
It transforms a standard Packed Array {\tt AP} of size $NT=N(N+1)/2$ to a full 2D array. This means
that performance of LAPACK's~\cite{laug} packed format routines becomes equal to
or better than their full array counterparts. RFPF is a
variant of Hybrid Full Packed (HFP) format~\cite{para04jgfg}. RFPF is a
rearrangement of a Standard full format rectangular Array {\tt SA} of size {\tt LDA*N} where ${\tt LDA} \ge N$. 
Array {\tt SA} holds a triangular part of a symmetric, triangular, or Hermitian matrix $A$ of order $N$.
The rearrangement of array {\tt SA} is equal to compact full format Rectangular
Array {\tt AR} of size ${\tt LDA1*N1}=NT$ and hence array {\tt AR} like array {\tt AP} uses minimal storage. 
(The specific values of {\tt LDA1} and {\tt N1} can vary depending on various cases and they will be specified
later during the text.)
Array {\tt AR} will hold a full rectangular matrix $A_R$ obtained from a triangle of matrix $A$. Note also that the
transpose of the rectangular matrix $A_R^T$ resides in the transpose of array {\tt AR} and hence also represents $A$.
Therefore, Level~3 BLAS~\cite{blas3,blas3alg} can be used on array {\tt AR} or its transpose. 
In fact, with the equivalent
LAPACK algorithm which uses the array {\tt AR} or its transpose, the performance is slightly better
than standard LAPACK algorithm which uses the array {\tt SA} or its transpose.
Therefore, this offers the possibility to replace all packed or full LAPACK routines with
equivalent LAPACK routines that work on array {\tt AR} or its transpose.  
For examples of transformations of a matrix $A$ to a matrix $A_R$ see the figures in
Section~\ref{algorithms}.

RFPF is 
closely related to HFP format, see~\cite{para04jgfg}, which 
represents $A$ as the concatenation of two
standard full arrays whose total size is also $NT$. A basic simple idea
leads to both formats. Let $A$ be an order $N$ symmetric matrix. Break $A$
into a block 2--by--2 form


\begin{equation}
A = \left[ \begin{array}{cc} A_{11} & A^T_{21} \\
    A_{21} & A_{22} \end{array} \right] \mbox{ or }
A = \left[ \begin{array}{cc} A_{11} & A_{12} \\
    A^T_{12} & A_{22} \end{array} \right] \label{equ:aa}
\end{equation}

\noindent
where $A_{11}$ and $A_{22}$ are symmetric. Clearly, we need only store the lower
triangles of $A_{11}$ and $A_{22}$ as well as the full matrix $A_{21}=A_{12}^T$ 
when we are interested in a lower triangular formulation. 

When $N= 2k$ is even, the lower triangle of $A_{11}$ and the upper triangle of $A^T_{22}$ can
be concatenated together along their main diagonals into a $(k+1)$--by--$k$ 
dense matrix (see the figures where N is even in
Section~\ref{algorithms}). This last operation is the crux of the basic simple
idea. The off-diagonal block $A_{21}$ is $k$--by--$k$, and so it can be appended
below the $(k+1)$--by--$k$ dense matrix. Thus, the lower triangle of $A$ can
be stored as a single $(N+1)$--by--$k$ dense matrix $A_R$. In effect, each
block matrix $A_{11}$, $A_{21}$ and $A_{22}$ is now stored in `full format'. This
means all entries of matrix $A_R$ in array {\tt AR} of size ${\tt LDA1}=N+1$ by ${\tt N1}=k$
can be accessed with constant row and column
strides. So, the full power of LAPACK's block Level~3 codes are now
available for RFPF which uses the minimum amount of storage.
Finally, matrix $A_R^T$ which has size $k$--by--$(N+1)$ is represented in the transpose of array {\tt AR} and
hence has the same desirable properties. 
There are eight representations of RFPF.
The matrix $A$ can have have either odd or even order $N$, 
or it can be represented either in standard lower or upper format 
or it can be represented by either matrix $A_R$ or its transpose $A_R^T$ 
giving $2^3 = 8$ representations in all.

All eight cases or representations are presented in
Section~\ref{algorithms}. 
The RFPF matrices are in the upper right part of the figures.
We have introduced colors and horizontal lines to try to visually delineate triangles
$T_1$, $T_2$ representing lower, upper triangles of symmetric matrices 
$A_{11}$, $A^T_{22}$ respectively and square or near square $S_1$ 
representing matrices $A_{21}$. For an upper triangle of $A$,
$T_1$, $T_2$ represents lower, upper triangles of symmetric matrices 
$A^T_{11}$, $A_{22}$ respectively and square or near square $S_1$ 
representing matrices $A_{12}$. For both lower and upper triangles of $A$ we have, after each $a_{i,j}$, added its 
position location in the arrays holding matrices $A$ and $A_R$. 

We now consider performance aspects of using RFPF in the context
of using LAPACK routines on triangular matrices stored in RFPF.
Let $X$ be a Level~3 LAPACK routine that operates either on
full format. $X$ has a full Level~3 LAPACK block 2--by--2
algorithm, call it
$FX$. We write a simple related partition algorithm (SRPA) with partition sizes
$n1$ and $n2$ where $n1+n2=N$. Apply the new SRPA using the new RFPF. The new
SRPA almost always has four major steps consisting entirely of calls to
existing full format LAPACK routines in two steps and calls to Level~3
BLAS in the remaining two steps, see Figure~\ref{int:rfp}.


\begin{figure}[h]
\caption{Simple related partition algorithm (SRPA) of RFPF}\label{int:rfp}
\resizebox*{1.00\textwidth}{!}{\centering \begin{tabular}{ccc}
\begin{tabular}{l}
  call X('L',n1,T1,ldt) ! step 1 \\
  call L3BLAS(n1,n2,'L',T1,ldt,S,lds) ! step 2
\end{tabular} & &
\begin{tabular}{l}
  call L3BLAS(n1,n2,S,lds,'U',T2,ldt) ! step 3 \\
  call X('U',n2,T2,ldt) ! step 4
\end{tabular}
\end{tabular}\par}
\end{figure}

Section~\ref{algorithms} shows $FX$ algorithms equal to factorization, solution and inversion 
algorithms on symmetric positive definite or Hermitian matrices.

\section{Cholesky Factorization using Rectangular Full Packed Format}
\label{fact}

The Cholesky factorization of a symmetric and positive definite matrix $A$
can be expressed as 
\begin{equation} 
\begin{array}{c} A = LL^T \mbox{ or } A = U^TU \mbox{(in the symmetric case)} \\
                 A = LL^H \mbox{ or } A = U^HU \mbox{(in the Hermitian case)} 
\end{array}\label{equ:alu} \end{equation} 
where $L$ and $U$ are lower triangular and upper triangular matrices.

Break the matrices $L$ and $U$ into 2--by--2 block form in the same way as
was done for the matrix $A$ in Equation~(\ref{equ:aa}):

\begin{equation}
L = \left[ \begin{array}{cc} L_{11} & 0 \\
    L_{21} & L_{22} \end{array} \right] \mbox{ and }
U = \left[ \begin{array}{cc} U_{11} & U_{12} \\
    0 & U_{22} \end{array} \right] \label{equ:lu}
\end{equation}

We now have

\begin{equation}
\begin{array}{c}
       \mbox{the symmetric case:} \\
LL^T = \left[ \begin{array}{cc} L_{11} & 0 \\ L_{21} & L_{22}
       \end{array} \right]
       \left[ \begin{array}{cc} L^T_{11} & L^T_{21} \\ 0 & L^T_{22}
       \end{array} \right] \mbox{ and }
U^TU = \left[ \begin{array}{cc} U_{11}^T & 0 \\ U_{12}^T & U_{22}^T
       \end{array} \right]
       \left[ \begin{array}{cc} U_{11} & U_{12} \\ 0 & U_{22}
       \end{array} \right] \\ \\
       \mbox{and the Hermitian case:} \\
LL^H = \left[ \begin{array}{cc} L_{11} & 0 \\ L_{21} & L_{22}
       \end{array} \right]
       \left[ \begin{array}{cc} L^H_{11} & L^H_{21} \\ 0 & L^H_{22}
       \end{array} \right] \mbox{ and }
U^HU = \left[ \begin{array}{cc} U_{11}^H & 0 \\ U_{12}^H & U_{22}^H
       \end{array} \right]
       \left[ \begin{array}{cc} U_{11} & U_{12} \\ 0 & U_{22}
       \end{array} \right]
       \end{array} \label{equ:lluu}
\end{equation}

\noindent
where $L_{11}$, $L_{22}$, $U_{11}$, and $U_{22}$ are lower and upper
triangular submatrices, and $L_{21}$ and $U_{12}$  are square or almost square 
submatrices.

Using Equations~(\ref{equ:alu}) and equating the blocks of Equations~(\ref{equ:aa}) and Equations~(\ref{equ:lluu})
gives us the basis of a 
2--by--2 block algorithm for Cholesky factorization using RFPF. We can now express each 
of these four block equalities by calls to 
existing LAPACK and Level~3 BLAS routines. 
An example, see Section~\ref{algorithms}, of this is the 
three block equations is $L_{11}L_{11}^T=A_{11}$, $L_{21}L_{11}^T=A_{21}$ and $L_{21}L_{21}^T+L_{22}L_{22}=A_{22}$. 
The first and second of these block equations are handled by calling LAPACK's POTRF routine $L_{11} \leftarrow A_{11}$ 
and by calling Level~3 BLAS TRSM via $L_{21} \leftarrow L_{21}L_{11}^{-T}$.
In both these block 
equations the Fortran equality of replacement ($\leftarrow$) is being used so that the lower triangle of $A_{11}$ is
being replaced $L_{11}$ and the nearly square matrix $A_{21}$ is being replaced by $L_{21}$. The third block equation
breaks into two parts: $A_{22} \leftarrow L_{21}L_{21}^T$ and $L_{22} \leftarrow A_{22}$ 
which are handled by calling Level~3
BLAS SYRK or HERK and by calling LAPACK's POTRF routine. At this point we can use the flexibility of the LAPACK library.
In RFPF $A_{22}$ is in upper format (upper triangle) while in standard format $A_{22}$ is in lower format
(lower triangle). Due to symmetry, both formats of $A_{22}$ contain equal values. 
This flexibility allows LAPACK to accommodate both
formats. Hence, in the calls to SYRK or HERK and POTRF we set uplo = 'U' even though the rectangular matrix of SYRK and HERK 
comes from a lower triangular formulation.
     
New LAPACK like routine PFTRF
performs these four computations.
PF 
was chosen to fit with LAPACK's use of PO and PP. 
The PFTRF routine covers the Cholesky Factorization algorithm for the eight cases of the RFPF.
Section~\ref{algorithms} 
has Figure~\ref{fig:lo:ntr:odd} with four subfigures. Here we are interested 
in the first and second subfigure. The first subfigure contains 
the layouts of matrices $A$ and $A_R$. The second subfigure has the Cholesky 
factorization algorithm obtained by simple algebraic manipulations of the three block equalities obtained above.

\section{Solution}\label{sol}

In Section~\ref{fact} we obtained the 2--by--2 Cholesky factorization~(\ref{equ:lu}) of matrix $A$. 
Now, we can solve the equation $AX = B$: 
\begin{itemize}
  \item[$\bullet$] If $A$ has lower triangular format then
  \begin{equation} 
    \begin{array}{c} LY = B \mbox{ and } L^TX = Y \mbox{(in the symmetric case)} \\
                     LY = B \mbox{ and } L^HX = Y \mbox{(in the Hermitian case)}
    \end{array} \label{equ:sol:l}\end{equation}
  \item[$\bullet$] If $A$ has an upper triangular format then
  \begin{equation} 
    \begin{array}{c} U^TY = B \mbox{ and } UX = Y \mbox{(in the symmetric case)} \\
              U^HY = B \mbox{ and } UX = Y \mbox{(in the Hermitian case)}
    \end{array} \label{equ:sol:u}\end{equation}
\end{itemize}

$B$, $X$ and $Y$ are either vectors or rectangular matrices. $B$ contains the
RHS values. $X$ and $Y$ contain the solution values. $B$, $X$ and $Y$ are vectors when there is
one RHS and matrices when there are many RHS. The values of $X$ and $Y$
are stored over the values of $B$.

Expanding~(\ref{equ:sol:l}) and~(\ref{equ:sol:u}) using~(\ref{equ:lu}) gives the forward
substitution equations
\begin{equation} 
  \begin{array}{c}
       \mbox{in the symmetric case:} \\
  \left[ \begin{array}{cc} L_{11} & 0 \\ L_{21} & L_{22} \end{array} \right]
  \left[ \begin{array}{c} Y_1 \\ Y_2 \end{array} \right] =
  \left[ \begin{array}{c} B_1 \\ B_2 \end{array} \right] \mbox{ and }
  \left[ \begin{array}{cc} U_{11}^T & 0 \\ U_{12}^T & U_{22}^T \end{array} \right]
  \left[ \begin{array}{c} Y_1 \\ Y_2 \end{array} \right] =
  \left[ \begin{array}{c} B_1 \\ B_2 \end{array} \right] \\ \\
       \mbox{and in the Hermitian case:} \\
  \left[ \begin{array}{cc} L_{11} & 0 \\ L_{21} & L_{22} \end{array} \right]
  \left[ \begin{array}{c} Y_1 \\ Y_2 \end{array} \right] =
  \left[ \begin{array}{c} B_1 \\ B_2 \end{array} \right] \mbox{ and }
  \left[ \begin{array}{cc} U_{11}^H & 0 \\ U_{12}^H & U_{22}^H \end{array} \right]
  \left[ \begin{array}{c} Y_1 \\ Y_2 \end{array} \right] =
  \left[ \begin{array}{c} B_1 \\ B_2 \end{array} \right]
  \end{array}, \label{equ:sol:forw}
\end{equation}
and the back substitution equations
\begin{equation} 
  \begin{array}{c}
       \mbox{in the symmetric case:} \\
  \left[ \begin{array}{cc} L_{11}^T & L_{21}^T \\ 0 & L_{22}^T \end{array} \right]
  \left[ \begin{array}{c} X_1 \\ X_2 \end{array} \right] =
  \left[ \begin{array}{c} Y_1 \\ Y_2 \end{array} \right] \mbox{ and }
  \left[ \begin{array}{cc} U_{11} & U_{12} \\ 0 & U_{22} \end{array} \right]
  \left[ \begin{array}{c} X_1 \\ X_2 \end{array} \right] =
  \left[ \begin{array}{c} Y_1 \\ Y_2 \end{array} \right] \\ \\
       \mbox{and in the Hermitian case:} \\
  \left[ \begin{array}{cc} L_{11}^H & L_{21}^H \\ 0 & L_{22}^H \end{array} \right]
  \left[ \begin{array}{c} X_1 \\ X_2 \end{array} \right] =
  \left[ \begin{array}{c} Y_1 \\ Y_2 \end{array} \right] \mbox{ and }
  \left[ \begin{array}{cc} U_{11} & U_{12} \\ 0 & U_{22} \end{array} \right]
  \left[ \begin{array}{c} X_1 \\ X_2 \end{array} \right] =
  \left[ \begin{array}{c} Y_1 \\ Y_2 \end{array} \right].\label{equ:sol:back}
  \end{array}. 
\end{equation}

The Equations~(\ref{equ:sol:forw}) and~(\ref{equ:sol:back}) gives the basis of a
$2 \times 2$ block algorithm for Cholesky solution using RFPF format. We can now express these
two sets of two block equalities by using existing Level~3 BLAS routines.
An example, see Section~\ref{algorithms}, of the first set of these two block equalities is
$L_{11}Y_1=B_1$ and $L_{21}Y_1+L_{22}Y_2=B_2$.
The first block equality is handled by Level~3 BLAS TRSM: $Y_1 \leftarrow L_{11}^{-1}B_1$. The second block equality is
handled by Level~3 BLAS GEMM and TRSM: $B_2 \leftarrow B_2-L_{21}Y_1$ and $Y_2 \leftarrow L_{22}^{-1}Y_2$.
The backsolution routines are similarly derived. One gets  $X_2 \leftarrow L_{22}^{-T}Y_2$,
$Y_1 \leftarrow Y_1-L_{21}^TX_2$ and $X_1 \leftarrow L_{11}^{-T}Y_1$.

New LAPACK like routine PFTRS performs these two solution computations for the eight cases of RFPF. PFTRS calls a new
Level~3 BLAS TFSM in the same way that POTRS calls TRSM.
The third subfigure in Section~\ref{algorithms} gives the Cholesky solution 
algorithm using RFPF obtained by simple 
algebraic manipulation of the block Equations~(\ref{equ:sol:forw}) 
and~(\ref{equ:sol:back}).


\section{Inversion}\label{inv}

Following LAPACK we consider the following three stage procedure:

\begin{enumerate}
  \item Factorize the matrix $A$ and overwrite $A$ with either $L$ or $U$ by calling PFTRF; see Section~\ref{fact}.
  \item Compute the inverse of either $L$ or $U$. Call these matrices $W$ or $V$ and overwrite either $L$ or $U$ with them.
This is done by calling new routine new LAPACK like TFTRI.
  \item Calculate either the product $W^TW$ or $VV^T$ and overwrite either $W$ or $V$ with them.
\end{enumerate}

As in Sections~\ref{fact} and~\ref{sol} we examine 2--by--2 block algorithms for the steps two and three above.
In Section~\ref{fact} we obtain either matrices $L$ or $U$ in RFPF. 
Like LAPACK inversion algorithms for POTRI and PPTRI, this is our
starting point for our LAPACK inversion algorithm using RFPF. The LAPACK inversion algorithms for POTRI and PPTRI
also follow from steps two and three above by first calling in the full case LAPACK TRTRI
and then calling LAPACK LAUUM. 

Take the inverse of Equation~(\ref{equ:alu}) and obtain        
\begin{equation} 
\begin{array}{c} A^{-1} = W^TW \mbox{ or } A^{-1} = VV^T \mbox{(in the symmetric case)} \\
                 A^{-1} = W^HW \mbox{ or } A^{-1} = VV^H \mbox{(in the Hermitian case)} 
\end{array}\label{equ:alum1} \end{equation} 
where $W$ and $V$ are lower and upper triangular matrices.

Using the 2--by--2 blocking for either $L$ or $U$ in Equation (\ref{equ:lu}) 
we obtain the following 2--by--2 blocking for $W$ and $V$:

\begin{equation}
W = \left[ \begin{array}{cc} W_{11} & 0 \\
    W_{21} & W_{22} \end{array} \right] \mbox{ and }
V = \left[ \begin{array}{cc} V_{11} & V_{12} \\
    0 & V_{22} \end{array} \right] \label{equ:wv}
\end{equation}

\noindent

From the identities $WL=LW=I$ and $VU=UV=I$ and the 2--by--2 block layouts of Equations (\ref{equ:lu})
and \ref{equ:lu}), we obtain
three block equations for $W$ and $V$ which can be solved using LAPACK routines for TRTRI and Level~3 BLAS TRMM. 
An example, see Figure 4, 
of these three block equations is $L_{11}W_{11}=I$, $L_{21}W_{11}+L_{22}W_{21}=0$ and $L_{22}W_{22}=I$. 
The first and third of these block equations are handled by LAPACK TRTRI routines as $W_{11} \leftarrow L_{11}^{-1}$ and
$V_{22} \leftarrow U_{22}^{-1}$. In the second inverse computation we use the fact that $L_{22}$ is equally represented by
it transpose $L_{22}^T$ which is $U_{22}$ in RFPF. 
The second block equation leads to
two calls to Level~3 BLAS TRMM via $L_{21} \leftarrow -L_{21}W_{11}$ and $W_{21}=W_{22}L_{21}$. In the last two block 
equations the Fortran equality of replacement ($\leftarrow$) is being used so that $W_{21}=-W_{22}L_{21}W_{11}$ is
replacing $L_{21}$. 

Now we turn to part three of the three stage LAPACK procedure above. For this we use the 2--by--2 blocks layouts of 
Equation (\ref{equ:wv}) and the matrix multiplications indicated by following block Equations (\ref{equ:lluum1}) giving
\begin{equation}
\begin{array}{c}
       \mbox{the symmetric case:} \\
W^TW = \left[ \begin{array}{cc} W^T_{11} & W^T_{21} \\ 0 &W^T_{22}
       \end{array} \right]
       \left[ \begin{array}{cc} W_{11} & 0 \\ W_{21} & W_{22}
       \end{array} \right] \mbox{ and }
VV^T = \left[ \begin{array}{cc} V_{11} & V_{12} \\ 0 & V_{22}
       \end{array} \right]
       \left[ \begin{array}{cc} V^T_{11} & 0 \\ V^T_{12} & V^T_{22}
       \end{array} \right] \\ \\
       \mbox{and the Hermitian case:} \\
W^HW = \left[ \begin{array}{cc} W^H_{11} & W^H_{21} \\ 0 &W^H_{22}
       \end{array} \right]
       \left[ \begin{array}{cc} W_{11} & 0 \\ W_{21} & W_{22}
       \end{array} \right] \mbox{ and }
VV^H = \left[ \begin{array}{cc} V_{11} & V_{12} \\ 0 & V_{22}
       \end{array} \right]
       \left[ \begin{array}{cc} V^H_{11} & 0 \\ V^H_{12} & V^H_{22}
       \end{array} \right]
       \end{array} \label{equ:lluum1}
\end{equation}
where $W_{11}$, $W_{22}$, $V_{11}$, and $V_{22}$ are lower and upper
triangular submatrices, and $W_{21}$ and $V_{12}$  are square or almost square 
submatrices. The values of the indicated block multiplications of $W$ or $V$ in Equation (\ref{equ:lluum1}) are stored 
over the block values of $W$ or $V$.  

Performing the indicated 2--by--2 block multiplications of Equation (\ref{equ:lluum1}) 
leads to three block matrix computations.
An example, see Section~\ref{algorithms}, of these three block computations 
is $W^T_{11}W_{11}+W^T_{21}W_{21}$,  $W^T_{22}W_{21}$
and $W^T_{22}W_{22}$. Additionally, we want to overwrite the values of these 
block multiplications on their original block operands. 
Block operand $W_{11}$ only occurs in the (1,1) block operand computation and 
hence can be overwritten by a call to LAPACK LAUUM, 
$W_{11} \leftarrow W^T_{11}W_{11}$, followed by a call to Level~3 BLAS SYRK or HERK, 
$W_{11} \leftarrow W_{11}+W^T_{21}W_{21}$. 
Block operand $W_{21}$ now only occurs in the (2,1) block computation and
hence can be overwritten by a call to Level~3 BLAS TRMM, $W_{21} \leftarrow W^T_{22}W_{21}$. 
Finally, block operand $W_{22}$ can be overwritten by a call to LAPACK LAUUM, $W_{22} \leftarrow W^T_{22}W_{22}$.  

The fourth subfigure in Section~\ref{algorithms}  has the Cholesky inversion 
algorithms using RFPF based on the results of this Section. New LAPACK routine, PFTRI, performs this computation
for the eight cases of RFPF.

\section{RFP Data Formats and Algorithms}\label{algorithms}

This section contains three figures. 
\begin{enumerate}
  \item The first figure describes the RFPF (Rectangular Full Packed Format)
and gives algorithms for Cholesky factorization, solution and
inversion of symmetric positive definite matrices, where $N$ is odd,
{\tt uplo} = {\tt 'lower'}, and {\tt trans} = {\tt 'no transpose'}. This figure has four
subfigures.
  \begin{enumerate}
    \item The first subfigure depicts the lower triangle of a 
symmetric positive definite matrix $A$ in {\bf standard full}
and its representation by the matrix $A_R$ in {\bf RFPF}. 
    \item The second subfigure gives the RFPF Cholesky factorization
algorithm and its calling sequences of the LAPACK and BLAS subroutines, 
see Section~\ref{fact}.
    \item The third subfigure gives the RFPF Cholesky solution
algorithm and its calling sequences to the LAPACK and BLAS subroutines,
see Section~\ref{sol}.
    \item The fourth subfigure in each figure gives the RFPF Cholesky inversion
algorithm and its calling sequences to the LAPACK and BLAS subroutines, 
see Section~\ref{inv}.
  \end{enumerate}
  \item The second figure shows the transformation from full to RFPF of all ``no transform'' cases.
  \item The third figure depicts all eight cases in RFPF.
\end{enumerate} 

The data format for $A$ has ${\tt LDA}=N$.  
Matrix $A_R$ has ${\tt LDAR} = N$ if $N$ is odd and ${\tt LDAR} = N+1$ if $N$ is even and 
$n1$ columns where $n1= \lceil N/2 \rceil$. Hence, matrix $A_R$ always has {\tt LDAR} rows and $n1$ columns.
Matrix $A_R^T$ always has $n1$ rows and {\tt LDAR} columns and its leading dimension is equal to $n1$.   
Matrix $A_R$ always has ${\tt LDAR} \times n1=NT=N(N+1)/2$ elements as does matrix $A_R^T$.

The order $N$ of matrix $A$ in the first figure is seven and six or seven in the remaining two figures.


\begin{figure}[ph]
\caption{\bf The Cholesky factorization algorithm using the Rectangular Full
 Packed Format (RFPF) if $N$ is odd,
{\tt uplo} = {\tt 'lower'}, and {\tt trans} = {\tt 'no transpose'}.}
 \label{fig:lo:ntr:odd}
{\bf \footnotesize
$$
\begin{array}{c}
\begin{array}{c} \mbox{$A$ of LAPACK full data format} \\
   \mbox{LDA=N} = 7, \mbox{memory needed} \\ \mbox{LDA}\times \mbox{N} = 49 \\
   \left( \begin{array}{cccc|ccc}
      a_{1,1_{1}}  & \diamond      & \diamond      & \diamond      & \diamond
   & \diamond      & \diamond \\
      a_{2,1_{2}}  & a_{2,2_{9}}   & \diamond      & \diamond      & \diamond
   & \diamond      & \diamond \\
      a_{3,1_{3}}  & a_{3,2_{10}}  & a_{3,3_{17}}  & \diamond      & \diamond
   & \diamond      & \diamond \\
      a_{4,1_{4}}  & a_{4,2_{11}}  & a_{4,3_{18}}  & a_{4,4_{25}}  & \diamond
   & \diamond      & \diamond \\
      a_{5,1_{5}}  & a_{5,2_{12}}  & a_{5,3_{19}}  & a_{5,4_{26}}  & a_{5,5_{33}
}  & \diamond      & \diamond \\
      a_{6,1_{6}}  & a_{6,2_{13}}  & a_{6,3_{20}}  & a_{6,4_{27}}  & a_{6,5_{34}
}  & a_{6,6_{41}}  & \diamond \\
      a_{7,1_{7}}  & a_{7,2_{14}}  & a_{7,3_{21}}  & a_{7,4_{28}}  & a_{7,5_{35}
}  & a_{7,6_{42}}  & a_{7,7_{49}}
     \end{array} \right) \\ \mbox{Matrix } $A$
\end{array} \hspace{0.25 cm}
\begin{array}{c} \mbox{$A_R$ of Rectangular full packed} \\
   \mbox{LDAR=N} = 7, \mbox{memory needed} \\ \mbox{LDAR}\times \mbox{n1} = 28 \\
   \left( \begin{array}{ccccccc}
     a_{1,1_{1}} & \multicolumn{1}{|c}{\color{red}a_{5,5_{8}}} & \color{red}a_{6,5_{15}} & \color{red}a_{7,5_{22}} \\ \cline{2-2}
     a_{2,1_{2}} & a_{2,2_{9}} & \multicolumn{1}{|c}{\color{red}a_{6,6_{16}}} & \color{red}a_{7,6_{23}} \\ \cline{3-3}
     a_{3,1_{3}} & a_{3,2_{10}} & a_{3,3_{17}} & \multicolumn{1}{|c}{\color{red}a_{7,7_{24}}} \\ \cline {4-4}
     a_{4,1_{4}} & a_{4,2_{11}} & a_{4,3_{18}} & a_{4,4_{25}} \\ \cline{1-4}
     \color{blue}a_{5,1_{5}} & \color{blue}a_{5,2_{12}} & \color{blue}a_{5,3_{19}} & \color{blue}a_{5,4_{26}} \\
     \color{blue}a_{6,1_{6}} & \color{blue}a_{6,2_{13}} & \color{blue}a_{6,3_{20}} & \color{blue}a_{6,4_{27}} \\
     \color{blue}a_{7,1_{7}} & \color{blue}a_{7,2_{14}} & \color{blue}a_{7,3_{21}} & \color{blue}a_{7,4_{28}}
   \end{array} \right) \\ \mbox{Matrix $A_{R}$}
\end{array} \\ \\ \vspace{0.25 cm}
\begin{array}{c} \mbox{Cholesky Factorization Algorithm } (n1=\lceil N/2 \rceil,n2=N-n1): \\
  \begin{array}{l}
        \mbox{1) factor } L_{11}L^T_{11} = A_{11}; \\
        \mbox{~~~ call~POTRF}('L',n1,AR,N,\& \\
        \mbox{~~~~~} info); \\
        \mbox{2) solve } L_{21}L^T_{11} = A_{21}; \\
        \mbox{~~~ call TRSM(}'R','L','T','N',n2,\& \\
        \mbox{~~~~      } n1,one,AR,N,AR(n1+1,1),N);
  \end{array}\hspace{0.15 cm}
  \begin{array}{l}
        \mbox{3) update } A_{22}:= A_{22} - L_{21}L^T_{21}; \\
        \mbox{~~~ call~SYRK/HERK}(`U','N',n2,n1, \& \\
        \mbox{~~~~~} -one,AR(n1+1,1),N,one,AR(1,2),N); \\
        \mbox{4) factor } U^T_{22}U_{22} = A_{22}; \\
        \mbox{~~~ call~POTRF}(`U',n2,AR(1,2),N,\& \\
        \mbox{~~~~~} info);
  \end{array}
\end{array} \\ \\ \vspace{0.25 cm}
\begin{array}{c} \mbox{Cholesky Solution Algorithm,} \\
                 \mbox{where $B(LDB,nr)$ and $LDB \ge N$ (here $LDB=N$)}: \\
  \begin{array}{l}
        \multicolumn{1}{c} {LY~=~B} \\
        \mbox{1) solve } L_{11}Y_1 = B_1; \\
        \mbox{~~~ call~TRSM}('L','L','N','N',n1, \& \\
        \mbox{~~~~~} nr,one,AR,N,B,N); \\
        \mbox{2) Multiply } B_2 = B_2 - L_{21}Y_1; \\
        \mbox{~~~ call GEMM(}'N','N',n2,nr,n1,-one, \& \\
        \mbox{~~~~      } AR(n1+1,1),N,B,N,one, \& \\
        \mbox{~~~~      } B(n1+1,1),N); \\
        \mbox{3) solve } L_{22}Y_2 = B_2; \\
        \mbox{~~~ call~TRSM}('L','U','T','N',n2, \& \\
        \mbox{~~~~~} nr,one,AR(1,2),N,B(n1+1,1),N);
  \end{array}\hspace{0.15 cm}
  \begin{array}{l}
        \multicolumn{1}{c} {L'X~=~Y} \\
        \mbox{1) solve } L_{22}^TX_2 = Y_2; \\
        \mbox{~~~ call~TRSM}('L','U','N','N',n2, \& \\
        \mbox{~~~~~} nr,one,AR(1,2),N,B(n1+1,1),N); \\
        \mbox{2) Multiply } Y_1 = Y_1 - L_{21}^TX_2; \\
        \mbox{~~~ call GEMM(}'T','N',n1,nr,n2,-one, \& \\
        \mbox{~~~~      } AR(n1+1,1),N,B(n1+1,1), \& \\
        \mbox{~~~~      } N,one,B,N); \\
        \mbox{3) solve } L_{22}^TX_1 = Y_1; \\
        \mbox{~~~ call~TRSM}('L','L','T','N',n1, \& \\
        \mbox{~~~~~} nr,one,AR,N,B,N);
  \end{array}
\end{array} \\ \\ \vspace{0.25 cm}
\begin{array}{c} \mbox{Cholesky Inversion Algorithm } : \\
  \begin{array}{l}
        \multicolumn{1}{c} {\mbox{Inversion}} \\
        \mbox{1) invert  } W_{11}=L_{11}^{-1}; \\
        \mbox{~~~ call~TRTRI}('L','N',n1,AR,N,info); \\
        \mbox{2) Multiply } L_{21} = -L_{21}W_{11}; \\
        \mbox{~~~ call~TRMM}('R','L','N','N',n2, \& \\
        \mbox{~~~~      } n1,-one,AR,N,AR(n1+1,1),N); \\
        \mbox{3) invert} V_{22}=U_{22}^{-1}; \\
        \mbox{~~~ call~TRTRI}('U','N',n2,AR(1,2), \& \\
        \mbox{~~~~      } N,info); \\
        \mbox{4) invert  } V_{22}=U_{22}^{-1}; \\
        \mbox{~~~ call~TRMM}('L','U','T','N',n2, \& \\
        \mbox{~~~~      } n1,one,AR(1,2),N,AR(n1+1,1),N);
  \end{array}\hspace{0.15 cm}
  \begin{array}{l}
        \multicolumn{1}{c} {\mbox{Triangular matrix multiplication}} \\
        \mbox{1) Triang. Mult. } W_{11}= W_{11}^TW_{11}; \\
        \mbox{~~~ call~LAUUM}('L',n1,AR,N,info); \\
        \mbox{2) update } W_{11}=W_{11}+ W_{21}^TW_{21}; \\
        \mbox{~~~ call~SYRK/HERK}('L','T',n1,n2, \& \\
        \mbox{~~~~      } one,AR(n1+1,1),N,one,AR,N); \\
        \mbox{3) Multiply } W_{21}=V_{22}W_{21}; \\
        \mbox{~~~ call~TRMM}('L','U','N','N',n2, \& \\
        \mbox{~~~~      } n1,one,AR(1,2),N,A(n1+1,1),N); \\
        \mbox{4) Triang. Mult. } V_{11}=V_{11}V_{11}^T; \\
        \mbox{~~~ call~LAUUM}('U',n2,AR(1,2),N,info);
  \end{array}
\end{array}
\end{array}
$$
}
\end{figure}


\begin{figure}[]
{\bf \small
\caption{\bf Eight two-dimensional arrays for storing the matrices $A$ and $A_R$ that are
needed by the LAPACK subroutine POTRF (full format)
and PFTRF RFPF respectively. The leading dimension LDA is $N$ for 
LAPACK, and LDAR for RFPF. LDAR = $N$ for $N$ odd, and $N+1$ for $N$ even. Here $N$ is
$7$ or $6$. The memory needed is 
$LDA{\times}N$ for full format and $LDAR \times n1=(N+1)N/2$ for RFPF 
Here $49$ and $36$ for full format and 28 and 21 for RFPF. 
The column size of RFPF is $n1=\lceil N/2\rceil$, here 4 and 3. }
\label{fig:full:rfp} \vspace{-0.20 cm} 
$$
\begin{array}{c}
  \mbox{\ref{fig:full:rfp}.1 The matrices $A$ of order N and $A_R$ of size $LDAR$ by $n1$, }
  \mbox{here $N=7$.} \\ \vspace{0.20 cm} 
\begin{array}{c} \mbox{\ref{fig:full:rfp}.1.1 Full Format} \\
   \left[ \begin{array}{cccc|ccc}
      a_{1,1}  & \diamond      & \diamond      & \diamond      & \diamond & \diamond      & \diamond \\
      a_{2,1}  & a_{2,2}   & \diamond      & \diamond      & \diamond & \diamond      & \diamond \\
      a_{3,1}  & a_{3,2}  & a_{3,3}  & \diamond      & \diamond & \diamond      & \diamond \\
      a_{4,1}  & a_{4,2}  & a_{4,3}  & a_{4,4}  & \diamond & \diamond      & \diamond \\ \cline{1-4}
      \color{blue}a_{5,1}  & \color{blue}a_{5,2}  & \color{blue}a_{5,3}  & \color{blue}a_{5,4}  & \color{red}a_{5,5}  & \diamond      & \diamond \\
      \color{blue}a_{6,1}  & \color{blue}a_{6,2}  & \color{blue}a_{6,3}  & \color{blue}a_{6,4}  & \color{red}a_{6,5}  & \color{red}a_{6,6}  & \diamond \\
      \color{blue}a_{7,1}  & \color{blue}a_{7,2}  & \color{blue}a_{7,3}  & \color{blue}a_{7,4}  & \color{red}a_{7,5}  & \color{red}a_{7,6}  & \color{red}a_{7,7} \\
     \end{array} \right],
\end{array} \hspace{0.25 cm}
\begin{array}{c} \mbox{\ref{fig:full:rfp}.1.2 RFPF} \\
   \left[ \begin{array}{ccccccc}
      a_{1,1}  & \multicolumn{1}{|c}{\color{red}a_{5,5}}   & \color{red}a_{6,5} & \color{red}a_{7,5}  \\ \cline{2-2}
      a_{2,1}  & a_{2,2}   & \multicolumn{1}{|c}{\color{red}a_{6,6}} & \color{red}a_{7,6}  \\ \cline{3-3}
      a_{3,1}  & a_{3,2}  & a_{3,3}  & \multicolumn{1}{|c}{\color{red}a_{7,7}}  \\ \cline {4-4}
      a_{4,1}  & a_{4,2}  & a_{4,3}  & a_{4,4}  \\ \cline{1-4}
      \color{blue}a_{5,1}  & \color{blue}a_{5,2}  & \color{blue}a_{5,3}  & \color{blue}a_{5,4}  \\
      \color{blue}a_{6,1}  & \color{blue}a_{6,2}  & \color{blue}a_{6,3}  & \color{blue}a_{6,4}  \\
      \color{blue}a_{7,1}  & \color{blue}a_{7,2}  & \color{blue}a_{7,3}  & \color{blue}a_{7,4}
     \end{array} \right]
\end{array} \\ \vspace{0.20 cm}
  \mbox{\ref{fig:full:rfp}.2 The matrices $A$ of order N and $A_R$ of size $LDAR$ by $n1$, }
  \mbox{here $N=6$.} \\
\begin{array}{c} \mbox{\ref{fig:full:rfp}.2.1 Full format} \\
   \left[ \begin{array}{ccc|cccc}
      a_{1,1}  & \diamond      & \diamond      & \diamond      & \diamond & \diamond \\
      a_{2,1}  & a_{2,2}   & \diamond      & \diamond      & \diamond & \diamond \\
      a_{3,1}  & a_{3,2}  & a_{3,3}  & \diamond      & \diamond & \diamond \\ \cline{1-3}
      \color{blue}a_{4,1}  & \color{blue}a_{4,2}  & \color{blue}a_{4,3}  & \color{red}a_{4,4}  & \diamond & \diamond \\
      \color{blue}a_{5,1}  & \color{blue}a_{5,2}  & \color{blue}a_{5,3}  & \color{red}a_{5,4}  & \color{red}a_{5,5}  & \diamond \\
      \color{blue}a_{6,1}  & \color{blue}a_{6,2}  & \color{blue}a_{6,3}  & \color{red}a_{6,4}  & \color{red}a_{6,5}  & \color{red}a_{6,6} \\
     \end{array} \right],
\end{array} \hspace{0.25 cm}
\begin{array}{c} \mbox{\ref{fig:full:rfp}.2.2 RFPF} \\
   \left[ \begin{array}{ccccccc}
      \color{red}a_{4,4}  & \color{red}a_{5,4}   & \color{red}a_{6,4}  \\ \cline{1-1}
      a_{1,1}  & \multicolumn{1}{|c}{\color{red}a_{5,5}}   & \color{red}a_{6,5}  \\ \cline{2-2}
      a_{2,1}  & a_{2,2}  & \multicolumn{1}{|c}{\color{red}a_{6,6}}  \\ \cline{3-3}
      a_{3,1}  & a_{3,2}  & a_{3,3}  \\ \cline{1-3}
      \color{blue}a_{4,1}  & \color{blue}a_{4,2}  & \color{blue}a_{4,3}  \\
      \color{blue}a_{5,1}  & \color{blue}a_{5,2}  & \color{blue}a_{5,3}  \\
      \color{blue}a_{6,1}  & \color{blue}a_{6,2}  & \color{blue}a_{6,3}
     \end{array} \right]
\end{array} \\ \vspace{0.20 cm}
  \mbox{\ref{fig:full:rfp}.3 The matrices $A$ of order N and $A_R$ of size $LDAR$ by $n1$, }
  \mbox{here $N=7$.} \\
\begin{array}{c} \mbox{\ref{fig:full:rfp}.3.1 Full format} \\
   \left[ \begin{array}{ccc|cccc}
      \color{red}a_{1,1} & \color{red}a_{1,2} & \color{red}a_{1,3} & \color{blue}a_{1,4} & \color{blue}a_{1,5} & \color{blue}a_{1,6} & \color{blue}a_{1,7} \\
      \diamond & \color{red}a_{2,2} & \color{red}a_{2,3} & \color{blue}a_{2,4} & \color{blue}a_{2,5} & \color{blue}a_{2,6} & \color{blue}a_{2,7} \\
      \diamond & \diamond & \color{red}a_{3,3} & \color{blue}a_{3,4} & \color{blue}a_{3,5} & \color{blue}a_{3,6} \color{blue}& \color{blue}a_{3,7} \\ \cline{4-7}
      \diamond & \diamond & \diamond & a_{4,4} & a_{4,5} & a_{4,6} & a_{4,7} \\
      \diamond & \diamond & \diamond & \diamond & a_{5,5}  & a_{5,6} & a_{5,7} \\
      \diamond & \diamond & \diamond & \diamond & \diamond & a_{6,6} & a_{6,7} \\
      \diamond & \diamond & \diamond & \diamond & \diamond & \diamond & a_{7,7} \\
     \end{array} \right],
\end{array} \hspace{0.25 cm}
\begin{array}{c} \mbox{\ref{fig:full:rfp}.3.2 RFPF} \\
  \left[ \begin{array}{cccc}
    \color{blue}a_{1,4} & \color{blue}a_{1,5} & \color{blue}a_{1,6} & \color{blue}a_{1,7} \\
    \color{blue}a_{2,4} & \color{blue}a_{2,5} & \color{blue}a_{2,6} & \color{blue}a_{2,7} \\
    \color{blue}a_{3,4} & \color{blue}a_{3,5} & \color{blue}a_{3,6} & \color{blue}a_{3,7} \\ \hline
    a_{4,4} & a_{4,5} & a_{4,6} & a_{4,7} \\ \cline{1-1}
    \multicolumn{1}{c|}{\color{red}a_{1,1}}  & a_{5,5} & a_{5,6} & a_{5,7} \\ \cline{2-2}
    \color{red}a_{1,2} & \multicolumn{1}{c|}{\color{red}a_{2,2}} & a_{6,6} & a_{6,7} \\ \cline{3-3}
    \color{red}a_{1,3} & \color{red}a_{2,3} & \multicolumn{1}{c|}{\color{red}a_{3,3}} & a_{7,7}
     \end{array} \right]
\end{array} \\ \vspace{0.20 cm}
  \mbox{\ref{fig:full:rfp}.4 The matrices $A$ of order N and $A_R$ of size $LDAR$ by $n1$, }
  \mbox{here $N=6$.} \\
\begin{array}{c} \mbox{\ref{fig:full:rfp}.4.1 Full format} \\
   \left[ \begin{array}{ccc|cccc}
      \color{red}a_{1,1} & \color{red}a_{1,2} & \color{red}a_{1,3} & \color{blue}a_{1,4} & \color{blue}a_{1,5} & \color{blue}a_{1,6} \\
      \diamond & \color{red}a_{2,2} & \color{red}a_{2,3} & \color{blue}a_{2,4} & \color{blue}a_{2,5} & \color{blue}a_{2,6} \\
      \diamond & \diamond & \color{red}a_{3,3} & \color{blue}a_{3,4} & \color{blue}a_{3,5} & \color{blue}a_{3,6} \color{blue}\\ \cline{4-6}
      \diamond & \diamond & \diamond & a_{4,4} & a_{4,5} & a_{4,6} \\
      \diamond & \diamond & \diamond & \diamond & a_{5,5}  & a_{5,6} \\
      \diamond & \diamond & \diamond & \diamond & \diamond & a_{6,6} \\
     \end{array} \right],
\end{array} \hspace{0.25 cm}
\begin{array}{c} \mbox{\ref{fig:full:rfp}.4.2 RFPF} \\
   \left[ \begin{array}{cccc|ccc}
     \color{blue}a_{1,4} & \color{blue}a_{1,5} & \color{blue}a_{1,6} \\
     \color{blue}a_{2,4} & \color{blue}a_{2,5} & \color{blue}a_{2,6} \\
     \color{blue}a_{3,4} & \color{blue}a_{3,5} & \color{blue}a_{3,6} \\ \hline
     a_{4,4} & a_{4,5_{11}} & a_{4,6} \\ \cline{1-1}
     \multicolumn{1}{c|}{\color{red}a_{1,1}}  & a_{5,5} & a_{5,6_{19}} \\ \cline{2-2}
     \color{red}a_{1,2} & \multicolumn{1}{c|}{\color{red}a_{2,2}} & a_{6,6} \\ \cline{3-3}
     \color{red}a_{1,3} & \color{red}a_{2,3} & \multicolumn{1}{c}{\color{red}a_{3,3}}
     \end{array} \right]
\end{array}
\end{array}
$$
}
\end{figure}


\begin{figure}[]
{\bf \small
\caption{\bf Eight two-dimensional arrays for storing the matrices $A_R$ and $A_R^T$ in
RFPF. The leading dimension $LDAR$ of $A_R$ is 
$N$ when $N$ is odd and $N+1$ when $N$ is even. For the matrix $A_R^T$ it
is $n1=\lceil N/2 \rceil$.
The memory needed for both $A_R$ and $A_R^T$ is 
\hbox{$(N+1)/2\times N$}. This amount is 28 for $N=7$ and 21 for $N=6$.}

\label{fig:rfp:all} \vspace{-0.20 cm}
$$
\begin{array}{c}
  \mbox{\ref{fig:rfp:all}.1 RFPF for the matrices of rank odd, here
  $N=7$ and $n1=4$} \\
\begin{array}{c}
\begin{array}{c}
   \mbox{Lower triangular} \\
\begin{array}{c}
   \mbox{$LDAR=N$} \\
   \left[ \begin{array}{ccccccc}
      a_{1,1}  & \multicolumn{1}{|c}{\color{red}a_{5,5}}   & \color{red}a_{6,5} & \color{red}a_{7,5}  \\ \cline{2-2}
      a_{2,1}  & a_{2,2}   & \multicolumn{1}{|c}{\color{red}a_{6,6}} & \color{red}a_{7,6}  \\ \cline{3-3}
      a_{3,1}  & a_{3,2}  & a_{3,3}  & \multicolumn{1}{|c}{\color{red}a_{7,7}}  \\ \cline {4-4}
      a_{4,1}  & a_{4,2}  & a_{4,3}  & a_{4,4}  \\ \cline{1-4}
      \color{blue}a_{5,1}  & \color{blue}a_{5,2}  & \color{blue}a_{5,3}  & \color{blue}a_{5,4}  \\
      \color{blue}a_{6,1}  & \color{blue}a_{6,2}  & \color{blue}a_{6,3}  & \color{blue}a_{6,4}  \\
      \color{blue}a_{7,1}  & \color{blue}a_{7,2}  & \color{blue}a_{7,3}  & \color{blue}a_{7,4}
     \end{array} \right]
\end{array}
\begin{array}{c}
  \mbox{transpose, $lda=n1$}  \\
  \left[ \begin{array}{cccc|ccc}
    a_{1,1} & a_{2,1} & a_{3,1} & a_{4,1} & \color{blue}a_{5,1} & \color{blue}a_{6,1} & \color{blue}a_{7,1} \\ \cline{1-1}
    \multicolumn{1}{c|}{\color{red}a_{5,5}} & a_{2,2} & a_{3,2} & a_{4,2} & \color{blue}a_{5,2} & \color{blue}a_{6,2} & \color{blue}a_{7,2} \\ \cline{2-2}
    \color{red}a_{6,5} & \multicolumn{1}{c|}{\color{red}a_{6,6}} & a_{3,3} & a_{4,3} & \color{blue}a_{5,3} & \color{blue}a_{6,3} & \color{blue}a_{7,3} \\ \cline{3-3}
    \color{red}a_{7,5} & \color{red}a_{7,6} & \multicolumn{1}{c|}{\color{red}a_{7,7}}& a_{4,4} & \color{blue}a_{5,4} & \color{blue}a_{6,4} & \color{blue}a_{7,4} \\
     \end{array} \right]
\end{array}
\end{array}
\end{array} \\ \vspace{0.25 cm}
\begin{array}{c}
\begin{array}{c}
   \mbox{Upper triangular} \\
\begin{array}{c}
   \mbox{$LDAR=N$} \\
  \left[ \begin{array}{cccc}
    \color{blue}a_{1,4} & \color{blue}a_{1,5} & \color{blue}a_{1,6} & \color{blue}a_{1,7} \\
    \color{blue}a_{2,4} & \color{blue}a_{2,5} & \color{blue}a_{2,6} & \color{blue}a_{2,7} \\
    \color{blue}a_{3,4} & \color{blue}a_{3,5} & \color{blue}a_{3,6} & \color{blue}a_{3,7} \\ \hline
    a_{4,4} & a_{4,5} & a_{4,6} & a_{4,7} \\ \cline{1-1}
    \multicolumn{1}{c|}{\color{red}a_{1,1}}  & a_{5,5} & a_{5,6} & a_{5,7} \\ \cline{2-2}
    \color{red}a_{1,2} & \multicolumn{1}{c|}{\color{red}a_{2,2}} & a_{6,6} & a_{6,7} \\ \cline{3-3}
    \color{red}a_{1,3} & \color{red}a_{2,3} & \multicolumn{1}{c|}{\color{red}a_{3,3}} & a_{7,7}
  \end{array} \right]
\end{array}
\begin{array}{c}
  \mbox{transpose, $lda=n1$}  \\
   \left[ \begin{array}{ccc|cccc}
     \color{blue}a_{1,4} & \color{blue}a_{2,4} & \color{blue}a_{3,4} & \multicolumn{1}{c|}{a_{4,4}} & \color{red}a_{1,1} & \color{red}a_{1,2} & \color{red}a_{1,3} \\ \cline{5-5}
     \color{blue}a_{1,5} & \color{blue}a_{2,5} & \color{blue}a_{3,5} & a_{4,5} & \multicolumn{1}{c|}{a_{5,5}} & \color{red}a_{2,2} & \color{red}a_{2,3} \\ \cline{6-6}
     \color{blue}a_{1,6} & \color{blue}a_{2,6} & \color{blue}a_{3,6} & a_{4,6} & a_{5,6} & \multicolumn{1}{c|}{a_{6,6}} & \color{red}a_{3,3} \\ \cline{7-7}
     \color{blue}a_{1,7} & \color{blue}a_{2,7} & \color{blue}a_{3,7} & a_{4,7} & a_{5,7} & a_{6,7} & a_{7,7}
     \end{array} \right]
\end{array}
\end{array}
\end{array} \\ \vspace{0.25 cm}
\begin{array}{c}
  \mbox{\ref{fig:rfp:all}.2 RFPF for the matrices of rank even, here $N=6$
   and $n1=3$.} \\
\begin{array}{c}
\begin{array}{c}
   \mbox{Lower triangular} \\
\begin{array}{c}
   \mbox{$LDAR=N+1$} \\
   \left[ \begin{array}{ccccccc}
      \color{red}a_{4,4}  & \color{red}a_{5,4}   & \color{red}a_{6,4_{15}}  \\ \cline{1-1}
      a_{1,1}  & \multicolumn{1}{|c}{\color{red}a_{5,5}}   & \color{red}a_{6,5_{16}}  \\ \cline{2-2}
      a_{2,1}  & a_{2,2}  & \multicolumn{1}{|c}{\color{red}a_{6,6}}  \\ \cline{3-3}
      a_{3,1}  & a_{3,2}  & a_{3,3}  \\ \cline{1-3}
      \color{blue}a_{4,1}  & \color{blue}a_{4,2}  & \color{blue}a_{4,3}  \\
      \color{blue}a_{5,1}  & \color{blue}a_{5,2}  & \color{blue}a_{5,3}  \\
      \color{blue}a_{6,1}  & \color{blue}a_{6,2}  & \color{blue}a_{6,3}
     \end{array} \right]
\end{array}
\begin{array}{c}
   \mbox{transpose, $lda=n1$} \\
  \left[ \begin{array}{cccc|ccc}
    \multicolumn{1}{c|}{\color{red}a_{4,4}} & a_{1,1} & a_{2,1} & a_{3,1} & \color{blue}a_{4,1} & \color{blue}a_{5,1} & \color{blue}a_{6,1} \\ \cline{2-2}
    \color{red}a_{5,4} & \multicolumn{1}{c|}{\color{red}a_{5,5}} & a_{2,2} & a_{3,2} & \color{blue}a_{4,2} & \color{blue}a_{5,2} & \color{blue}a_{6,2} \\ \cline{3-3}
    \color{red}a_{6,4} & \color{red}a_{6,5} & \multicolumn{1}{c|}{\color{red}a_{6,6}} & a_{3,3} & \color{blue}a_{4,3} & \color{blue}a_{5,3} & \color{blue}a_{6,3} \\
     \end{array} \right]
\end{array}
\end{array}
\end{array} \\ \vspace{0.25 cm}
\begin{array}{c}
   \mbox{Upper triangular} \\
\begin{array}{c}
   \mbox{$LDAR=N+1$} \\
   \left[ \begin{array}{cccc|ccc}
     \color{blue}a_{1,4} & \color{blue}a_{1,5} & \color{blue}a_{1,6} \\
     \color{blue}a_{2,4} & \color{blue}a_{2,5} & \color{blue}a_{2,6} \\
     \color{blue}a_{3,4} & \color{blue}a_{3,5} & \color{blue}a_{3,6} \\ \hline
     a_{4,4} & a_{4,5_{11}} & a_{4,6} \\ \cline{1-1}
     \multicolumn{1}{c|}{\color{red}a_{1,1}}  & a_{5,5} & a_{5,6_{19}} \\ \cline{2-2}
     \color{red}a_{1,2} & \multicolumn{1}{c|}{\color{red}a_{2,2}} & a_{6,6} \\ \cline{3-3}
     \color{red}a_{1,3} & \color{red}a_{2,3} & \multicolumn{1}{c}{\color{red}a_{3,3}}
     \end{array} \right]
\end{array}
\begin{array}{c}
\begin{array}{c}
   \mbox{transpose, $lda=n1$} \\
   \left[ \begin{array}{ccc|cccc}
     \color{blue}a_{1,4} & \color{blue}a_{2,4} & \color{blue}a_{3,4} & \multicolumn{1}{c|}{a_{4,4}} & \color{red}a_{1,1} & \color{red}a_{1,2} & \color{red}a_{1,3} \\ \cline{5-5}
     \color{blue}a_{1,5} & \color{blue}a_{2,5} & \color{blue}a_{3,5} & a_{4,5} & \multicolumn{1}{c|}{a_{5,5}} & \color{red}a_{2,2} & \color{red}a_{2,3} \\ \cline{6-6}
     \color{blue}a_{1,6} & \color{blue}a_{2,6} & \color{blue}a_{3,6} & a_{4,6} & a_{5,6} & \multicolumn{1}{c|}{a_{6,6}} & \color{red}a_{3,3}
     \end{array} \right]
\end{array}
\end{array}
\end{array}
\end{array}
\end{array}
$$
}
\end{figure}

\section{Stability of the RFPF Algorithm} \label{sec:stability}
The RFPF Cholesky factorization 
(Section~\ref{fact}), Cholesky solution (Section~\ref{sol}), and Cholesky inversion
(Section~\ref{inv}) algorithms are equivalent to the traditional
algorithms in the
books~\cite{HPCdongduffsorvorst98,demmel97,GV98,Trefethen97}. The 
whole theory of the traditional Cholesky factorization, solution, 
inversion and BLAS algorithms
carries over to this three Cholesky and BLAS
algorithms described in Sections~\ref{fact}, \ref{sol}, and~\ref{inv}.
The error analysis and stability of these algorithms is very well
described in the book of~\cite{Higham96}.
The difference between LAPACK algorithms PO, PP and
RFPF\footnote{full, packed and rectangular full packed.} is how inner
products are accumulated. In each case a different order is used.
They are all mathematically equivalent, and, stability analysis shows
that any summation order is stable.

\section{A Performance Study using RFP Format}\label{performance}


The LAPACK library~\cite{laug} routines POTRF/PPTRF, POTRI/PPTRI, and
POTRS/PPTRS are compared with the RFPF routines PFTRF, PFTRI, and PFTRS for
Cholesky factorization (PxTRF), Cholesky inverse (PxTRI) and Cholesky solution
(PxTRS) respectively.
In the previous sentence, the character 'x' can be 'O'
(full format), 'P' (packed format), or 'F' (RFPF).
In all cases long real precision arithmetic (also called double
precision) is used. Sometimes we also show results for long complex precision
(also called complex*16). Results were obtained on several different
computers using everywhere the vendor Level~3 and Level~2 BLAS.  The sequential
performance results were done on the following computers: 

\begin{description}

\item[$\bullet$ Sun Fire E25K (newton):] 72 UltraSPARC IV+ dual-core CPUs (1800
MHz/ 2 MB shared L2-cache, 32 MB shared L3-cache), 416 GB memory (120 CPUs/368
GB). Further information at ``http://www.gbar.dtu.dk/index.php/Hardware''. 


\item[$\bullet$ SGI Altix 3700 (Freke):] 64 CPUs - Intel Itanium2 1.5 GHz/6 MB
L3-cache. 256 GB memory. Peak performance: 384 GFlops.  Further information at
``http://www.cscaa.dk/freke/''.

\item[$\bullet$ Intel Tigerton computer (zoot):] quad-socket quad-core Intel
Tigerton 2.4GHz (16 total cores) with 32 GB of memory. We use Intel MKL
10.0.1.014.

\item[$\bullet$ DMI Itanium:] CPU Intel Itanium2: 1.3 GHz, cache: 3 MB on-chip
L3 cache.

\item[$\bullet$ DMI NEC SX-6 computer:] 8 CPU's, per CPU peak: 8 Gflops, per
node peak: 64 Gflops, vector register length: 256.

\end{description}

The performance results are given in Figures~\ref{fig:sun_rea} to
\ref{fig:last}. In Appendix~\ref{appendix-table}, we give the table data used
in the figures, see Tables~\ref{tab:sunreafac} to~\ref{tab:last}.  We also give
speedup numbers, see Tables~\ref{tab:sun_rea_speedup}
to~\ref{tab:zoot_speedup_vendor}.

The figures from \ref{fig:sun_rea} to \ref{fig:sgi_com} are paired.  Figure
\ref{fig:sun_rea} (double precision) and Figure \ref{fig:sun_com} (double
complex precision) present results for the Sun UltraSPARC IV+ computer.
Figure \ref{fig:sgi_rea} (double precision) and Figure \ref{fig:sgi_com}
(double complex precision) present results for the SGI Altix 3700 computer.
Figure~\ref{fig:ita} (double precision) presents results for the Intel Itanium2
computer.  Figure~\ref{fig:nec} (double precision) presents results for the
NEC SX-6 computer.  Figure~\ref{fig:zoot_reference} (double precision) presents
results for the quad-socket quad-core Intel Tigerton computer using reference
LAPACK-3.2.0 (from netlib.org).  Figure~\ref{fig:zoot_vendor} (double
precision) presents results for the quad-socket quad-core Intel Tigerton
computer using vendor LAPACK library (MKL-10.0.1.14).

Figure \ref{fig:last} shows the SMP parallelism of these subroutines on the IBM
Power4 (clock rate: 1300 MHz; two CPUs per chip; L1 cache: 128 KB (64 KB per
CPU) instruction, 64 KB 2-way (32 KB per CPU) data; L2 cache: 1.5 MB 8-way
shared between the two CPUs; L3 cache: 32 MB 8-way  shared (off-chip); TLB:
1024 entries) and SUN UltraSPARC-IV (clock rate: 1350 MHz; L1 cache: 64 kB
4-way data, 32 kB 4-way instruction, and 2 kB Write, 2 kB Prefetch; L2 cache: 8
MB; TLB: 1040 entries) computers respectively. They compare SMP times of PFTRF,
vendor POTRF and reference PPTRF.

The RFPF packed results greatly outperform the packed and more often than not are
better than the full results.  Note that our timings do {\it not} include the
cost of sorting any LAPACK data formats to RFPF data formats and vice versa.
We think that users will input their matrix data using RFPF.  Hence, this is
our rationale for not including the data transformation times.

For all our experiments, we use vendor Level 3 and Level 2 BLAS.  For all our
experiments except Figure~\ref{fig:zoot_reference}
and Figure~\ref{fig:last}, we use the provided vendor
library for LAPACK and BLAS.


We include comparisons with reference LAPACK for the quad-socket quad-core
Intel Tigerton machine in Figure~\ref{fig:zoot_reference}. In this case,
the vendor LAPACK library packed storage routines significantly
outperform the LAPACK reference implementation from netlib.  In
Figure~\ref{fig:zoot_vendor}, you find the same experiments on the same machine
but, this time, using the vendor library (MKL-10.0.1.014). We think that MKL is
using the reference implementation for Inverse Cholesky (packed and full
format). For Cholesky factorization, we see that both packed and full format
routines (PPTRF and POTRF) are tuned. But even, in this case, our RFPF storage
format results are better.

When we compare RFPF with full storage, results are mixed.  However, both codes
are rarely far apart.  Most of the performance ratios are between 0.95 to 1.05
overall.  But, note that the RFPF performance is more uniform over its
versions (four presented; the other four are for n odd ). For LAPACK full (two
versions ), the performance variation is greater. Moreover, in the case
of the inversion on quad-socket quad-core Tigerton
(Figure~\ref{fig:zoot_reference} and Figure~\ref{fig:zoot_vendor})
RFPF clearly outperforms both variants of the full format.

\begin{figure}
 \includegraphics[width=0.33\textwidth]{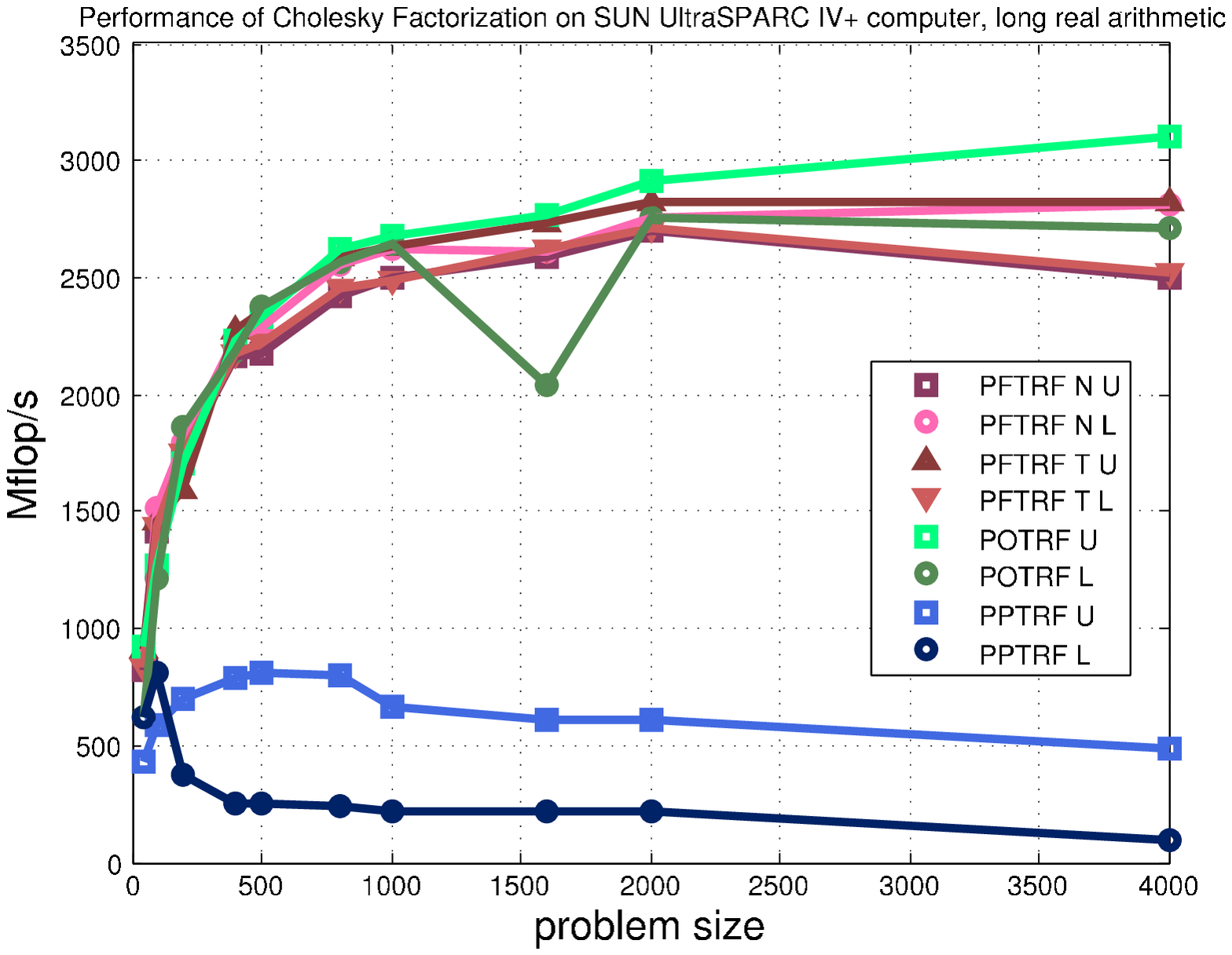}
 \includegraphics[width=0.33\textwidth]{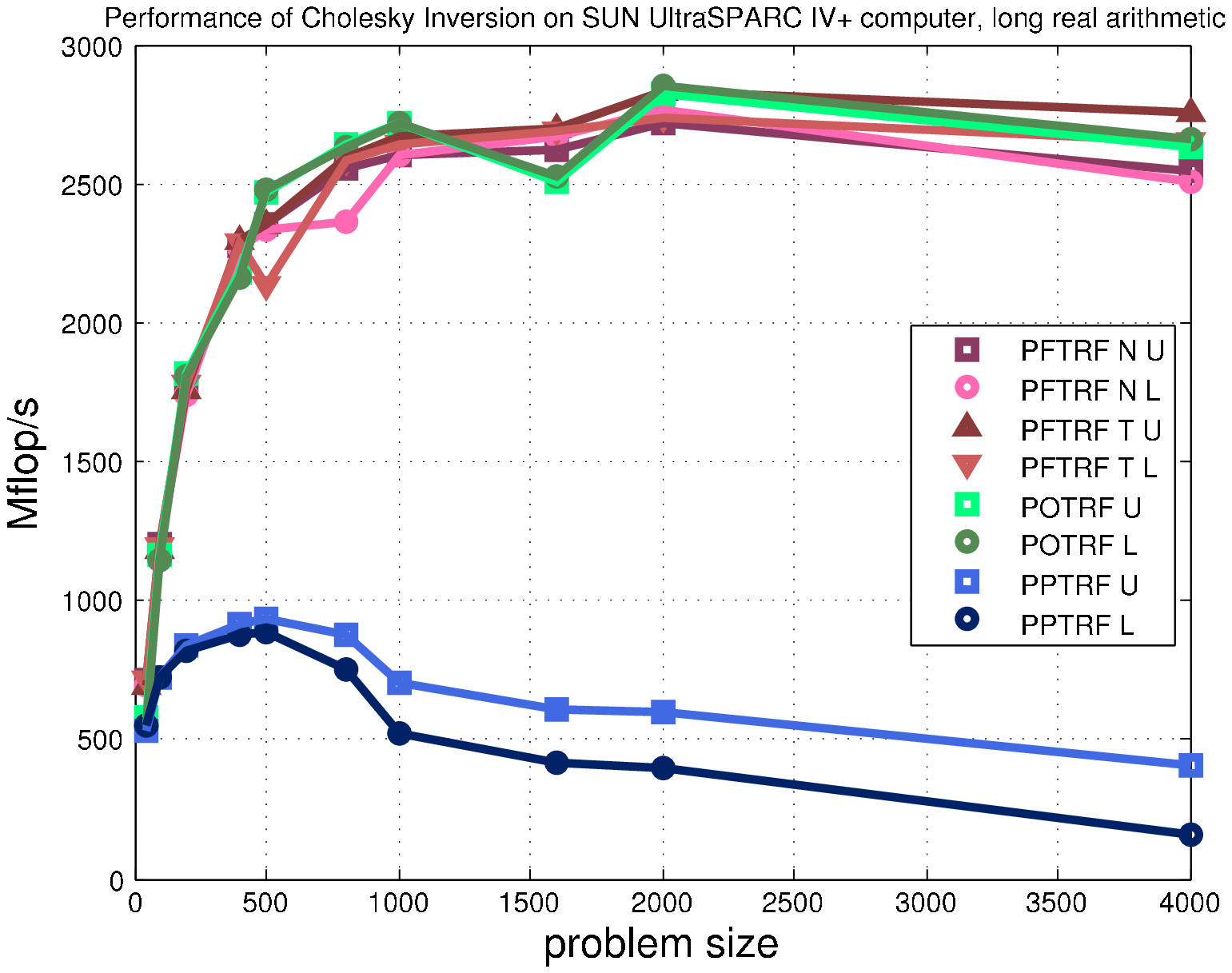}
 \includegraphics[width=0.33\textwidth]{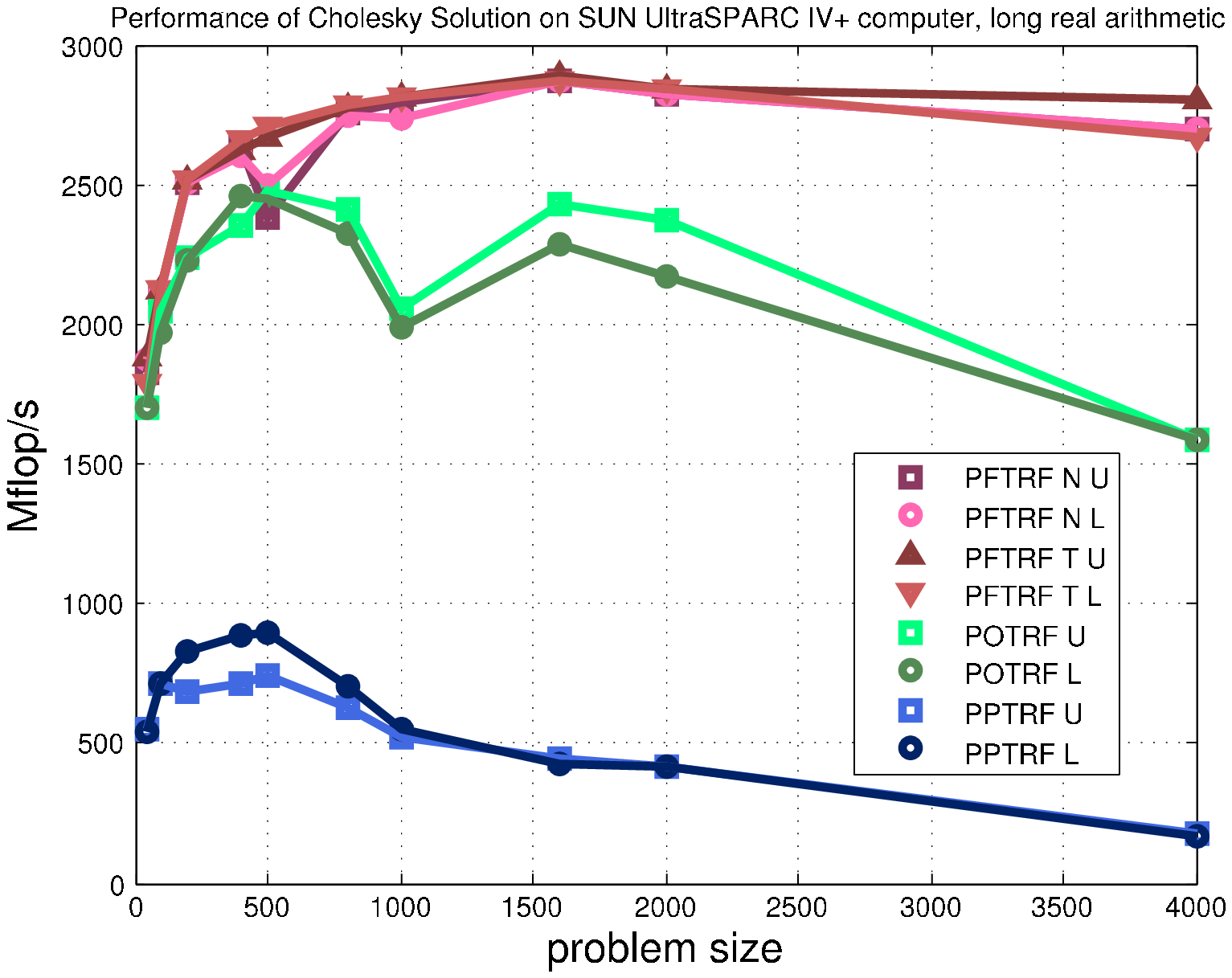}
\caption{\label{fig:sun_rea}
Performance in Mflop/s of Cholesky Factorization/Inversion/Solution on
SUN UltraSPARC IV+ computer, long real  arithmetic.  This is the
same data as presented in Appendix~\ref{appendix-table}  in Tables
\ref{tab:sunreafac}, \ref{tab:sunreainv} and \ref{tab:sunreasol}.
For PxTRF, $nrhs= \max(100, n/10)$.
}
\end{figure}

\begin{figure}
 \includegraphics[width=0.33\textwidth]{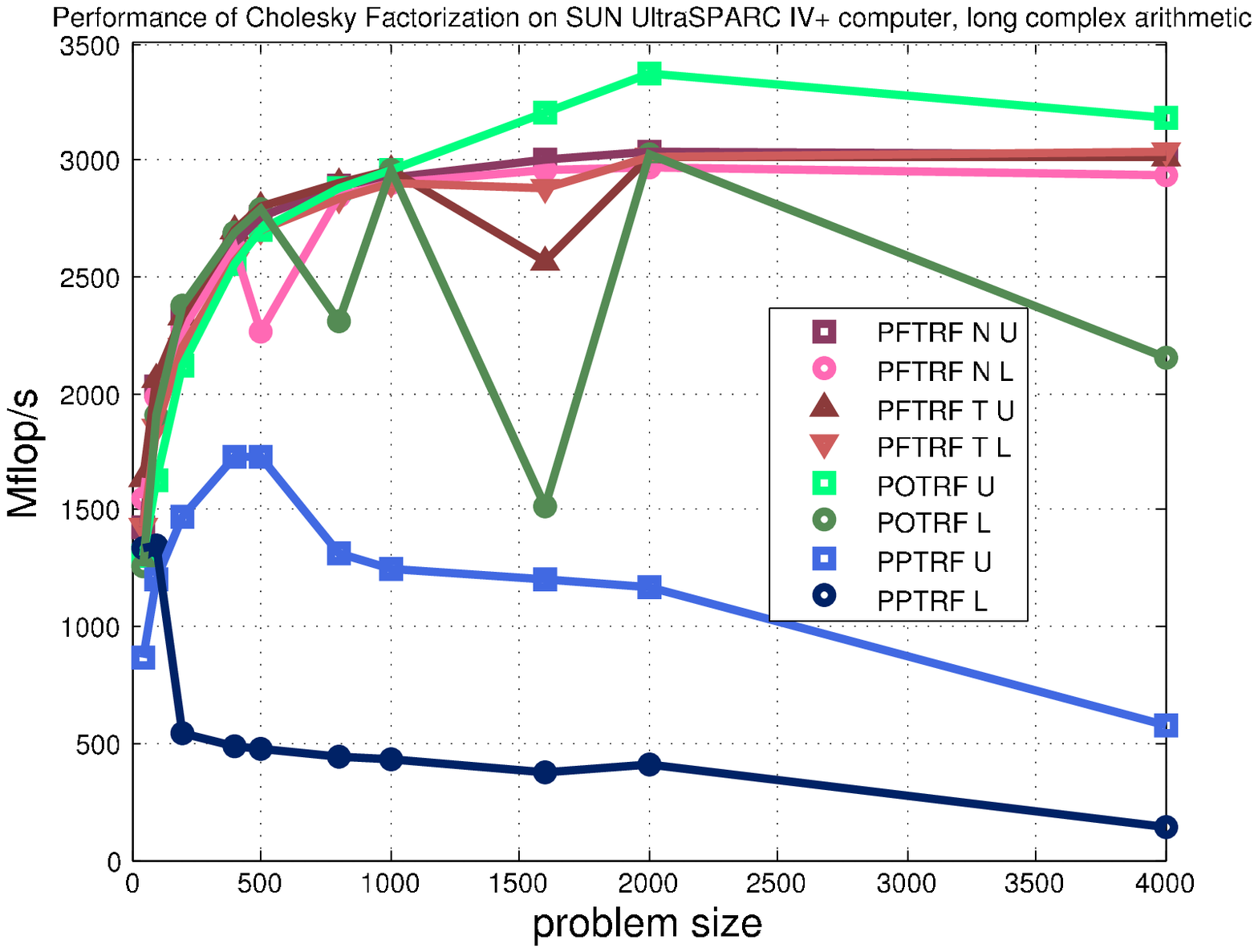}
 \includegraphics[width=0.33\textwidth]{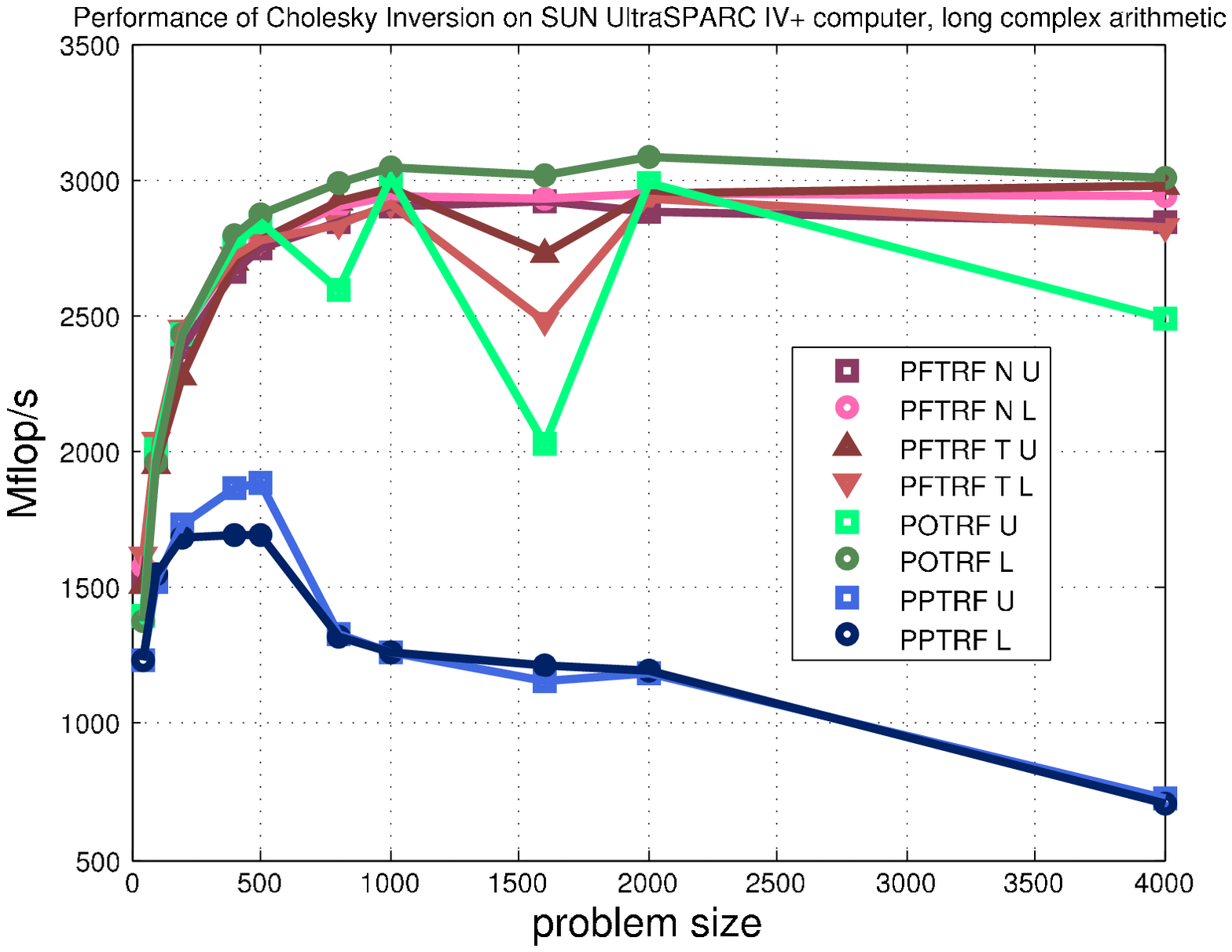}
 \includegraphics[width=0.33\textwidth]{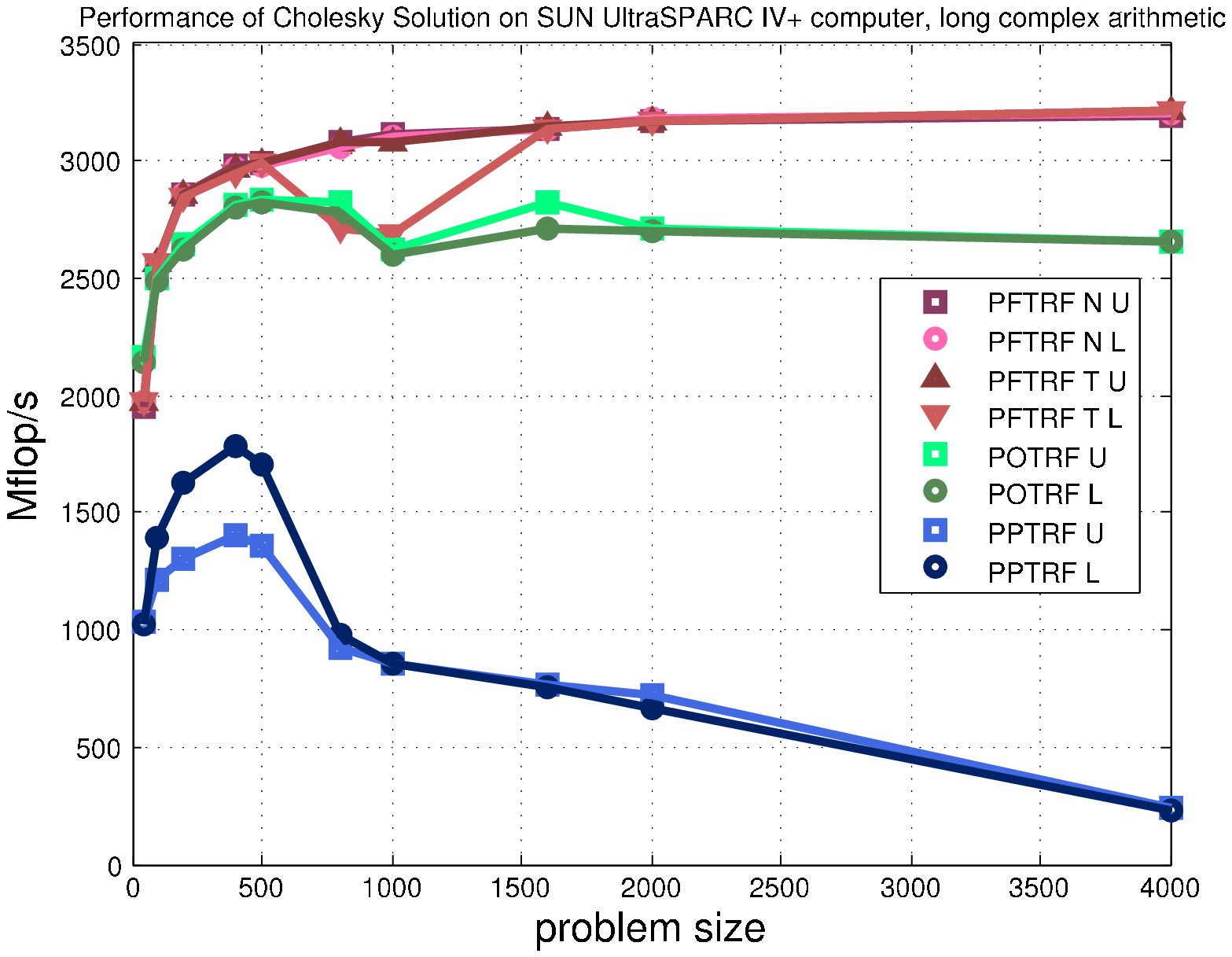}
\caption{\label{fig:sun_com}
Performance in Mflop/s of Cholesky Factorization/Inversion/Solution on
SUN UltraSPARC IV+ computer, long complex  arithmetic.  This is the
same data as presented in Appendix~\ref{appendix-table}  in Tables
\ref{tab:suncomfac}, \ref{tab:suncominv} and \ref{tab:suncomsol}.
For PxTRF, $nrhs= \max(100, n/10)$.
}
\end{figure}

%

\begin{figure}
 \includegraphics[width=0.33\textwidth]{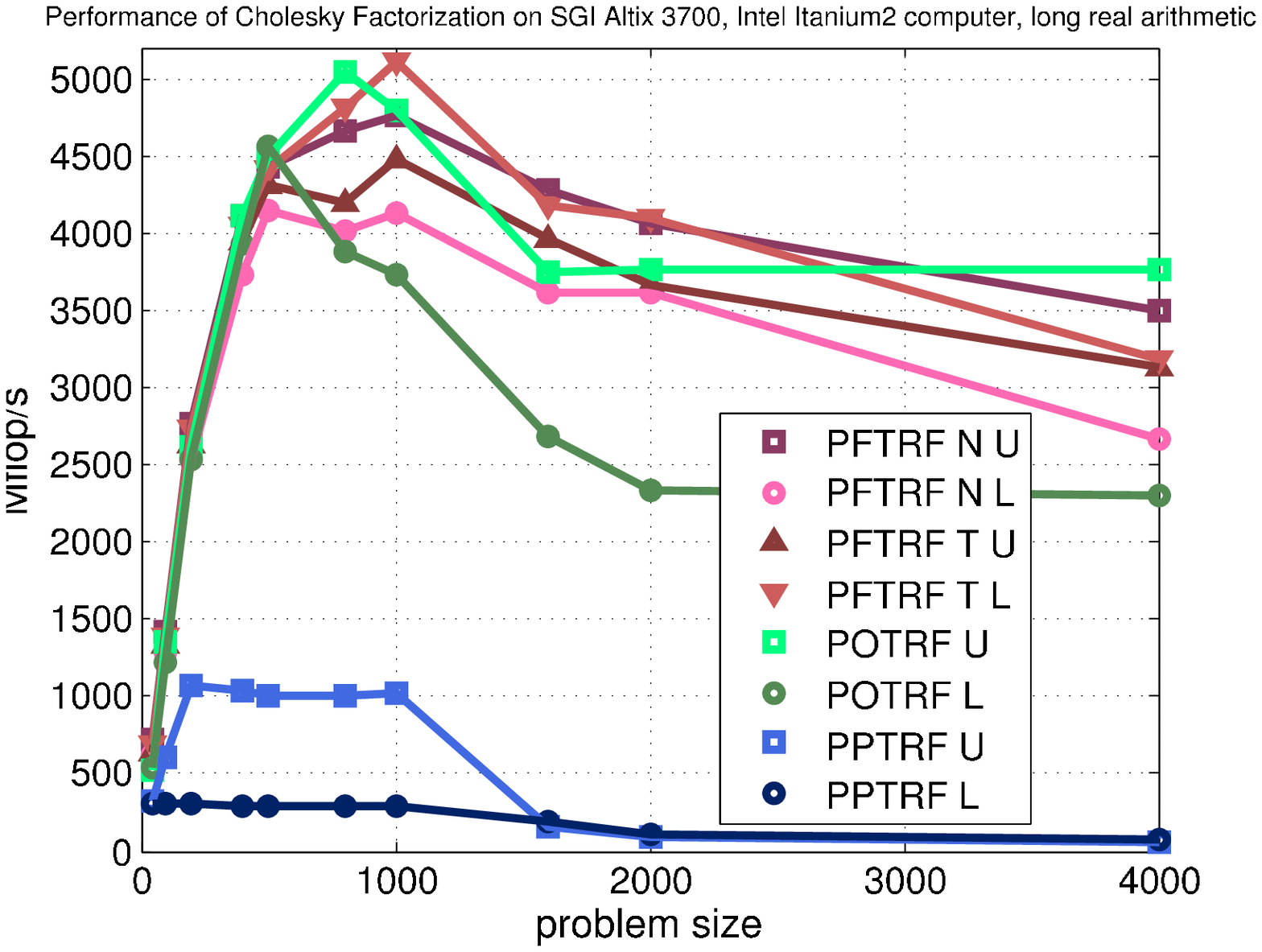}
 \includegraphics[width=0.33\textwidth]{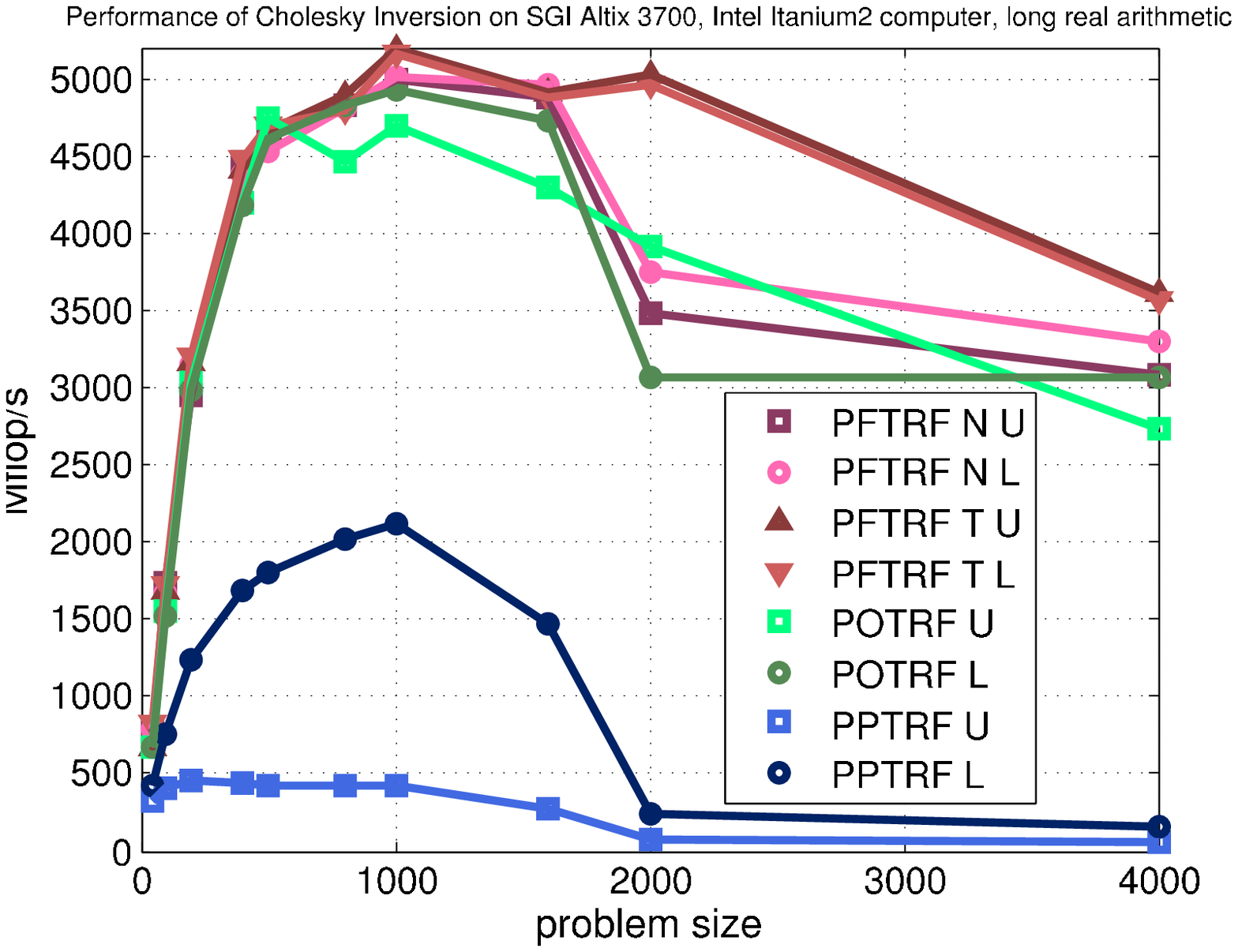}
 \includegraphics[width=0.33\textwidth]{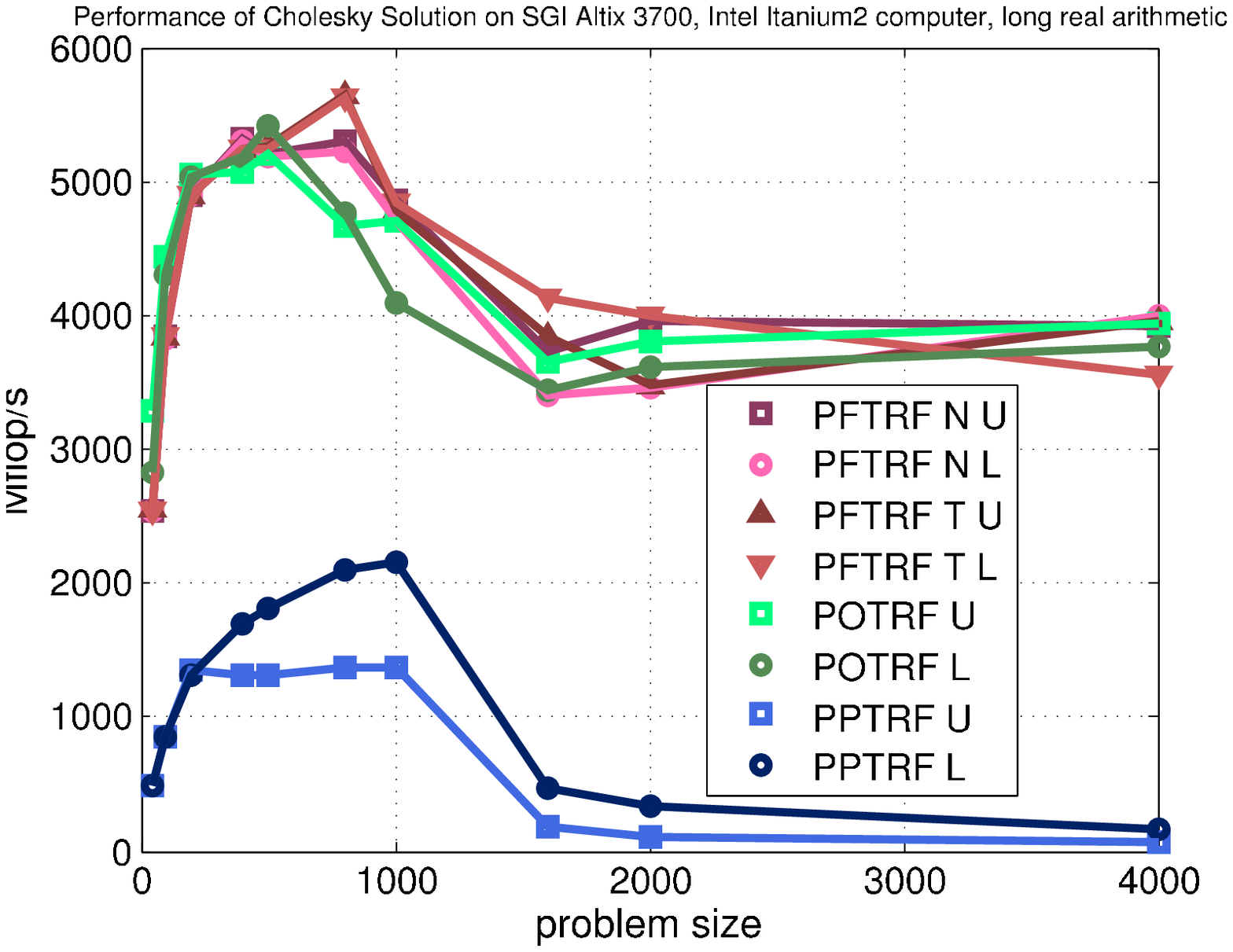}
\caption{\label{fig:sgi_rea}
Performance in Mflop/s of Cholesky Factorization/Inversion/Solution on SGI
Altix 3700, Intel Itanium 2 computer, long real arithmetic.  This is the
same data as presented in Appendix~\ref{appendix-table}  in Tables
\ref{tab:sgireafac}, \ref{tab:sgireainv} and \ref{tab:sgireasol}.
For PxTRF, $nrhs= \max(100, n/10)$.
}
\end{figure}

\begin{figure}
 \includegraphics[width=0.33\textwidth]{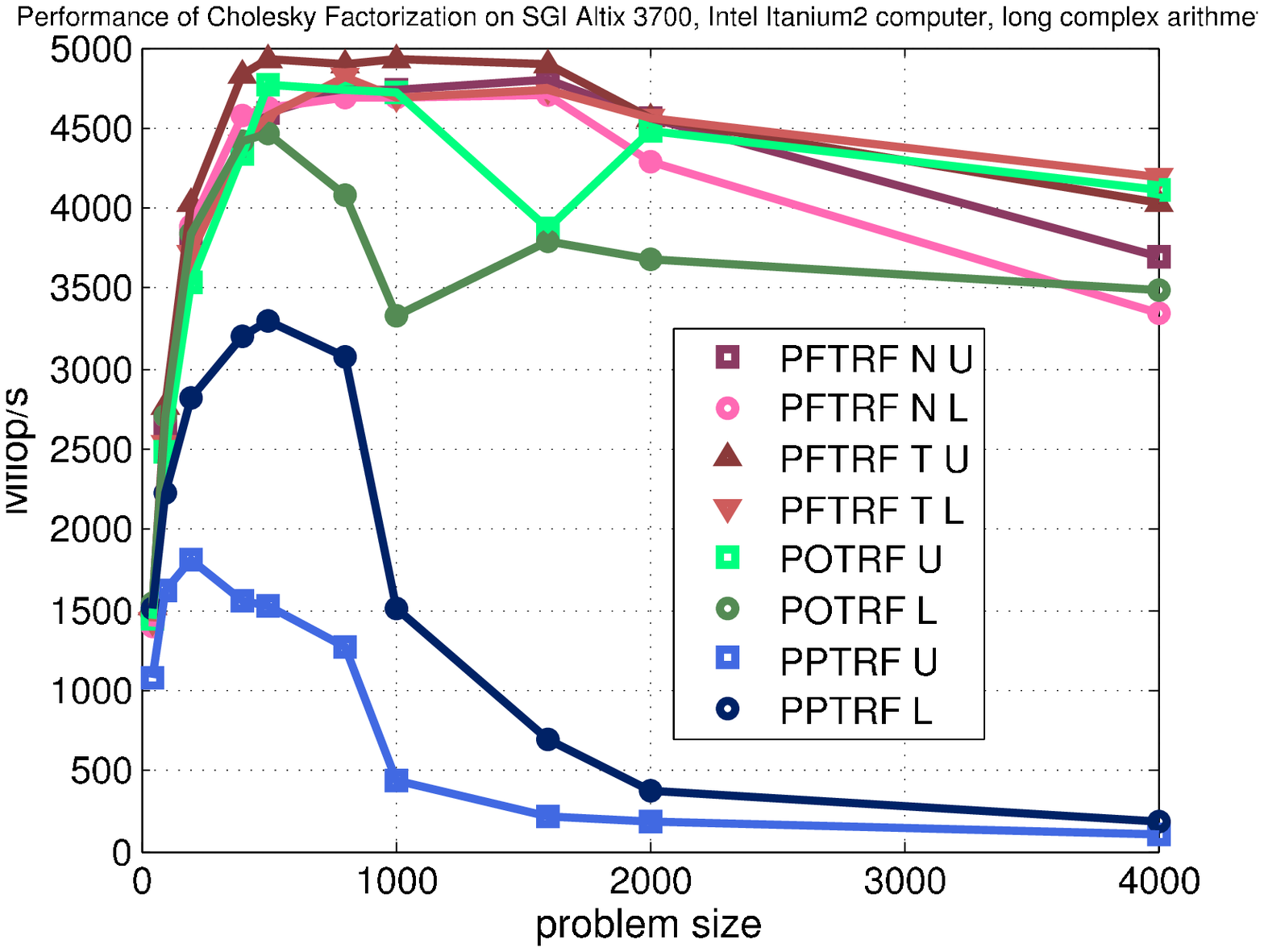}
 \includegraphics[width=0.33\textwidth]{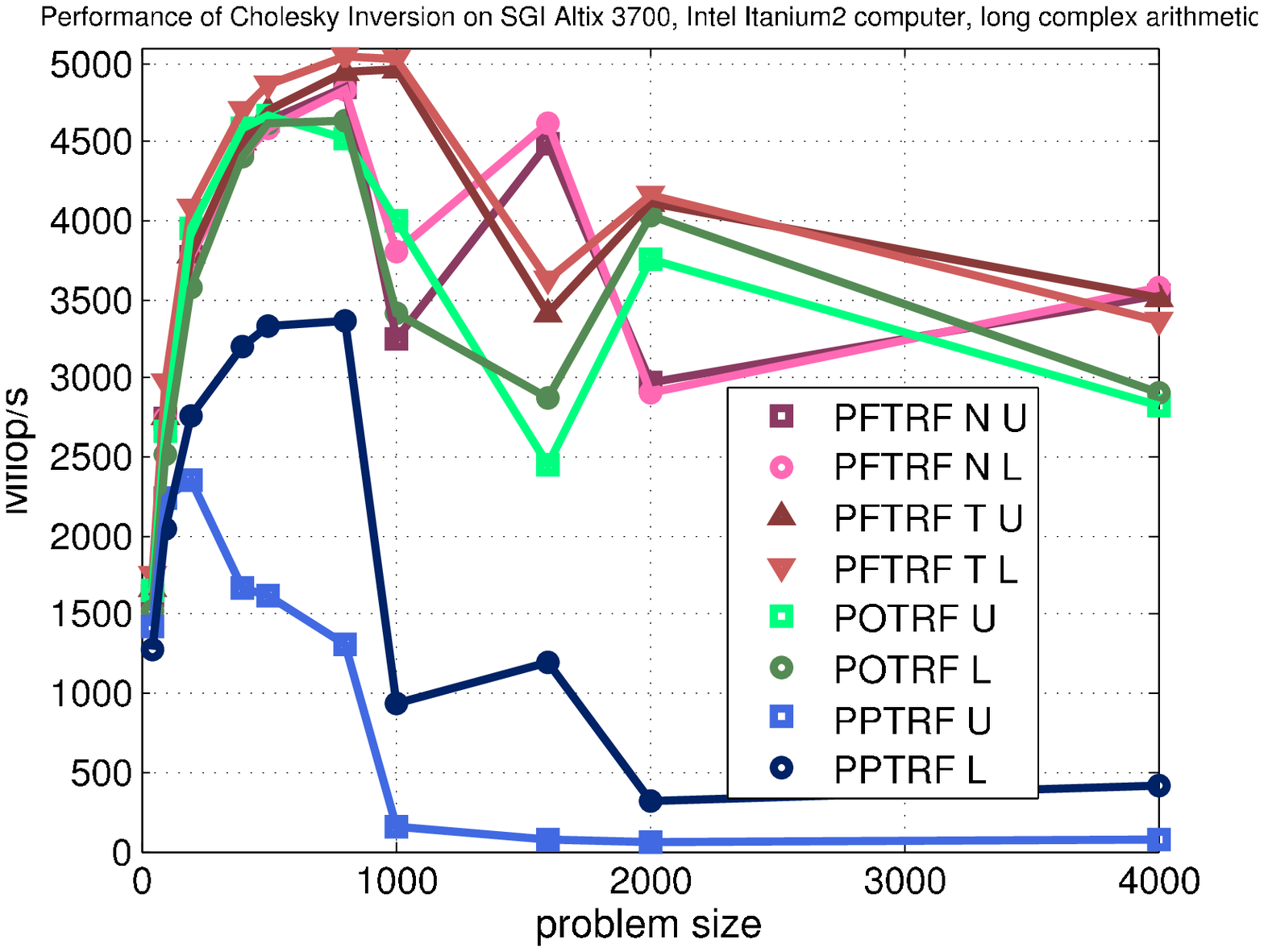}
 \includegraphics[width=0.33\textwidth]{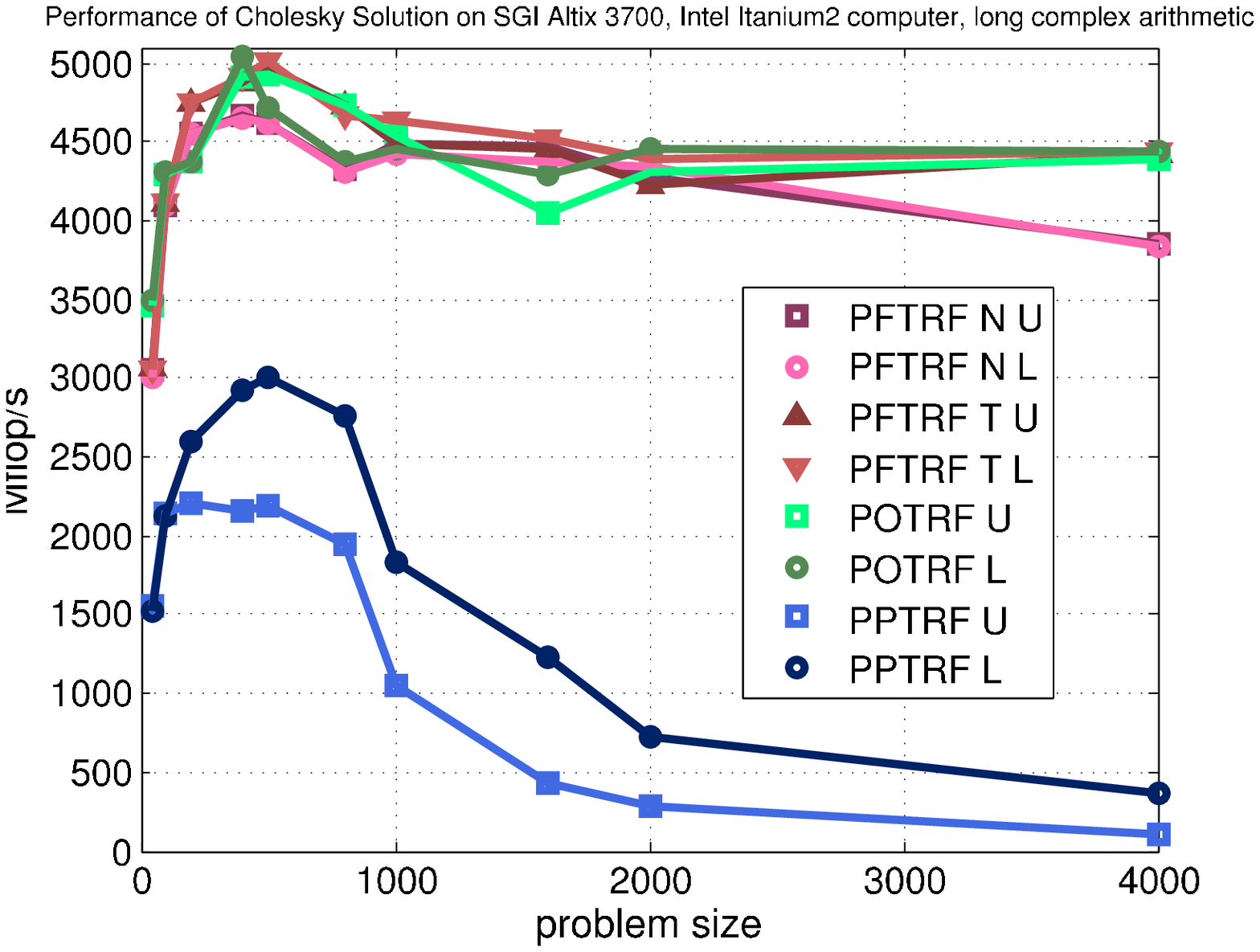}
\caption{\label{fig:sgi_com}
Performance in Mflop/s of Cholesky Factorization/Inversion/Solution on SGI
Altix 3700, Intel Itanium 2 computer, long complex  arithmetic.  This is the
same data as presented in Appendix~\ref{appendix-table}  in Tables
\ref{tab:sgicomfac}, \ref{tab:sgicominv} and \ref{tab:sgicomsol}.
For PxTRF, $nrhs= \max(100, n/10)$.
}
\end{figure}

\begin{figure}
 \includegraphics[width=0.33\textwidth]{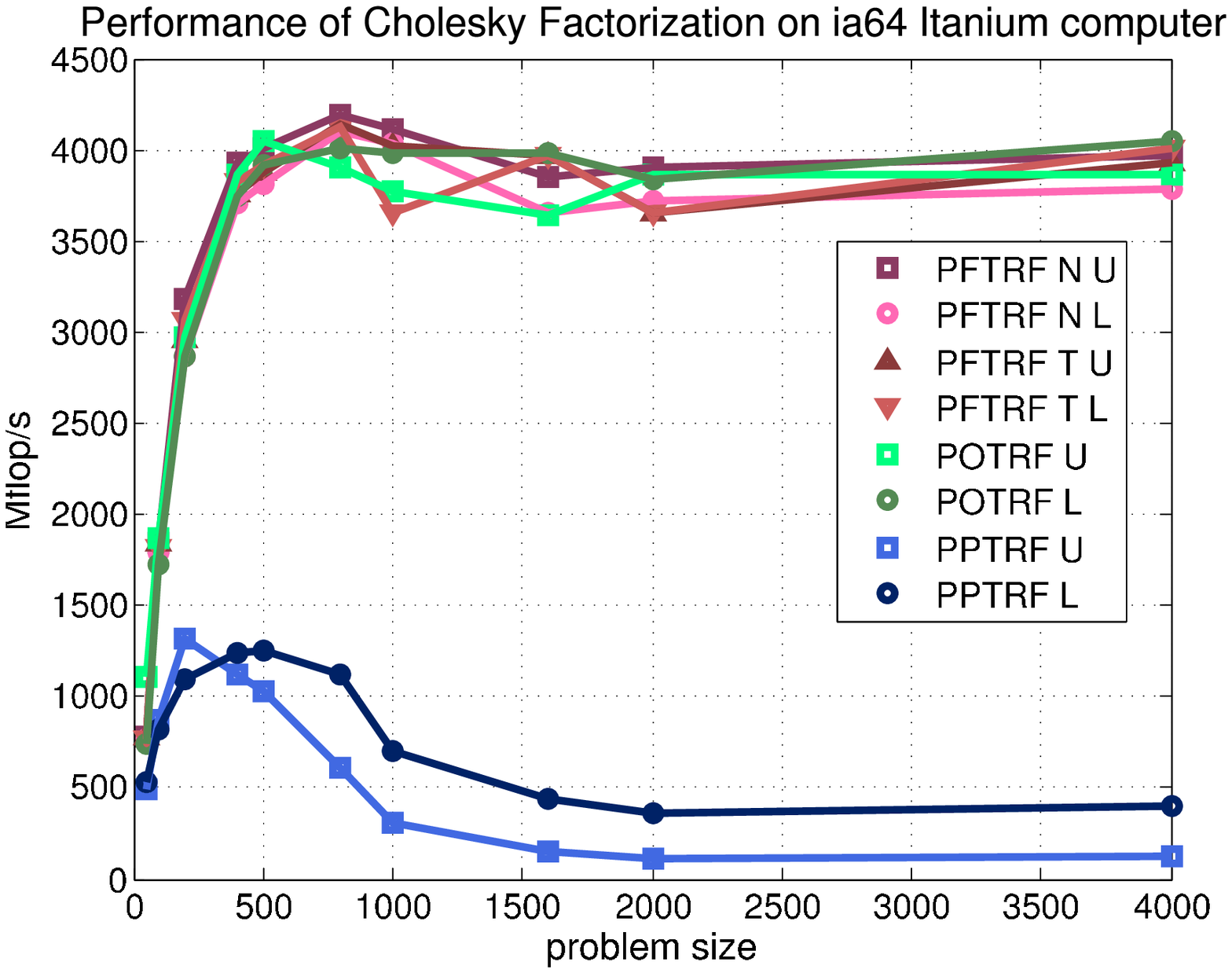}
 \includegraphics[width=0.33\textwidth]{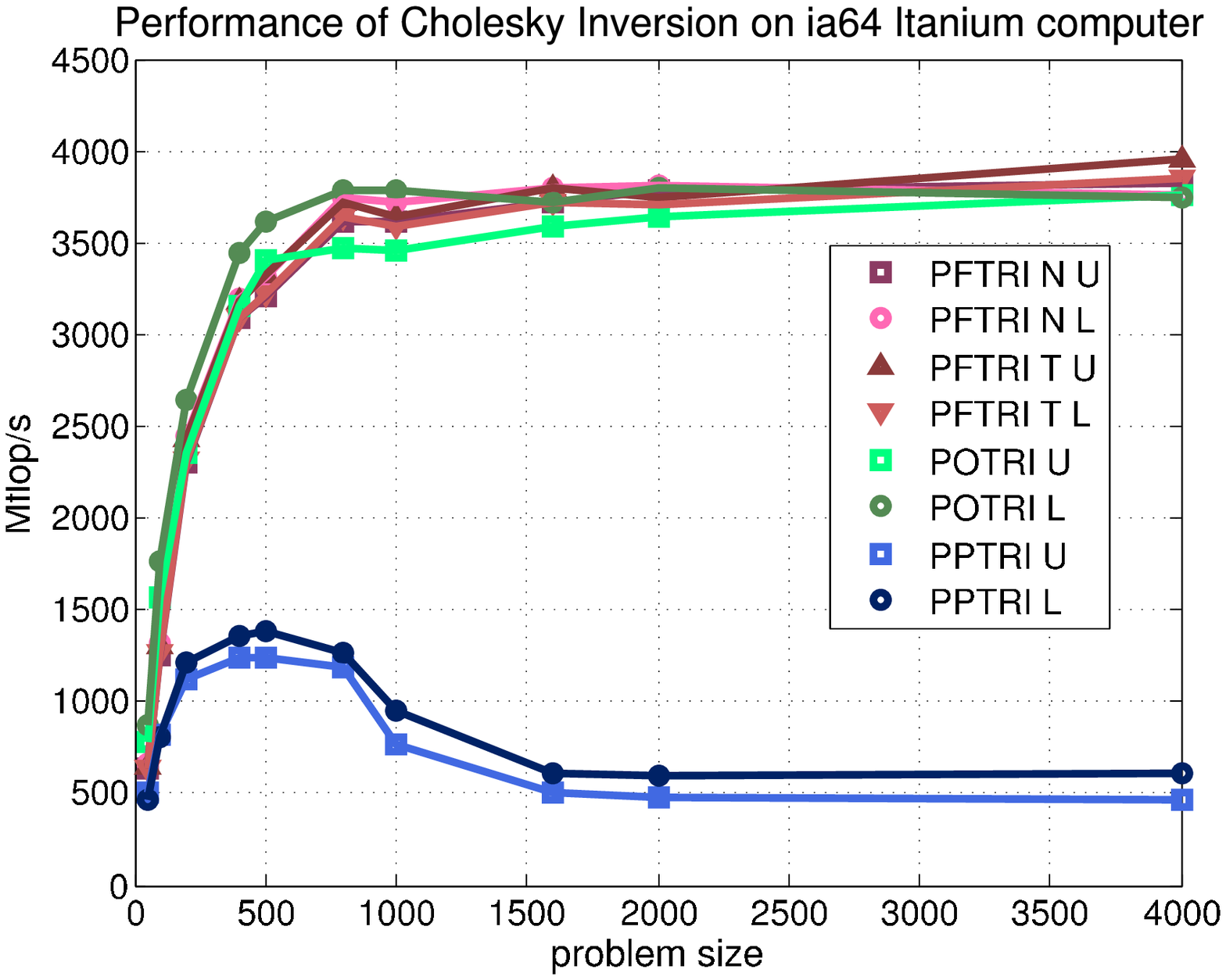}
 \includegraphics[width=0.33\textwidth]{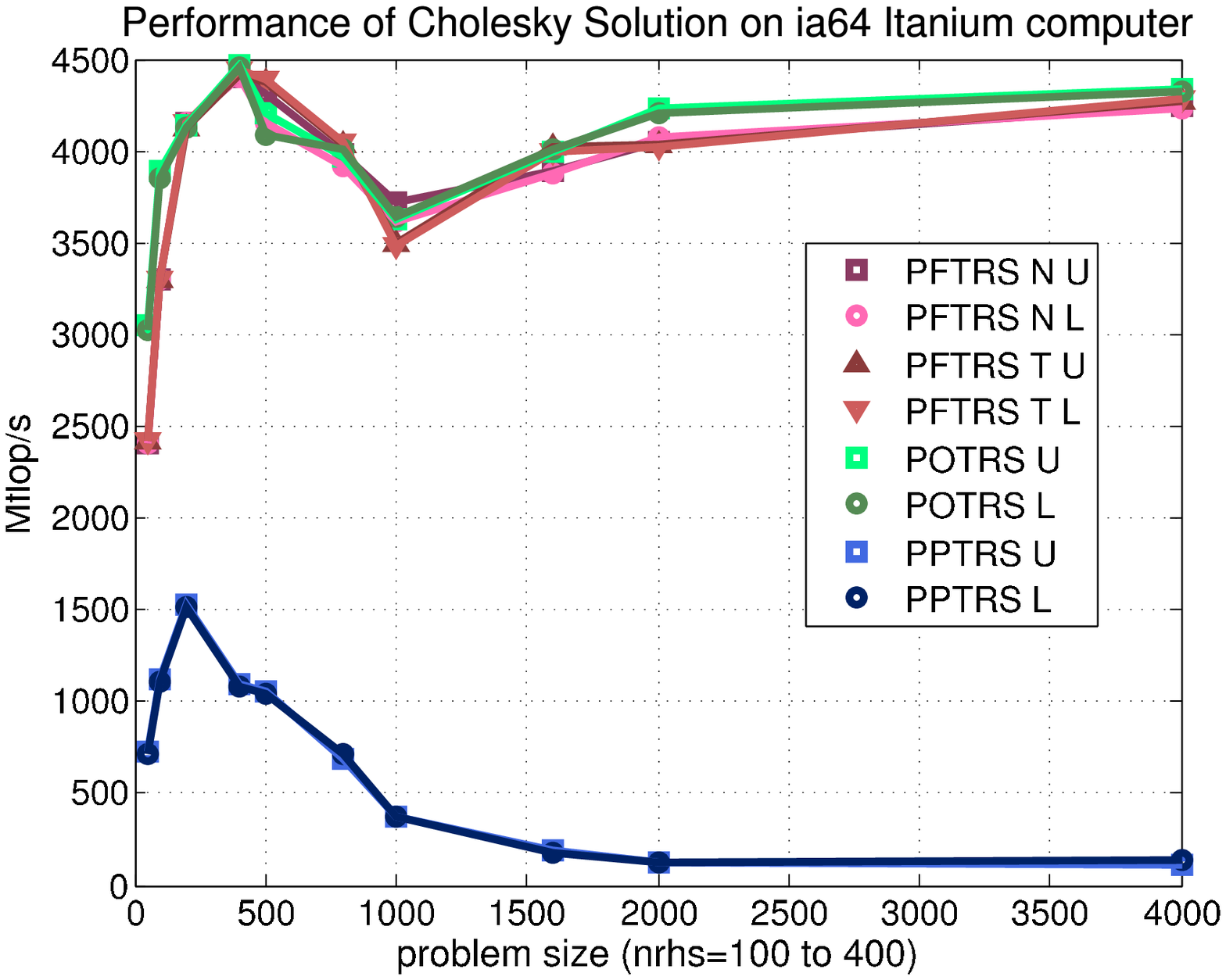}
\caption{\label{fig:ita}
Performance in Mflop/s of Cholesky Factorization/Inversion/Solution
on ia64 Itanium computer, long real arithmetic. This is the same data as presented  in Appendix~\ref{appendix-table}  in Tables
\ref{tab:itafac}, \ref{tab:itainv} and \ref{tab:itasol}.
For PxTRF, $nrhs= \max(100, n/10)$.
}
\end{figure}

\begin{figure}
 \includegraphics[width=0.33\textwidth]{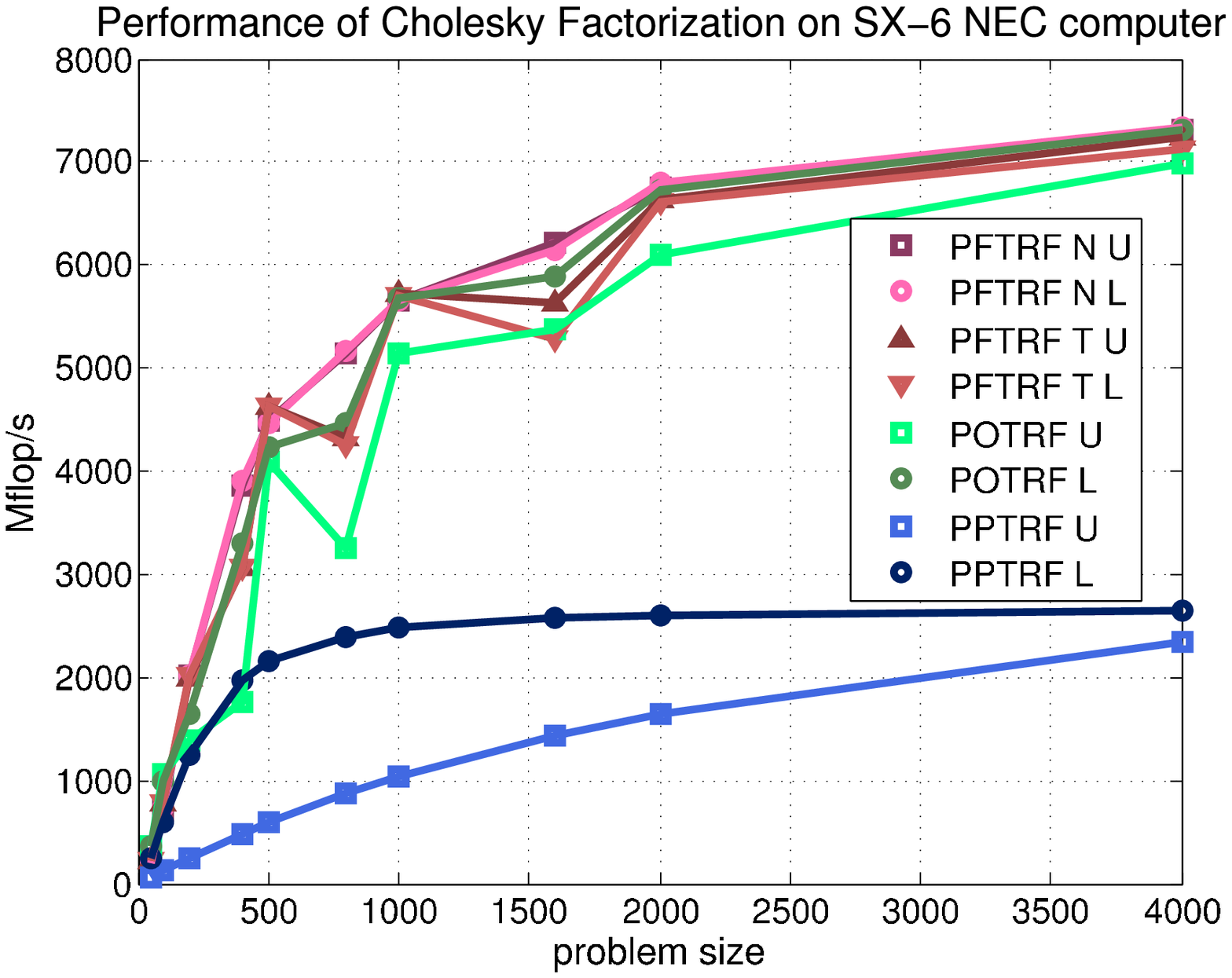}
 \includegraphics[width=0.33\textwidth]{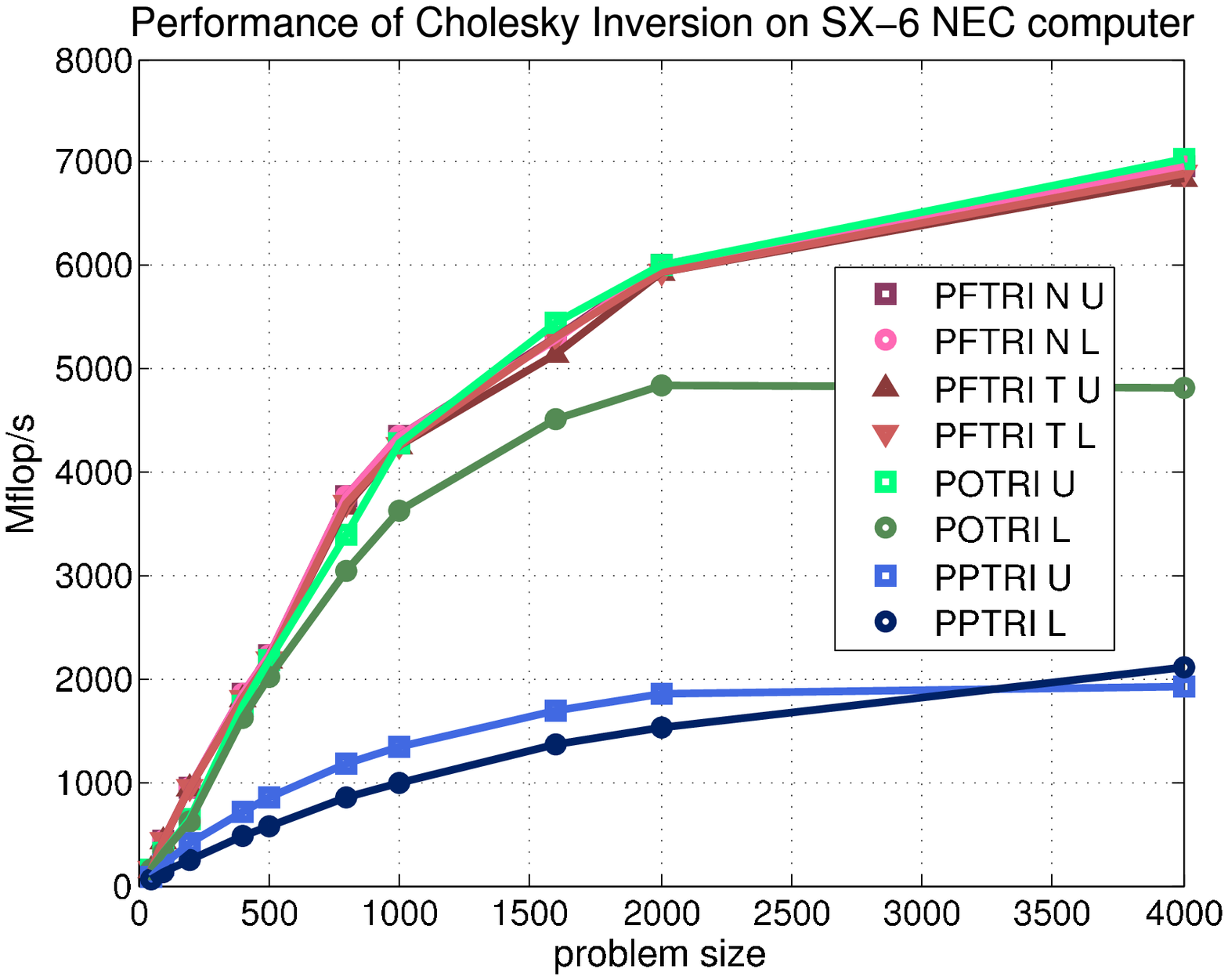}
 \includegraphics[width=0.33\textwidth]{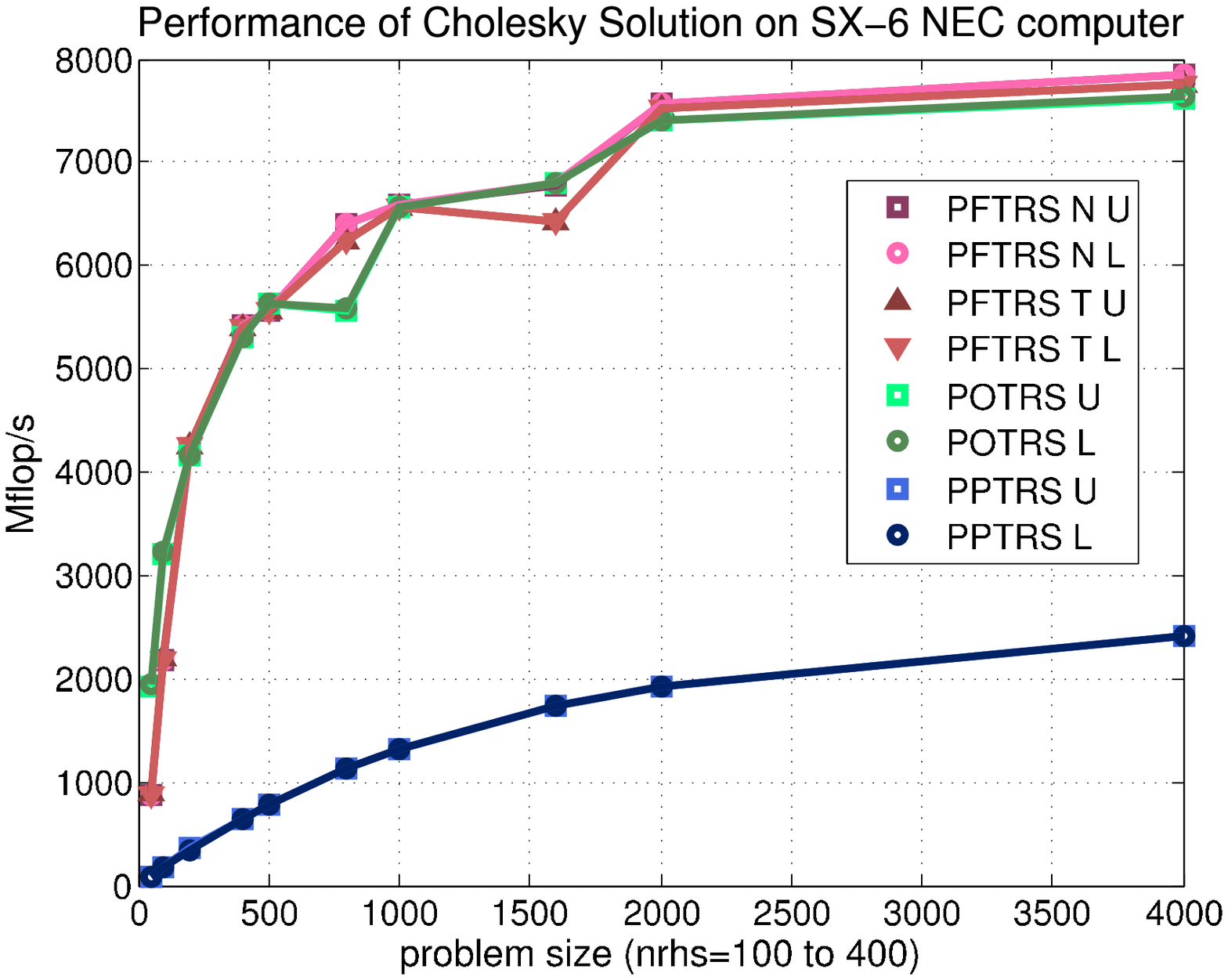}
\caption{\label{fig:nec}
Performance in Mflop/s of Cholesky Factorization/Inversion/Solution
on SX-6 NEC computer, long real arithmetic. This is the same data as presented  in Appendix~\ref{appendix-table}  in Tables
\ref{tab:necvfac}, \ref{tab:necvinv} and \ref{tab:necvsol}.
For PxTRF, $nrhs= \max(100, n/10)$.
}
\end{figure}

\begin{figure}
 \includegraphics[width=0.33\textwidth]{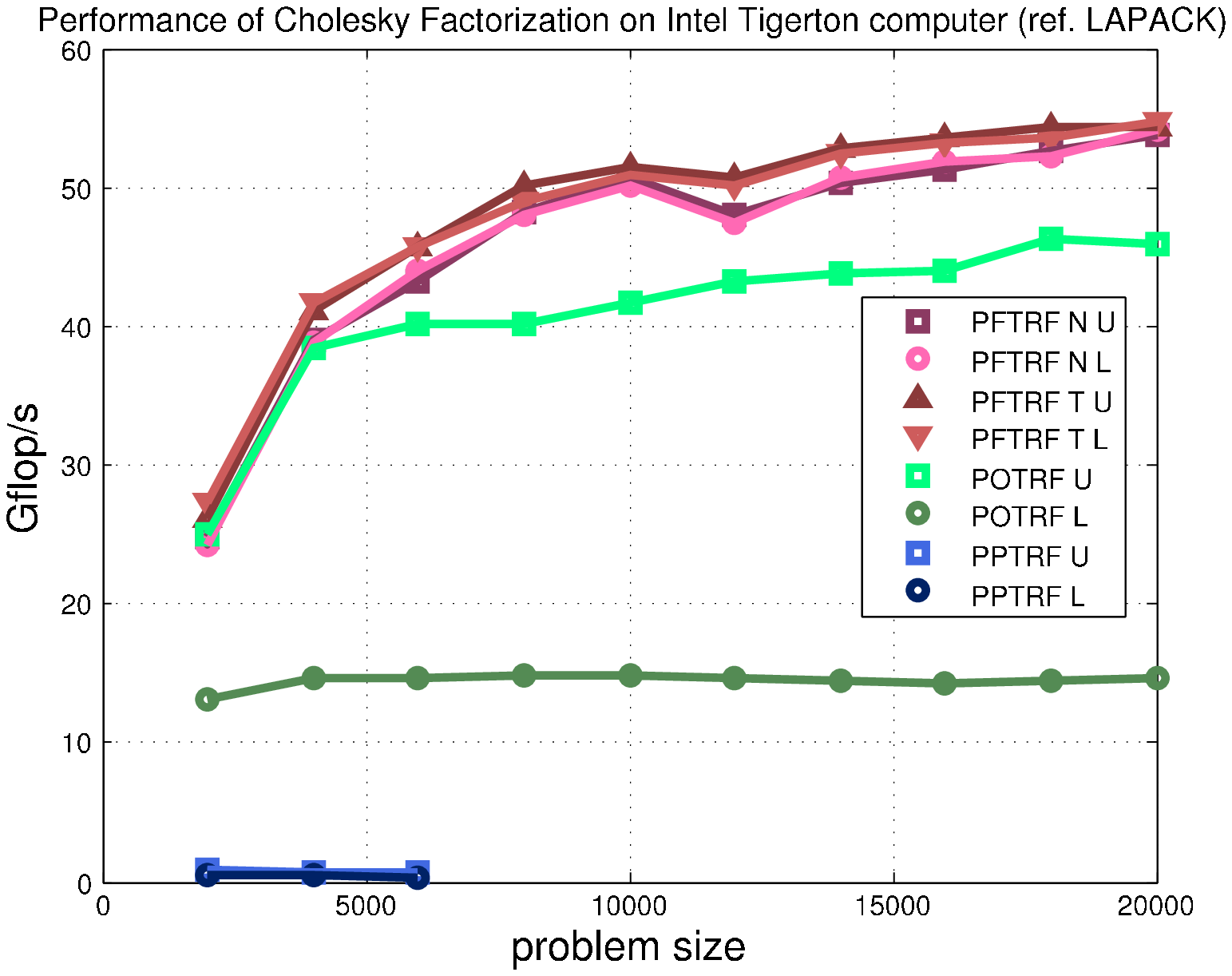}
 \includegraphics[width=0.33\textwidth]{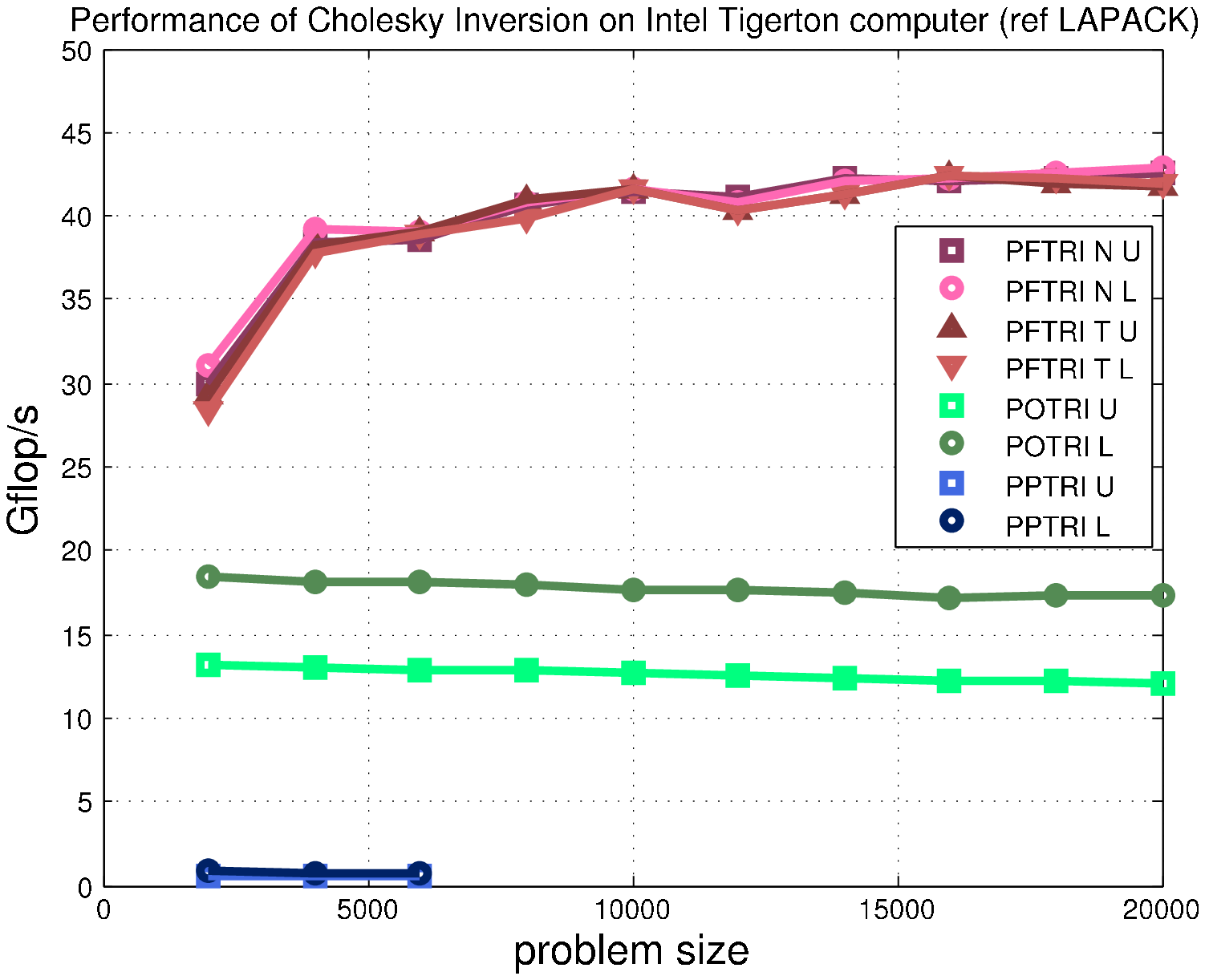}
 \includegraphics[width=0.33\textwidth]{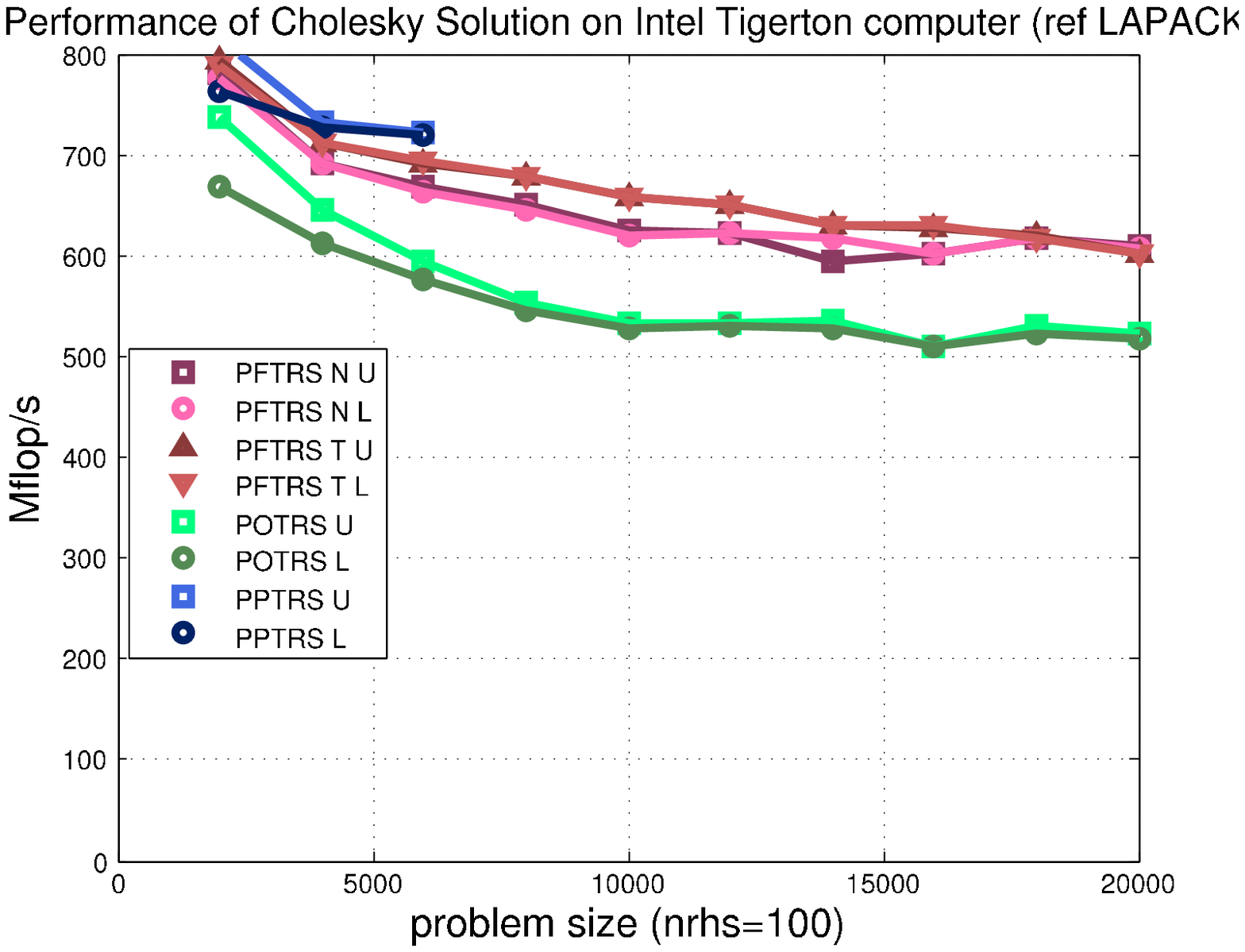}
\caption{\label{fig:zoot_reference}
Performance of Cholesky Factorization/Inversion/Solution
on quad-socket quad-core Intel Tigerton computer, long real arithmetic.
We use reference LAPACK-3.2.0 (from netlib) and MKL-10.0.1.014 multithreaded BLAS.
This is the same data as presented  in Appendix~\ref{appendix-table}  in Tables
\ref{tab:zootfac_reference}, \ref{tab:zootinv_reference} and \ref{tab:zootsol_reference}.
For the solution phase, $nrhs$ is fixed to 100 for any $n$.
Due to time limitation, 
the experiment was stopped for the packed storage format inversion at $n=4000$.
}
\end{figure}

\begin{figure}
 \includegraphics[width=0.33\textwidth]{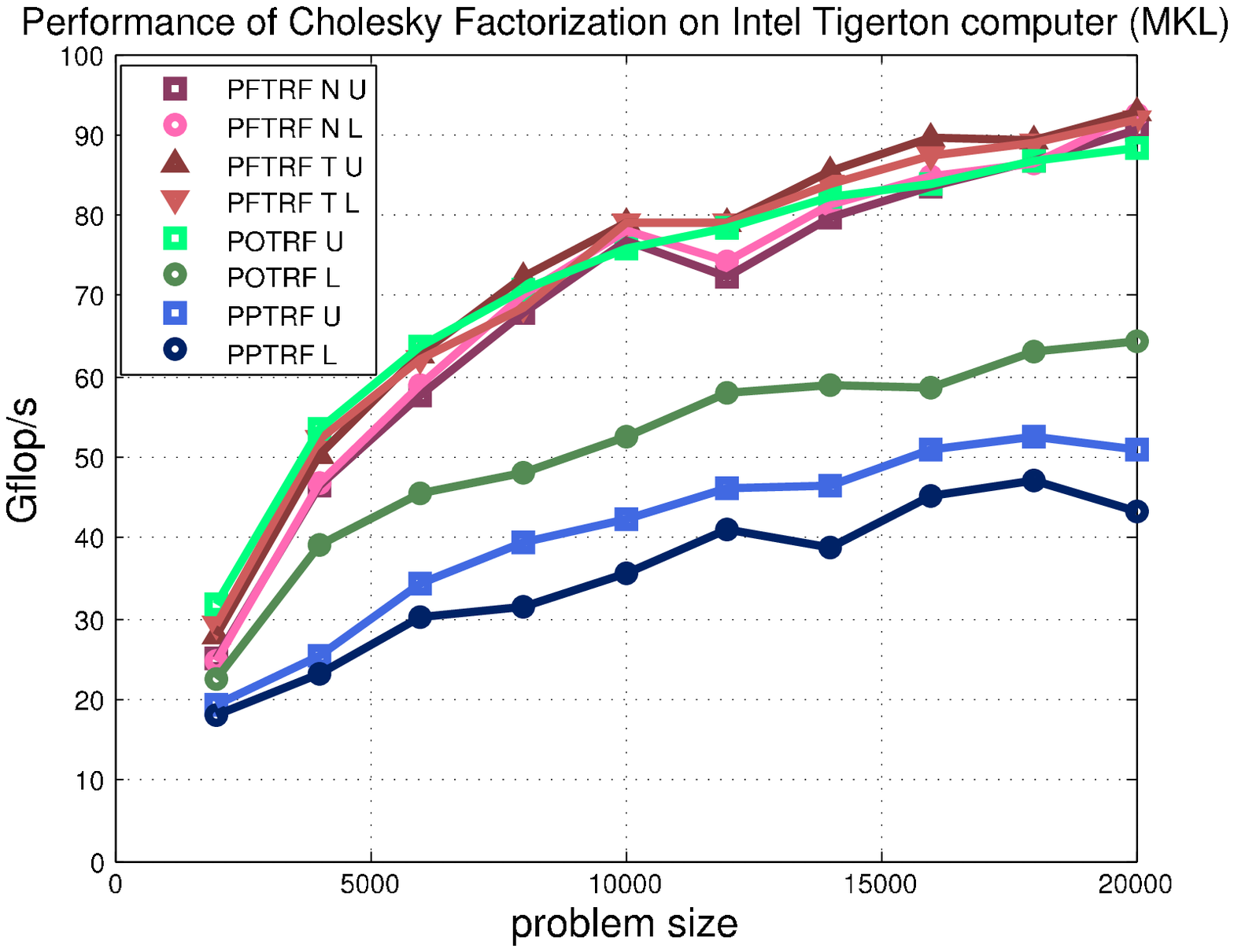}
 \includegraphics[width=0.33\textwidth]{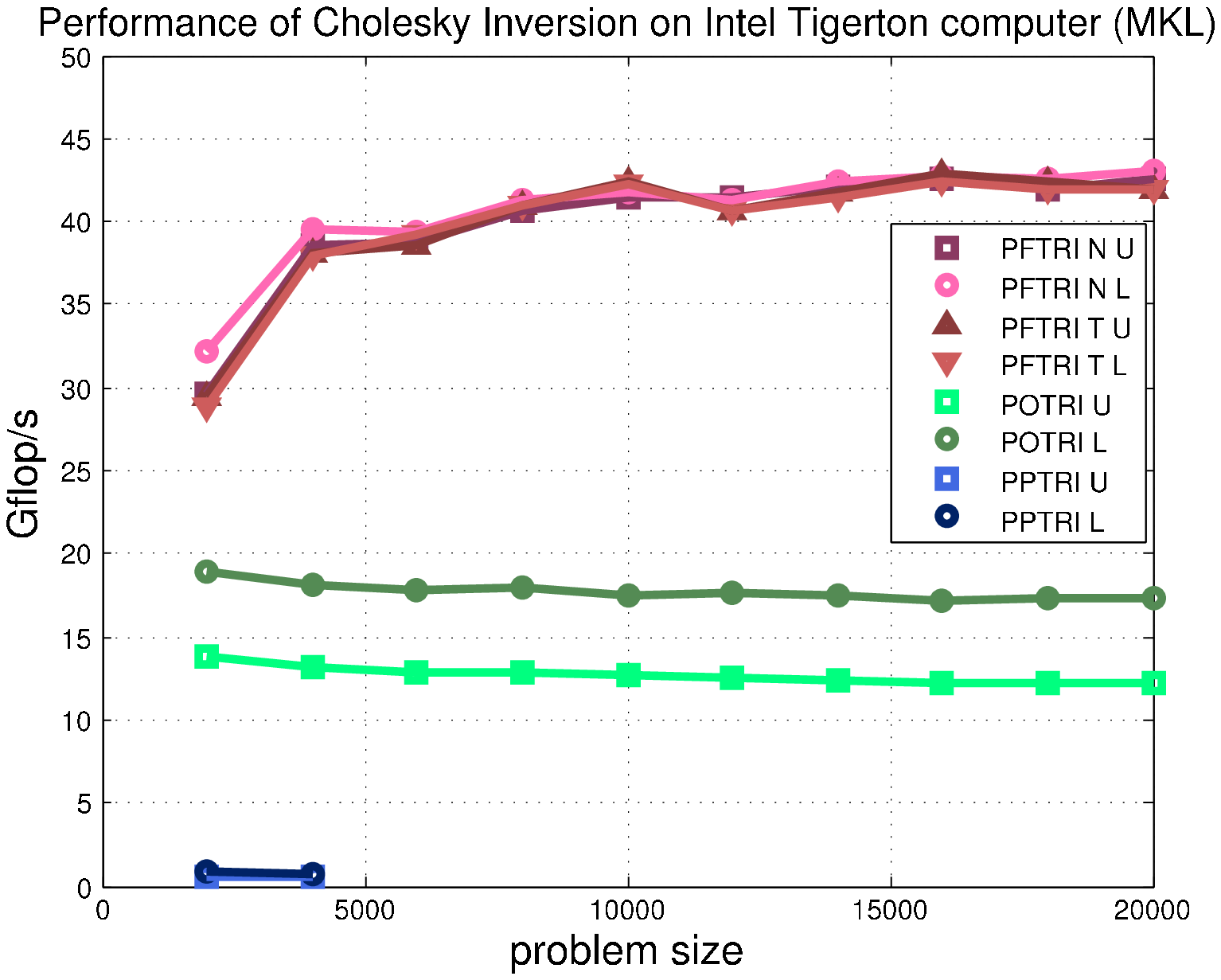}
 \includegraphics[width=0.33\textwidth]{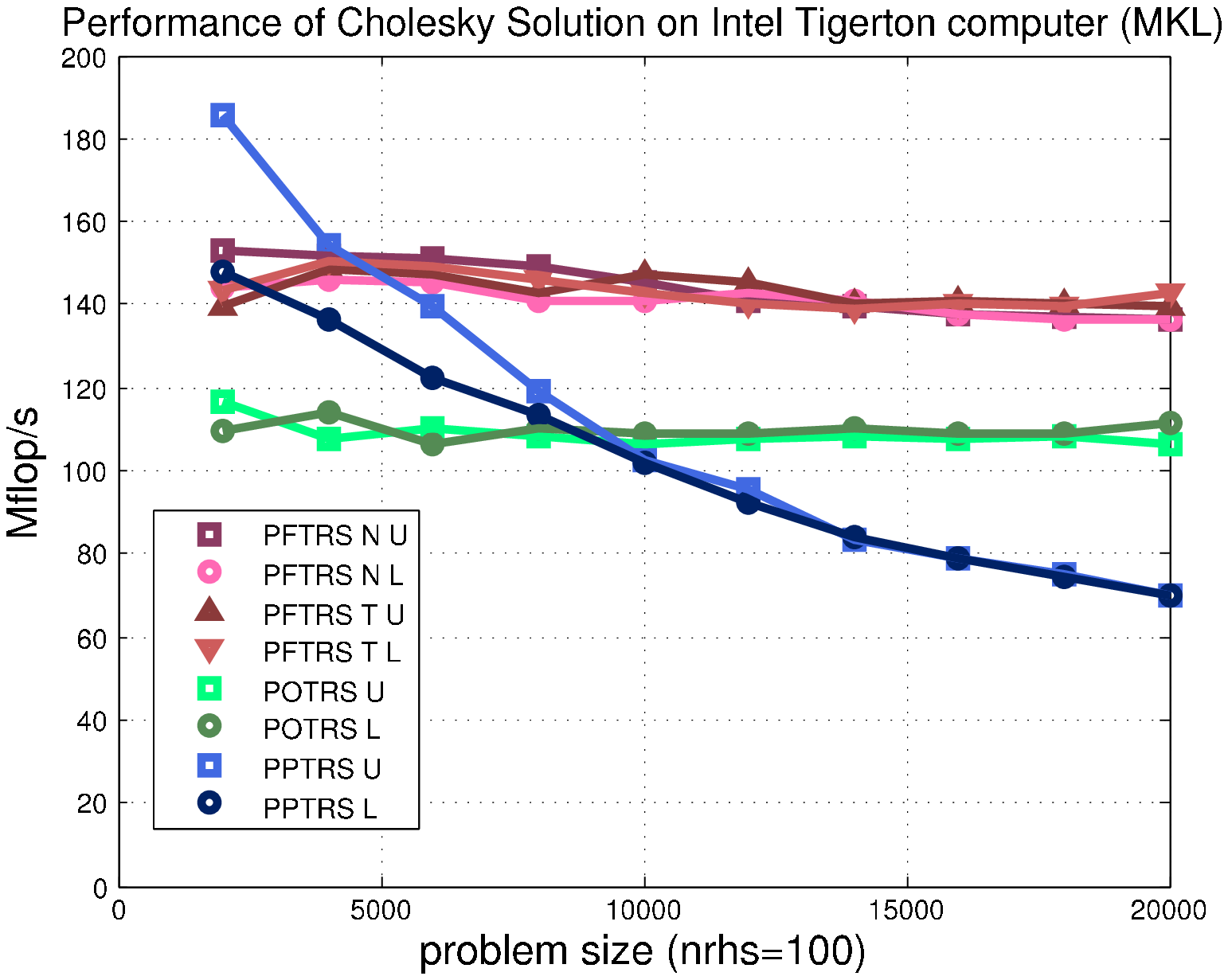}
\caption{\label{fig:zoot_vendor}
Performance of Cholesky Factorization/Inversion/Solution
on quad-socket quad-core Intel Tigerton computer, long real arithmetic.
We use MKL-10.0.1.014 multithreaded LAPACK and BLAS.
This is the same data as presented  in Appendix~\ref{appendix-table}  in Tables
\ref{tab:zootfac_vendor}, \ref{tab:zootinv_vendor} and \ref{tab:zootsol_vendor}.
For the solution phase, $nrhs$ is fixed to 100 for any $n$.
Due to time limitation, 
the experiment was stopped for the packed storage format inversion at $n=4000$.
}
\end{figure}

\begin{figure}
 \includegraphics[width=0.50\textwidth]{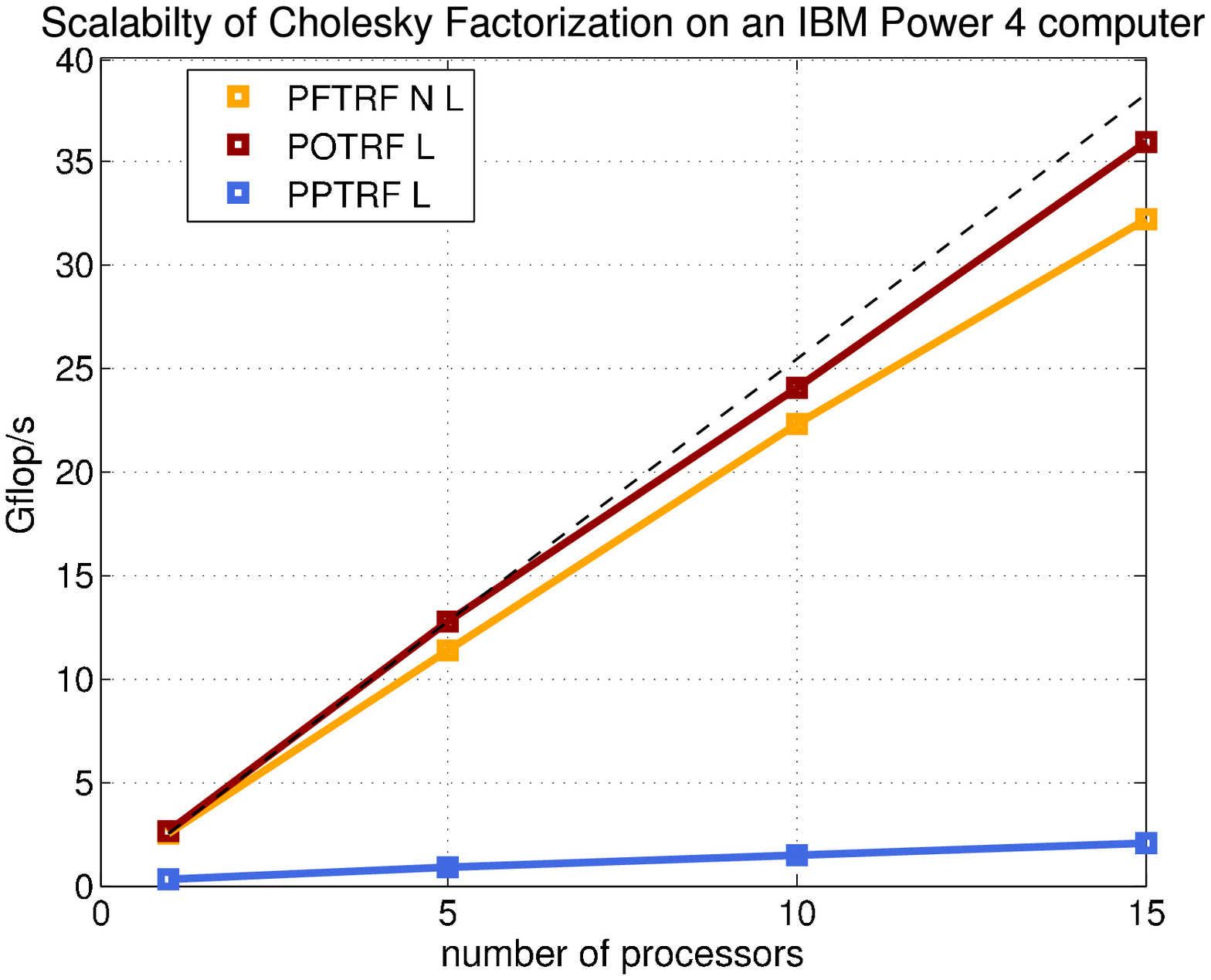}
 \includegraphics[width=0.50\textwidth]{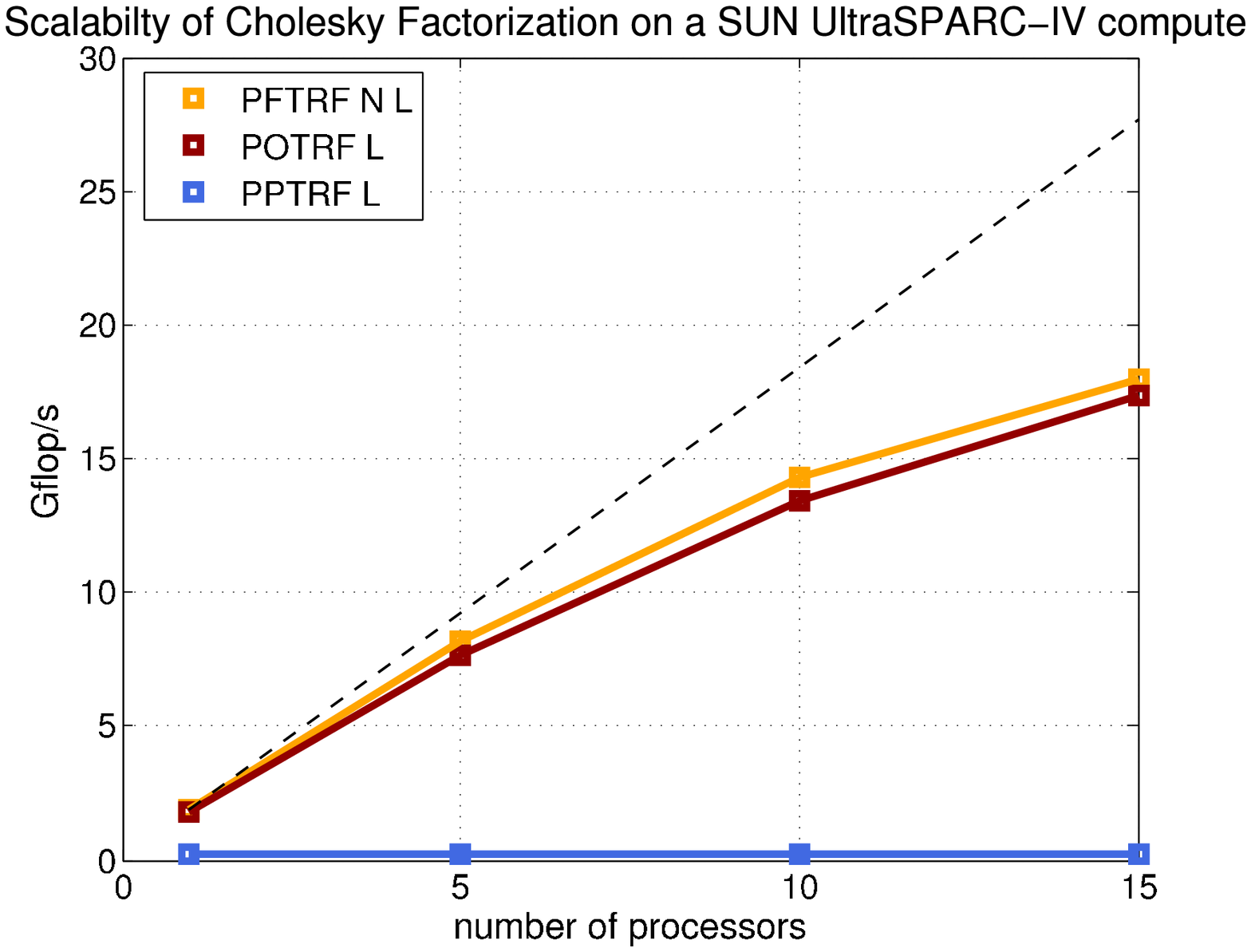}
\caption{\label{fig:last}
Performance in Gflop/s of Cholesky Factorization on IBM Power 4 (left)
and 
SUN UltraSPARC-IV (right)
computer, long real arithmetic,
with a different number of Processors,
testing the SMP Parallelism.
The implementation of PPTRF of sunperf does not show any SMP  
parallelism.
UPLO = 'L'. $N=5,000$ (strong scaling experiment).
This is the same data as presented  in Appendix~\ref{appendix-table}  in Tables
\ref{tab:ibmpar} and
\ref{tab:sunpar}
}
\end{figure}


\section{Integration in LAPACK}
\label{sec:lapack}

As mentioned in the introduction, as of release 3.2 (November 2008), LAPACK
supports a preliminary version of RFPF. Ultimately, the goal would be for RFPF
to support as many functionnalities as full format or standard packed format
does.  The 44 routines included in release 3.2 for RFPF are given in
Table~\ref{tab:lapack}. The names for the RFPF routines follow the naming
nomenclature used by LAPACK.  We have added the format description letters: PF
for Symmetric/Hermitian Positive Definite RFPF (PO for full, PP for packed), SF
for Symmetric RFPF (SY for full, SP for packed), HF for Hermitian RFPF (HE for
full, HP for packed), and TF for Triangular RFPF (TR for full, TP for packed).

Currently, for the complex case, we assume that the transpose complex-conjugate
part is stored whenever the transpose part is stored in the real case.  This
corresponds to the theory developed in this present manuscript.  In the
future, we will want to have the flexibility to store the transpose part
(as opposed to transpose complex conjugate) whenever the transpose part is
stored in the real case. In particular, this feature will be useful for
complex symmetric matrices.

\begin{table}
\begin{tabular}{llllll}
\hline
functionality      & \multicolumn{4}{l}{ routine names and calling sequence } \\
\hline
\hline
Cholesky factorization      & CPFTRF & DPFTRF & SPFTRF & ZPFTRF \\ & \multicolumn{4}{l}{ (TRANSR,UPLO,N,A,INFO) } \\
\hline
Multiple solve after PFTRF  & CPFTRS & DPFTRS & SPFTRS & ZPFTRS \\ & \multicolumn{4}{l}{ (TRANSR,UPLO,N,NR,A,B,LDB,INFO) } \\
\hline
Inversion after PFTRF       & CPFTRI & DPFTRI & SPFTRI & ZPFTRI \\ & \multicolumn{4}{l}{ (TRANSR,UPLO,N,A,INFO) } \\
\hline
Triangular inversion        & CTRTRI & DTRTRI & STRTRI & ZTRTRI \\ & \multicolumn{4}{l}{ (TRANSR,UPLO,DIAG,N,A,INFO) } \\
\hline
Sym/Herm matrix norm        & CLANHF & DLANSF & SLANSF & ZLANHF \\ & \multicolumn{4}{l}{ (NORM,TRANSR,UPLO,N,A,WORK) } \\
\hline
Triangular solve            & CTFSM  & DTFSM  & STFSM  & ZTFSM  \\ & \multicolumn{4}{l}{ (TRANSR,SIDE,UPLO,TRANS,DIAG,M,N,ALPHA,A,B,LDB) } \\
\hline
Sym/Herm rank-$k$ update    & CHFRK  & DSFRK  & SSFRK  & ZHFRK  \\ & \multicolumn{4}{l}{ (TRANSR,UPLO,TRANS,N,K,ALPHA,A,LDA,BETA,C) } \\
\hline
Conv. from TP to TF         & CTPTTF & DTPTTF & STPTTF & ZTPTTF \\ & \multicolumn{4}{l}{ (TRANSR,UPLO,N,AP,ARF,INFO) } \\
\hline
Conv. from TR to TF         & CTRTTF & DTRTTF & STRTTF & ZTRTTF \\ & \multicolumn{4}{l}{ (TRANSR,UPLO,N,A,LDA,ARF,INFO) } \\
\hline
Conv. from TF to TP         & CTFTTP & DTFTTP & STFTTP & ZTFTTP \\ & \multicolumn{4}{l}{ (TRANSR,UPLO,N,ARF,AP,INFO) } \\
\hline
Conv. from TF to TR         & CTFTTR & DTFTTR & STFTTR & ZTFTTR \\ & \multicolumn{4}{l}{ (TRANSR,UPLO,N,ARF,A,LDA,INFO) }  \\
\hline
\end{tabular}
\caption{\label{tab:lapack} LAPACK 3.2 RFPF routines.
}
\end{table}

\section{Summary and Conclusions}
\label{sec:summary}

This paper describes RFPF as a standard minimal full format for representing
both symmetric and triangular matrices. Hence, from a user point of view, these matrix layouts are a
replacement for both the standard formats of DLA, namely full and packed
storage. These new layouts possess three good features: they are efficient, they are supported by
Level~3 BLAS and LAPACK full format routines, and they require minimal storage.

\section{Acknowledgments}

The results in this paper were obtained on seven computers, an IBM, a SGI, two SUNs,
\hbox{Itanium}, NEC, and Intel Tigerton computers. The IBM machine belongs to
the Center for Scientific Computing at Aarhus, the SUN machines to the Danish
Technical University, the Itanium and NEC machines to the Danish Meteorological
Institute, and the Intel Tigerton machine to the Innovative Computing
Laboratory at the University of Tennessee.

We would like to thank Bernd~Dammann for consulting on the SUN systems;
Niels~Carl~W.~Hansen for consulting on the IBM and SGI systems; and Bjarne~Stig~Andersen
for obtaining the results on the Itanium and NEC computers. We thank IBMers
John Gunnels who worked earlier on the HFPF format and JP Fasano who was
instrumental in getting the source code released by the IBM Open Source
Committee. We thank Allan Backer for discussions about an older version of this
manuscript.

\bibliography{tomsrfp}
\begin{appendix}
\section{Performance results}\label{appendix-table}

\clearpage

\begin{table}
{\normalsize \center
\begin{tabular}{|c|cc|cc|cc|cc|} \hline
 n & \multicolumn{4}{c|}{RFPF} & \multicolumn{4}{c|}{LAPACK} \\ \cline{2-9}
  & \multicolumn{2}{c|}{NO TRANS}&\multicolumn{2}{c|}{TRANS}&\multicolumn{2}{c|}{POTRF}&\multicolumn{2}{c|}{PPTRF} \\ \cline{2-9}
      &   U  &   L  &   U  &   L  &   U  &   L  &  U  &  L  \\ \hline
   50 &  827 &  898 &  915 &  834 &  924 &  622 & 435 & 622 \\
  100 & 1420 & 1517 & 1464 & 1434 & 1264 & 1218 & 592 & 811 \\
  200 & 1734 & 1795 & 1590 & 1746 & 1707 & 1858 & 703 & 378 \\
  400 & 2165 & 2242 & 2275 & 2177 & 2234 & 2182 & 791 & 257 \\
  500 & 2175 & 2292 & 2358 & 2221 & 2337 & 2378 & 809 & 251 \\
  800 & 2426 & 2550 & 2585 & 2455 & 2618 & 2567 & 795 & 240 \\
 1000 & 2498 & 2617 & 2636 & 2485 & 2677 & 2650 & 668 & 217 \\
 1600 & 2590 & 2609 & 2739 & 2626 & 2764 & 2044 & 614 & 217 \\
 2000 & 2703 & 2758 & 2829 & 2711 & 2912 & 2753 & 606 & 216 \\
 4000 & 2502 & 2810 & 2822 & 2517 & 3100 & 2708 & 485 &  91 \\ \hline
\end{tabular}
\caption{Performance in Mflop/s of Cholesky Factorization on
SUN UltraSPARC IV+ dual-core CPUs computer, long real arithmetic}
\label{tab:sunreafac}
}
\end{table}

\begin{table}
{\normalsize \center
\begin{tabular}{|c|cc|cc|cc|cc|} \hline
 n & \multicolumn{4}{c|}{RFPF} & \multicolumn{4}{c|}{LAPACK} \\ \cline{2-9}
  & \multicolumn{2}{c|}{NO TRANS}&\multicolumn{2}{c|}{TRANS}&\multicolumn{2}{c|}{POTRF}&\multicolumn{2}{c|}{PPTRF} \\ \cline{2-9}
      &   U  &   L  &   U  &   L  &   U  &   L  &  U  &  L  \\ \hline
   50 &  716 &  699 &  698 &  714 &  581 &  549 & 535 & 554 \\
  100 & 1199 & 1185 & 1183 & 1197 & 1163 & 1148 & 719 & 721 \\
  200 & 1768 & 1742 & 1756 & 1774 & 1821 & 1806 & 840 & 822 \\
  400 & 2277 & 2262 & 2293 & 2289 & 2179 & 2159 & 919 & 881 \\
  500 & 2354 & 2334 & 2357 & 2130 & 2468 & 2479 & 931 & 891 \\
  800 & 2551 & 2361 & 2593 & 2584 & 2636 & 2629 & 880 & 755 \\
 1000 & 2599 & 2600 & 2668 & 2639 & 2717 & 2717 & 708 & 520 \\
 1600 & 2621 & 2665 & 2702 & 2693 & 2507 & 2529 & 610 & 419 \\
 2000 & 2717 & 2767 & 2831 & 2740 & 2818 & 2854 & 599 & 401 \\
 4000 & 2542 & 2506 & 2757 & 2652 & 2635 & 2661 & 412 & 158 \\ \hline
\end{tabular}
\caption{Performance in Mflop/s of Cholesky Inversion on SUN
UltraSPARC IV+ dual-core CPUs computer, long real arithmetic}
\label{tab:sunreainv}
}
\end{table}

\begin{table}
{\normalsize \center
\begin{tabular}{|c|c|cc|cc|cc|cc|} \hline
 r & n & \multicolumn{4}{c|}{RFPF} & \multicolumn{4}{c|}{LAPACK} \\ \cline{3-10}
 h & & \multicolumn{2}{c|}{NO TRANS}&\multicolumn{2}{c|}{TRANS}&\multicolumn{2}
{c|}{POTRS}&\multicolumn{2}{c|}{PPTRS} \\ \cline{3-10}
 s &    &   U  &   L  &   U  &   L  &   U  &   L  &  U  &  L  \\ \hline
 100 &   50 & 1829 & 1877 & 1883 & 1792 & 1698 & 1705 & 549 & 545 \\
 100 &  100 & 2118 & 2117 & 2121 & 2123 & 2042 & 1968 & 713 & 711 \\
 100 &  200 & 2505 & 2511 & 2515 & 2515 & 2242 & 2231 & 689 & 828 \\
 100 &  400 & 2638 & 2598 & 2626 & 2664 & 2356 & 2456 & 715 & 888 \\
 100 &  500 & 2386 & 2499 & 2669 & 2706 & 2479 & 2451 & 743 & 895 \\
 100 &  800 & 2759 & 2746 & 2776 & 2781 & 2410 & 2326 & 626 & 704 \\
 100 & 1000 & 2795 & 2739 & 2811 & 2817 & 2052 & 1987 & 525 & 554 \\
 160 & 1600 & 2870 & 2873 & 2886 & 2875 & 2431 & 2289 & 447 & 429 \\
 200 & 2000 & 2825 & 2825 & 2845 & 2838 & 2371 & 2167 & 416 & 416 \\
 400 & 4000 & 2701 & 2700 & 2808 & 2667 & 1589 & 1588 & 175 & 168 \\ \hline
\end{tabular}
\caption{Performance in Mflop/s of Cholesky Solution on SUN
UltraSPARC IV+ computer, long real arithmetic}\label{tab:sunreasol}
}
\end{table}

\clearpage

\begin{table}
{\normalsize \center
\begin{tabular}{|c|cc|cc|cc|cc|} \hline
 n & \multicolumn{4}{c|}{RFPF} & \multicolumn{4}{c|}{LAPACK} \\ \cline{2-9}
  & \multicolumn{2}{c|}{NO TRANS}&\multicolumn{2}{c|}{TRANS}&\multicolumn{2}{c|}{POTRF}&\multicolumn{2}{c|}{PPTRF} \\ \cline{2-9}
      &   U  &   L  &   U  &   L  &   U  &   L  &  U  &  L   \\ \hline
   50 & 1423 & 1552 & 1633 & 1423 & 1301 & 1259 &  872 & 1333 \\
  100 & 2032 & 1986 & 2067 & 1854 & 1624 & 1905 & 1199 & 1353 \\
  200 & 2329 & 2277 & 2337 & 2198 & 2117 & 2374 & 1465 &  542 \\
  400 & 2646 & 2624 & 2698 & 2561 & 2556 & 2684 & 1725 &  482 \\
  500 & 2760 & 2264 & 2801 & 2699 & 2695 & 2793 & 1731 &  476 \\
  800 & 2890 & 2851 & 2897 & 2839 & 2874 & 2310 & 1315 &  441 \\
 1000 & 2929 & 2899 & 2954 & 2900 & 2958 & 2958 & 1244 &  435 \\
 1600 & 3002 & 2962 & 2563 & 2874 & 3204 & 1519 & 1202 &  379 \\
 2000 & 3031 & 2971 & 3016 & 3011 & 3372 & 3021 & 1173 &  411 \\
 4000 & 3022 & 2930 & 3011 & 3036 & 3185 & 2148 &  572 &  139 \\ \hline
\end{tabular}
\caption{Performance in Mflop/s of Cholesky Factorization
on SUN UltraSPARC IV+ computer, long complex arithmetic.}
\label{tab:suncomfac}
}
\end{table}

\begin{table}
{\normalsize \center
\begin{tabular}{|c|cc|cc|cc|cc|} \hline
 n & \multicolumn{4}{c|}{RFPF} & \multicolumn{4}{c|}{LAPACK} \\ \cline{2-9}
  & \multicolumn{2}{c|}{NO TRANS}&\multicolumn{2}{c|}{TRANS}&\multicolumn{2}{c|}{POTRF}&\multicolumn{2}{c|}{PPTRF} \\ \cline{2-9}
      &   U  &   L  &   U  &   L  &   U  &   L  &  U  &  L  \\ \hline
   50 & 1525 & 1575 & 1515 & 1620 & 1400 & 1378 & 1230 & 1232 \\
  100 & 1968 & 2001 & 1948 & 2042 & 2012 & 1959 & 1525 & 1548 \\
  200 & 2388 & 2438 & 2277 & 2447 & 2428 & 2431 & 1731 & 1687 \\
  400 & 2665 & 2715 & 2700 & 2715 & 2758 & 2793 & 1867 & 1698 \\
  500 & 2748 & 2779 & 2777 & 2773 & 2840 & 2870 & 1885 & 1697 \\
  800 & 2841 & 2898 & 2917 & 2837 & 2599 & 2985 & 1330 & 1319 \\
 1000 & 2897 & 2943 & 2971 & 2914 & 3005 & 3040 & 1264 & 1258 \\
 1600 & 2920 & 2925 & 2724 & 2482 & 2031 & 3015 & 1153 & 1212 \\
 2000 & 2883 & 2948 & 2946 & 2931 & 2990 & 3079 & 1186 & 1193 \\
 4000 & 2839 & 2939 & 2975 & 2823 & 2485 & 3007 &  723 &  706 \\ \hline
\end{tabular}
\caption{Performance in Mflop/s of Cholesky Inversion
on SUN UltraSPARC IV+ computer, long complex arithmetic}
\label{tab:suncominv}
}
\end{table}

\begin{table}
{\normalsize \center
\begin{tabular}{|c|c|cc|cc|cc|cc|} \hline
 r & n & \multicolumn{4}{c|}{RFPF} & \multicolumn{4}{c|}{LAPACK} \\ \cline{3-10}
 h & & \multicolumn{2}{c|}{NO TRANS}&\multicolumn{2}{c|}{TRANS}&\multicolumn{2}
{c|}{POTRS}&\multicolumn{2}{c|}{PPTRS} \\ \cline{3-10}
 s &    &   U  &   L  &   U  &   L  &   U  &   L  &  U  &  L  \\ \hline
 100 &   50 & 1949 & 1972 & 1971 & 1978 & 2161 & 2138 & 1029 & 1028 \\
 100 &  100 & 2552 & 2550 & 2562 & 2562 & 2501 & 2484 & 1212 & 1393 \\
 100 &  200 & 2858 & 2859 & 2860 & 2847 & 2646 & 2620 & 1303 & 1629 \\
 100 &  400 & 2982 & 2972 & 2972 & 2949 & 2811 & 2803 & 1398 & 1780 \\
 100 &  500 & 2991 & 2983 & 2987 & 2994 & 2835 & 2821 & 1364 & 1700 \\
 100 &  800 & 3083 & 3062 & 3083 & 2717 & 2819 & 2784 &  921 &  973 \\
 100 & 1000 & 3112 & 3100 & 3085 & 2694 & 2626 & 2604 &  853 &  855 \\
 160 & 1600 & 3141 & 3140 & 3149 & 3137 & 2820 & 2715 &  762 &  752 \\
 200 & 2000 & 3172 & 3182 & 3174 & 3171 & 2714 & 2698 &  718 &  667 \\
 400 & 4000 & 3193 & 3201 & 3214 & 3211 & 2656 & 2661 &  240 &  230 \\ \hline
\end{tabular}
\caption{Performance in Mflop/s of Cholesky Solution on SUN
UltraSPARC IV+ computer, long complex arithmetic}
\label{tab:suncomsol}
}
\end{table}

%

\clearpage

\begin{table}
{\normalsize \center
\begin{tabular}{|c|cc|cc|cc|cc|} \hline
 n & \multicolumn{4}{c|}{RFPF} & \multicolumn{4}{c|}{LAPACK} \\ \cline{2-9}
  & \multicolumn{2}{c|}{NO TRANS}&\multicolumn{2}{c|}{TRANS}&\multicolumn{2}{c|}{POTRF}&\multicolumn{2}{c|}{PPTRF} \\ \cline{2-9}
      &   U  &   L  &   U  &   L  &   U  &   L  &  U  &  L  \\ \hline
   50 &  721 &  616 &  642 &  694 &  519 &  537 &  331 & 300 \\
  100 & 1419 & 1280 & 1337 & 1386 & 1347 & 1216 &  612 & 303 \\
  200 & 2764 & 2526 & 2637 & 2732 & 2621 & 2526 & 1072 & 300 \\
  400 & 4120 & 3728 & 3943 & 4053 & 4116 & 3932 & 1041 & 292 \\
  500 & 4430 & 4142 & 4313 & 4410 & 4495 & 4568 &  997 & 291 \\
  800 & 4663 & 4009 & 4198 & 4804 & 5034 & 3873 & 1007 & 290 \\
 1000 & 4764 & 4134 & 4485 & 5107 & 4789 & 3732 & 1029 & 289 \\
 1600 & 4278 & 3612 & 3956 & 4178 & 3740 & 2680 &  153 & 188 \\
 2000 & 4061 & 3611 & 3657 & 4087 & 3771 & 2335 &   85 & 107 \\
 4000 & 3493 & 2660 & 3126 & 3185 & 3769 & 2307 &   53 &  81 \\ \hline
\end{tabular}
\caption{Performance in Mflop/s of Cholesky Factorization
on SGI Altix 3700, Intel Itanium2 computer, long real arithmetic}
\label{tab:sgireafac}
}
\end{table}

\begin{table}
{\normalsize \center
\begin{tabular}{|c|cc|cc|cc|cc|} \hline
 n & \multicolumn{4}{c|}{RFPF} & \multicolumn{4}{c|}{LAPACK} \\ \cline{2-9}
  & \multicolumn{2}{c|}{NO TRANS}&\multicolumn{2}{c|}{TRANS}&\multicolumn{2}{c|}{POTRF}&\multicolumn{2}{c|}{PPTRF} \\ \cline{2-9}
      &   U  &   L  &   U  &   L  &   U  &   L  &  U  &  L  \\ \hline
   50 &  774 &  797 &  665 &  818 &  676 &  675 & 317 &  416 \\
  100 & 1731 & 1669 & 1681 & 1723 & 1561 & 1528 & 404 &  754 \\
  200 & 2945 & 3140 & 3169 & 3195 & 3034 & 2975 & 461 & 1246 \\
  400 & 4466 & 4383 & 4403 & 4476 & 4198 & 4176 & 439 & 1686 \\
  500 & 4648 & 4531 & 4662 & 4685 & 4740 & 4605 & 429 & 1795 \\
  800 & 4827 & 4815 & 4891 & 4799 & 4463 & 4833 & 422 & 2024 \\
 1000 & 4992 & 5016 & 5194 & 5155 & 4699 & 4931 & 421 & 2121 \\
 1600 & 4882 & 4957 & 4908 & 4874 & 4293 & 4733 & 267 & 1474 \\
 2000 & 3482 & 3749 & 5031 & 4967 & 3916 & 3072 &  70 &  238 \\
 4000 & 3080 & 3290 & 3613 & 3560 & 2725 & 3063 &  59 &  152 \\ \hline
\end{tabular}
\caption{Performance in Mflop/s of Cholesky Inversion
on SGI Altix 3700, Intel Itanium2 computer, long real arithmetic}
\label{tab:sgireainv}
}
\end{table}


\begin{table}
{\normalsize \center
\begin{tabular}{|c|c|cc|cc|cc|cc|} \hline
 r & n & \multicolumn{4}{c|}{RFPF} & \multicolumn{4}{c|}{LAPACK} \\ \cline{3-10}
 h & & \multicolumn{2}{c|}{NO TRANS}&\multicolumn{2}{c|}{TRANS}&\multicolumn{2}
{c|}{POTRS}&\multicolumn{2}{c|}{PPTRS} \\ \cline{3-10}
 s &    &   U  &   L  &   U  &   L  &   U  &   L  &  U  &  L  \\ \hline
 100 &   50 & 2535 & 2535 & 2552 & 2543 & 3283 & 2826 &  496 &  488 \\
 100 &  100 & 3838 & 3831 & 3853 & 3848 & 4438 & 4301 &  860 &  844 \\
 100 &  200 & 4898 & 4894 & 4894 & 4892 & 5045 & 5029 & 1357 & 1307 \\
 100 &  400 & 5311 & 5298 & 5251 & 5246 & 5067 & 5185 & 1312 & 1695 \\
 100 &  500 & 5214 & 5192 & 5259 & 5248 & 5195 & 5417 & 1319 & 1814 \\
 100 &  800 & 5300 & 5222 & 5645 & 5634 & 4666 & 4773 & 1369 & 2095 \\
 100 & 1000 & 4851 & 4712 & 4775 & 4846 & 4699 & 4098 & 1378 & 2159 \\
 160 & 1600 & 3721 & 3406 & 3850 & 4127 & 3658 & 3441 &  180 &  474 \\
 200 & 2000 & 3957 & 3469 & 3482 & 3998 & 3799 & 3620 &   97 &  338 \\
 400 & 4000 & 3913 & 3994 & 3957 & 3555 & 3945 & 3768 &   68 &  167 \\ \hline
\end{tabular}
\caption{Performance in Mflop/s of Cholesky Solution on SGI
Altix 3700, Intel Itanium2 computer, long real arithmetic}
\label{tab:sgireasol}
}
\end{table}

\clearpage

\begin{table}
{\normalsize \center
\begin{tabular}{|c|cc|cc|cc|cc|} \hline
 n & \multicolumn{4}{c|}{RFPF} & \multicolumn{4}{c|}{LAPACK} \\ \cline{2-9}
  & \multicolumn{2}{c|}{NO TRANS}&\multicolumn{2}{c|}{TRANS}&\multicolumn{2}{c|}{POTRF}&\multicolumn{2}{c|}{PPTRF} \\ \cline{2-9}
      &   U  &   L  &   U  &   L  &   U  &   L  &  U  &  L  \\ \hline
   50 & 1477 & 1401 & 1532 & 1431 & 1449 & 1548 & 1084 & 1510 \\
  100 & 2651 & 2713 & 2765 & 2537 & 2492 & 2712 & 1628 & 2234 \\
  200 & 3828 & 3889 & 4040 & 3718 & 3532 & 3837 & 1812 & 2822 \\
  400 & 4374 & 4581 & 4829 & 4402 & 4343 & 4410 & 1550 & 3205 \\
  500 & 4592 & 4621 & 4933 & 4570 & 4776 & 4463 & 1521 & 3294 \\
  800 & 4729 & 4688 & 4897 & 4815 & 4737 & 4085 & 1277 & 3073 \\
 1000 & 4735 & 4694 & 4928 & 4689 & 4727 & 3334 &  441 & 1504 \\
 1600 & 4796 & 4701 & 4901 & 4737 & 3872 & 3801 &  223 &  693 \\
 2000 & 4560 & 4295 & 4553 & 4560 & 4476 & 3681 &  180 &  368 \\
 4000 & 3705 & 3341 & 4039 & 4200 & 4108 & 3487 &  101 &  186 \\ \hline
\end{tabular}
\caption{Performance in Mflop/s of Cholesky Factorization
on SGI Altix 3700, Intel Itanium2 computer, long complex arithmetic}
\label{tab:sgicomfac}
}
\end{table}


\begin{table}
{\normalsize \center
\begin{tabular}{|c|cc|cc|cc|cc|} \hline
 n & \multicolumn{4}{c|}{RFPF} & \multicolumn{4}{c|}{LAPACK} \\ \cline{2-9}
  & \multicolumn{2}{c|}{NO TRANS}&\multicolumn{2}{c|}{TRANS}&\multicolumn{2}{c|}{POTRF}&\multicolumn{2}{c|}{PPTRF} \\ \cline{2-9}
      &   U  &   L  &   U  &   L  &   U  &   L  &  U  &  L  \\ \hline
   50 & 1618 & 1633 & 1666 & 1744 & 1652 & 1529 & 1424 & 1284 \\
  100 & 2750 & 2744 & 2762 & 2968 & 2661 & 2523 & 2241 & 2037 \\
  200 & 3766 & 3780 & 3787 & 4085 & 3951 & 3582 & 2359 & 2764 \\
  400 & 4489 & 4404 & 4509 & 4708 & 4587 & 4408 & 1671 & 3205 \\
  500 & 4642 & 4594 & 4699 & 4860 & 4667 & 4618 & 1627 & 3340 \\
  800 & 4854 & 4826 & 4949 & 5044 & 4522 & 4634 & 1315 & 3366 \\
 1000 & 3246 & 3804 & 4958 & 5019 & 4001 & 3420 &  148 &  939 \\
 1600 & 4491 & 4623 & 3420 & 3620 & 2446 & 2881 &   69 & 1204 \\
 2000 & 2978 & 2912 & 4119 & 4158 & 3756 & 4040 &   62 &  325 \\
 4000 & 3532 & 3573 & 3514 & 3365 & 2829 & 2911 &   70 &  412 \\ \hline
\end{tabular}
\caption{Performance in Mflop/s of Cholesky Inversion on SGI 
Altix 3700, Intel Itanium2 computer, long complex arithmetic}
\label{tab:sgicominv}
}
\end{table}


\begin{table}
{\normalsize \center
\begin{tabular}{|c|c|cc|cc|cc|cc|} \hline
 r & n & \multicolumn{4}{c|}{RFPF} & \multicolumn{4}{c|}{LAPACK} \\ \cline{3-10}
 h & & \multicolumn{2}{c|}{NO TRANS}&\multicolumn{2}{c|}{TRANS}&\multicolumn{2}
{c|}{POTRS}&\multicolumn{2}{c|}{PPTRS} \\ \cline{3-10}
 s &    &   U  &   L  &   U  &   L  &   U  &   L  &  U  &  L  \\ \hline
 100 &   50 & 3062 & 3009 & 3067 & 3054 & 3465 & 3500 & 1551 & 1528 \\
 100 &  100 & 4106 & 4106 & 4114 & 4109 & 4287 & 4314 & 2146 & 2129 \\
 100 &  200 & 4562 & 4559 & 4750 & 4748 & 4369 & 4381 & 2200 & 2605 \\
 100 &  400 & 4662 & 4647 & 4885 & 4920 & 4920 & 5044 & 2163 & 2927 \\
 100 &  500 & 4612 & 4612 & 4970 & 5007 & 4925 & 4717 & 2193 & 3005 \\
 100 &  800 & 4332 & 4313 & 4729 & 4675 & 4726 & 4376 & 1951 & 2765 \\
 100 & 1000 & 4487 & 4430 & 4492 & 4639 & 4542 & 4454 & 1046 & 1838 \\
 160 & 1600 & 4469 & 4369 & 4450 & 4524 & 4057 & 4287 &  428 & 1225 \\
 200 & 2000 & 4284 & 4335 & 4225 & 4385 & 4315 & 4464 &  290 &  726 \\
 400 & 4000 & 3847 & 3845 & 4420 & 4434 & 4398 & 4445 &  110 &  373 \\ \hline
\end{tabular}
\caption{Performance in Mflop/s of Cholesky Solution on SGI
Altix 3700, Intel Itanium2 computer, long complex arithmetic}
\label{tab:sgicomsol}
}
\end{table}

\clearpage
\begin{table}
{\normalsize \center
\begin{tabular}{|c|cc|cc|cc|cc|} \hline
 n & \multicolumn{4}{c|}{RFPF} & \multicolumn{4}{c|}{LAPACK} \\ \cline{2-9}
  & \multicolumn{2}{c|}{NO TRANS}&\multicolumn{2}{c|}{TRANS}&\multicolumn{2}{c|}{POTRF}&\multicolumn{2}{c|}{PPTRF} \\ \cline{2-9}
      &   U  &   L  &   U  &   L  &   U  &   L  &  U  &  L  \\ \hline
   50 &  781 &  771 &  784 &  771 & 1107 &  739 &  495 &  533 \\
  100 & 1843 & 1788 & 1848 & 1812 & 1874 & 1725 &  879 &  825 \\
  200 & 3178 & 2869 & 2963 & 3064 & 2967 & 2871 & 1323 & 1100 \\
  400 & 3931 & 3709 & 3756 & 3823 & 3870 & 3740 & 1121 & 1236 \\
  500 & 4008 & 3808 & 3883 & 3914 & 4043 & 3911 & 1032 & 1257 \\
  800 & 4198 & 4097 & 4145 & 4126 & 3900 & 4009 &  612 & 1127 \\
 1000 & 4115 & 4038 & 4015 & 3649 & 3769 & 3983 &  305 &  697 \\
 1600 & 3851 & 3652 & 3967 & 3971 & 3640 & 3987 &  147 &  437 \\
 2000 & 3899 & 3716 & 3660 & 3660 & 3865 & 3835 &  108 &  358 \\
 4000 & 3966 & 3791 & 3927 & 4011 & 3869 & 4052 &  119 &  398 \\ \hline
\end{tabular}
\caption{Performance in Mflop/s of Cholesky Factorization
on ia64 Itanium computer, long real arithemic}\label{tab:itafac}
}
\end{table}

\begin{table}
{\normalsize \center
\begin{tabular}{|c|cc|cc|cc|cc|} \hline
 n & \multicolumn{4}{c|}{RFPF} & \multicolumn{4}{c|}{LAPACK} \\ \cline{2-9}
  & \multicolumn{2}{c|}{NO TRANS}&\multicolumn{2}{c|}{TRANS}&\multicolumn{2}{c|}{POTRI}&\multicolumn{2}{c|}{PPTRI} \\ \cline{2-9}
      &   u  &   l  &   u  &   l  &   u  &   l  &  u  &  l  \\ \hline
   50 &  633 &  659 &  648 &  640 &  777 &  870 &  508 &  460 \\
  100 & 1252 & 1323 & 1300 & 1272 & 1573 & 1760 &  815 &  810 \\
  200 & 2305 & 2442 & 2431 & 2314 & 2357 & 2639 & 1118 & 1211 \\
  400 & 3084 & 3199 & 3188 & 3094 & 3152 & 3445 & 1234 & 1363 \\
  500 & 3204 & 3316 & 3329 & 3218 & 3400 & 3611 & 1239 & 1382 \\
  800 & 3617 & 3741 & 3720 & 3640 & 3468 & 3786 & 1182 & 1268 \\
 1000 & 3611 & 3716 & 3637 & 3590 & 3456 & 3790 &  767 &  946 \\
 1600 & 3721 & 3802 & 3795 & 3714 & 3589 & 3713 &  500 &  609 \\
 2000 & 3784 & 3812 & 3745 & 3704 & 3636 & 3798 &  473 &  596 \\
 4000 & 3822 & 3762 & 3956 & 3851 & 3760 & 3750 &  467 &  614 \\ \hline
\end{tabular}
\caption{Performance in Mflop/s of Cholesky Inversion
on ia64 Itanium computer, long real arithemic}\label{tab:itainv}
}
\end{table}

\begin{table}
{\normalsize \center
\begin{tabular}{|c|c|cc|cc|cc|cc|} \hline
 r & n & \multicolumn{4}{c|}{RFPF} & \multicolumn{4}{c|}{LAPACK} \\ \cline{3-10}
 h & & \multicolumn{2}{c|}{NO TRANS}&\multicolumn{2}{c|}{TRANS}&\multicolumn{2}{c|}
 {POTRS}&\multicolumn{2}{c|}{PPTRS} \\ \cline{3-10}
s &    &   u  &   l  &   u  &   l  &   u  &   l  &  u  &  l  \\ \hline
 100 &   50 & 2409 & 2412 & 2414 & 2422 & 3044 & 3018 &  725 &  714 \\
 100 &  100 & 3305 & 3301 & 3303 & 3303 & 3889 & 3855 & 1126 & 1109 \\
 100 &  200 & 4149 & 4154 & 4127 & 4146 & 4143 & 4127 & 1526 & 1512 \\
 100 &  400 & 4398 & 4403 & 4416 & 4444 & 4469 & 4451 & 1097 & 1088 \\
 100 &  500 & 4313 & 4155 & 4374 & 4394 & 4203 & 4093 & 1054 & 1045 \\
 100 &  800 & 3979 & 3919 & 4040 & 4051 & 3969 & 4011 &  692 &  720 \\
 100 & 1000 & 3716 & 3608 & 3498 & 3477 & 3630 & 3645 &  376 &  372 \\
 160 & 1600 & 3892 & 3874 & 4020 & 3994 & 4001 & 4011 &  188 &  182 \\
 200 & 2000 & 4052 & 4073 & 4040 & 4020 & 4231 & 4203 &  119 &  119 \\
 400 & 4000 & 4245 & 4225 & 4275 & 4287 & 4330 & 4320 &  115 &  144 \\ \hline
\end{tabular}
\caption{Performance in Mflop/s of Cholesky Solution
on ia64 Itanium computer, long real arithemic}\label{tab:itasol}
}
\end{table}

\clearpage

\begin{table}
\renewcommand{\arraystretch}{1.25}
{\normalsize \center
\begin{tabular}{|c|cc|cc|cc|cc|} \hline
 n & \multicolumn{4}{c|}{RFPF} & \multicolumn{4}{c|}{LAPACK} \\ \cline{2-9}
  & \multicolumn{2}{c|}{NO TRANS}&\multicolumn{2}{c|}{TRANS}&\multicolumn{2}{c|}{POTRF}&\multicolumn{2}{c|}{PPTRF} \\ \cline{2-9}
      &   u  &   l  &   u  &   l  &   u  &   l  &  u  &  l  \\ \hline
   50 &  206 &  200 &  225 &  225 &  365 &  353 &   57 &  238 \\
  100 &  721 &  728 &  789 &  788 & 1055 &  989 &  120 &  591 \\
  200 & 2028 & 2025 & 2005 & 2015 & 1380 & 1639 &  246 & 1250 \\
  400 & 3868 & 3915 & 3078 & 3073 & 1763 & 3311 &  479 & 1975 \\
  500 & 4483 & 4470 & 4636 & 4636 & 4103 & 4241 &  585 & 2149 \\
  800 & 5154 & 5168 & 4331 & 4261 & 3253 & 4469 &  870 & 2399 \\
 1000 & 5666 & 5654 & 5725 & 5703 & 5144 & 5689 & 1035 & 2474 \\
 1600 & 6224 & 6145 & 5644 & 5272 & 5375 & 5895 & 1441 & 2572 \\
 2000 & 6762 & 6788 & 6642 & 6610 & 6088 & 6732 & 1654 & 2598 \\
 4000 & 7321 & 7325 & 7236 & 7125 & 6994 & 7311 & 2339 & 2641 \\ \hline
\end{tabular}
\caption{Performance in Mflop/s of Cholesky Factorization
on SX-6 NEC computer with Vector Option, long real arithemic}\label{tab:necvfac}
}
\end{table}

\begin{table}
{\normalsize        \center
\begin{tabular}{|c|cc|cc|cc|cc|} \hline
 n & \multicolumn{4}{c|}{RFPF} & \multicolumn{4}{c|}{LAPACK} \\ \cline{2-9}
  & \multicolumn{2}{c|}{NO TRANS}&\multicolumn{2}{c|}{TRANS}&\multicolumn{2}{c|}{POTRI}&\multicolumn{2}{c|}{PPTRI} \\ \cline{2-9}
      &   u  &   l  &   u  &   l  &   u  &   l  &  u  &  l  \\ \hline
   50 &  152 &  152 &  150 &  152 &  148 &  145 &   91 &   61 \\
  100 &  430 &  432 &  428 &  432 &  313 &  310 &  194 &  126 \\
  200 &  950 &  956 &  940 &  941 &  636 &  627 &  404 &  249 \\
  400 & 1850 & 1852 & 1804 & 1806 & 1734 & 1624 &  722 &  470 \\
  500 & 2227 & 2228 & 2174 & 2181 & 2180 & 2029 &  856 &  572 \\
  800 & 3775 & 3775 & 3668 & 3686 & 3405 & 3052 & 1186 &  842 \\
 1000 & 4346 & 4346 & 4254 & 4263 & 4273 & 3638 & 1342 &  985 \\
 1600 & 5313 & 5294 & 5137 & 5308 & 5438 & 4511 & 1690 & 1361 \\
 2000 & 6006 & 6006 & 5930 & 5931 & 5997 & 4832 & 1854 & 1536 \\
 4000 & 6953 & 6953 & 6836 & 6888 & 7041 & 4814 & 1921 & 2122 \\ \hline
\end{tabular}
\caption{Performance in Mflop/s of Cholesky Inversion
on SX-6 NEC computer with Vector Option, long real arithemic}\label{tab:necvinv}
}
\end{table}

\begin{table}
{\normalsize        \center
\begin{tabular}{|c|c|cc|cc|cc|cc|} \hline
 r & n & \multicolumn{4}{c|}{RFPF} & \multicolumn{4}{c|}{LAPACK} \\ \cline{3-10}
 h & & \multicolumn{2}{c|}{NO TRANS}&\multicolumn{2}{c|}
{TRANS}&\multicolumn{2}{c|}{POTRS}&\multicolumn{2}{c|}{PPTRS} \\ \cline{3-10}
 s &    &   U  &   L  &   U  &   L  &   U  &   L  &  U  &  L  \\ \hline
 100 &   50 &  873 &  870 &  889 &  886 & 1933 & 1941 &   88 &   88 \\
 100 &  100 & 2173 & 2171 & 2200 & 2189 & 3216 & 3236 &  181 &  179 \\
 100 &  200 & 4236 & 4230 & 4253 & 4245 & 4165 & 4166 &  352 &  347 \\
 100 &  400 & 5431 & 5431 & 5410 & 5408 & 5302 & 5303 &  648 &  644 \\
 100 &  500 & 5563 & 5562 & 5568 & 5567 & 5629 & 5632 &  783 &  779 \\
 100 &  800 & 6407 & 6407 & 6240 & 6240 & 5569 & 5593 & 1132 & 1128 \\
 100 & 1000 & 6578 & 6578 & 6559 & 6558 & 6554 & 6566 & 1325 & 1320 \\
 160 & 1600 & 6781 & 6805 & 6430 & 6430 & 6799 & 6809 & 1732 & 1727 \\
 200 & 2000 & 7568 & 7569 & 7519 & 7519 & 7406 & 7407 & 1920 & 1914 \\
 400 & 4000 & 7858 & 7858 & 7761 & 7761 & 7626 & 7627 & 2414 & 2410 \\ \hline
\end{tabular}
\caption{Performance in Mflop/s of Cholesky Solution
on SX-6 NEC computer with Vector Option, long real arithemic}\label{tab:necvsol}
}
\end{table}

\clearpage

\begin{table}
{\normalsize
 \center
\begin{tabular}{|c|cc|cc|cc|cc|} \hline
 n & \multicolumn{4}{c|}{RFPF} & \multicolumn{4}{c|}{LAPACK} \\ \cline{2-9}
  & \multicolumn{2}{c|}{NO TRANS}&\multicolumn{2}{c|}{TRANS}&\multicolumn{2}{c|}{POTRF}&\multicolumn{2}{c|}{PPTRF} \\ \cline{2-9}
      &   U  &   L  &   U  &   L  &   U  &   L  &  U  &  L  \\ \hline
 2000 &  24.8460 & 24.2070 & 26.2493 & 27.3279& 24.9569& 13.0685&  0.9389&   0.4790  \\
 4000 &  39.0849 & 38.8042 & 41.1537 & 41.7441& 38.4284& 14.6297&  0.7378&   0.3879  \\
 6000 &  43.2940 & 43.9028 & 45.7611 & 45.6911& 40.1301& 14.6023&  0.7212&   0.3800  \\
 8000 &  48.2928 & 48.0530 & 50.1546 & 48.9082& 40.1865& 14.9028&   --   &    --     \\
10000 &  50.6669 & 50.0472 & 51.5198 & 50.8383& 41.7279& 14.9236&   --   &    --     \\
12000 &  47.9860 & 47.5107 & 50.6640 & 50.2138& 43.1972& 14.6511&   --   &    --     \\
14000 &  50.3806 & 50.6969 & 52.7881 & 52.3719& 43.7816& 14.5463&   --   &    --     \\
16000 &  51.2309 & 51.9454 & 53.5924 & 53.2322& 44.0667& 14.2067&   --   &    --     \\
18000 &  52.6901 & 52.2244 & 54.2978 & 53.5869& 46.2805& 14.5523&   --   &    --     \\
20000 &  53.6790 & 54.1209 & 54.3555 & 54.7896& 45.8757& 14.6236&   --   &    --     \\ \hline
\end{tabular}
\caption{Performance in Gflop/s of Cholesky Factorization
on Intel Tigerton computer, long real arithmetic.
We use reference LAPACK-3.2.0 (from netlib) and MKL-10.0.1.014 multithreaded BLAS.
Due to time limitation, 
the experiment was stopped for the packed storage format at $n=6000$.
}
\label{tab:zootfac_reference}
}
\end{table}

\begin{table}
{\normalsize 
 \center
\begin{tabular}{|c|cc|cc|cc|cc|} \hline
 n & \multicolumn{4}{c|}{RFPF} & \multicolumn{4}{c|}{LAPACK} \\ \cline{2-9}
  & \multicolumn{2}{c|}{NO TRANS}&\multicolumn{2}{c|}{TRANS}&\multicolumn{2}{c|}{POTRF}&\multicolumn{2}{c|}{PPTRF} \\ \cline{2-9}
      &   U  &   L  &   U  &   L  &   U  &   L  &  U  &  L  \\ \hline
 2000 &29.9701 &31.0403& 29.2714 &28.4205& 13.2510& 18.5249 &0.6338 &0.9229 \\
 4000 &38.4338 &39.1702& 38.3199 &37.7938& 13.0367& 18.1662 &0.5080 &0.7301 \\
 6000 &38.6324 &39.1249& 39.0177 &38.9534& 12.8468& 18.0594 &0.4972 &0.7149 \\
 8000 &40.6770 &40.7352& 40.9032 &39.8398& 12.8871& 17.9491 & --    & -- \\
10000 &41.3971 &41.5932& 41.6892 &41.6400& 12.6654& 17.5897 & --    & -- \\
12000 &41.1646 &40.8424& 40.2776 &40.4129& 12.4705& 17.5883 & --    & -- \\
14000 &42.1946 &42.1400& 41.2174 &41.3633& 12.4050& 17.4173 & --    & -- \\
16000 &42.0274 &42.2826& 42.4457 &42.3624& 12.1912& 17.2090 & --    & -- \\
18000 &42.2909 &42.4922& 41.9356 &42.2480& 12.1616& 17.3289 & --    & -- \\ \hline
\end{tabular}
\caption{Performance in Gflop/s of Cholesky Inversion
on Intel Tigerton computer, long real arithmetic
We use reference LAPACK-3.2.0 (from netlib) and MKL-10.0.1.014 multithreaded BLAS.
Due to time limitation, 
the experiment was stopped for the packed storage format at $n=6000$.
}
\label{tab:zootinv_reference}
}
\end{table}

\begin{table}
{\normalsize
 \center
\begin{tabular}{|c|c|cc|cc|cc|cc|} \hline
 r & n & \multicolumn{4}{c|}{RFPF} & \multicolumn{4}{c|}{LAPACK} \\ \cline{3-10}
 h & & \multicolumn{2}{c|}{NO TRANS}&\multicolumn{2}{c|}{TRANS}&\multicolumn{2}
{c|}{POTRS}&\multicolumn{2}{c|}{PPTRS} \\ \cline{3-10}
 s &    &   U  &   L  &   U  &   L  &   U  &   L  &  U  &  L  \\ \hline
 100 & 2000 & 0.7802 & 0.7759 & 0.7947 & 0.7897 & 0.7365 & 0.6691 & 0.8177 & 0.7628  \\
 100 & 4000 & 0.6925 & 0.6918 & 0.7130 & 0.7113 & 0.6462 & 0.6120 & 0.7310 & 0.7261  \\
 100 & 6000 & 0.6672 & 0.6639 & 0.6921 & 0.6937 & 0.5955 & 0.5773 & 0.7214 & 0.7193  \\
 100 & 8000 & 0.6494 & 0.6457 & 0.6787 & 0.6791 & 0.5524 & 0.5463 &  --    &  --     \\
 100 &10000 & 0.6247 & 0.6194 & 0.6594 & 0.6579 & 0.5329 & 0.5269 &  --    &  --     \\
 100 &12000 & 0.6228 & 0.6230 & 0.6512 & 0.6506 & 0.5336 & 0.5291 &  --    &  --     \\
 100 &14000 & 0.5933 & 0.6181 & 0.6291 & 0.6309 & 0.5356 & 0.5271 &  --    &  --     \\
 100 &16000 & 0.6020 & 0.6018 & 0.6265 & 0.6295 & 0.5095 & 0.5088 &  --    &  --     \\
 100 &18000 & 0.6175 & 0.6164 & 0.6196 & 0.6184 & 0.5310 & 0.5232 &  --    &  --     \\
 100 &20000 & 0.6092 & 0.6063 & 0.6022 & 0.6024 & 0.5221 & 0.5163 &  --    &  --     \\ \hline
\end{tabular}
\caption{Performance in Gflop/s of Cholesky Solution on 
Intel Tigerton computer, long real arithmetic.
We use reference LAPACK-3.2.0 (from netlib) and MKL-10.0.1.014 multithreaded BLAS.
Due to time limitation, 
the experiment was stopped for the packed storage format at $n=6000$.}
\label{tab:zootsol_reference}
}
\end{table}

\clearpage

\begin{table}
{\normalsize
 \center
\begin{tabular}{|c|cc|cc|cc|cc|} \hline
 n & \multicolumn{4}{c|}{RFPF} & \multicolumn{4}{c|}{LAPACK} \\ \cline{2-9}
  & \multicolumn{2}{c|}{NO TRANS}&\multicolumn{2}{c|}{TRANS}&\multicolumn{2}{c|}{POTRF}&\multicolumn{2}{c|}{PPTRF} \\ \cline{2-9}
      &   U  &   L  &   U  &   L  &   U  &   L  &  U  &  L  \\ \hline
 2000 & 25.0114  &   24.7273   &  27.9415  &   29.2117  &  31.7156  & 22.4987  & 19.1706 &18.1686 \\
 4000 & 46.5472  &   46.6683   &  50.4646  &   52.1384  &  53.5300  & 39.1913  & 25.5211 &23.0137 \\
 6000 & 57.7951  &   59.0809   &  62.8870  &   62.2730  &  63.7367  & 45.6812  & 34.4061 &30.1288 \\
 8000 & 67.8673  &   70.0423   &  72.3038  &   68.6783  &  70.7858  & 48.2404  & 39.5816 &31.3558 \\
10000 & 76.6851  &   78.1704   &  78.9962  &   79.0753  &  75.8030  & 52.6184  & 42.2241 &35.7579 \\
12000 & 72.2916  &   74.1424   &  79.1635  &   78.9553  &  78.4410  & 57.9543  & 46.0673 &41.0530 \\
14000 & 79.5957  &   81.4214   &  85.3673  &   84.0138  &  82.1996  & 59.0167  & 46.5374 &38.7725 \\
16000 & 83.6760  &   84.8718   &  89.7696  &   87.4224  &  83.8289  & 58.7681  & 50.8717 &45.3575 \\
18000 & 86.6604  &   86.5750   &  89.3476  &   88.8508  &  86.7870  & 62.9814  & 52.5077 &47.1880 \\
20000 & 90.7187  &   92.3898   &  92.9467  &   91.9760  &  88.2639  & 64.3982  & 51.0705 &43.1419 \\ \hline
\end{tabular}
\caption{Performance in Gflop/s of Cholesky Factorization
on Intel Tigerton computer, long real arithmetic.
We use MKL-10.0.1.014 multithreaded LAPACK and BLAS.
}
\label{tab:zootfac_vendor}
}
\end{table}

\begin{table}
{\normalsize 
 \center
\begin{tabular}{|c|cc|cc|cc|cc|} \hline
 n & \multicolumn{4}{c|}{RFPF} & \multicolumn{4}{c|}{LAPACK} \\ \cline{2-9}
  & \multicolumn{2}{c|}{NO TRANS}&\multicolumn{2}{c|}{TRANS}&\multicolumn{2}{c|}{POTRF}&\multicolumn{2}{c|}{PPTRF} \\ \cline{2-9}
      &   U  &   L  &   U  &   L  &   U  &   L  &  U  &  L  \\ \hline
 2000 & 29.7015 &  32.2611  & 29.5448 & 28.7837 & 13.8077  & 18.8739  &  0.6367   &  0.9238     \\
 4000 & 38.6352 &  39.5333  & 38.1630 & 37.9292 & 13.1999  & 18.1173  &  0.5069   &  0.7288     \\
 6000 & 38.7001 &  39.3848  & 38.5245 & 39.1682 & 12.8651  & 17.8524  &  --       &  --         \\
 8000 & 40.6456 &  41.2400  & 41.0437 & 40.9830 & 12.8791  & 17.9160  &  --       &  --         \\
10000 & 41.5013 &  41.7725  & 42.4129 & 42.3191 & 12.7119  & 17.4713  &  --       &  --         \\
12000 & 41.4199 &  41.2636  & 40.6651 & 40.5983 & 12.4945  & 17.6937  &  --       &  --         \\
14000 & 42.0461 &  42.4899  & 41.8353 & 41.4583 & 12.4234  & 17.5004  &  --       &  --         \\
16000 & 42.5350 &  42.7828  & 42.9538 & 42.4658 & 12.2014  & 17.2031  &  --       &  --         \\
18000 & 42.0039 &  42.5616  & 42.3765 & 41.9941 & 12.2800  & 17.3990  &  --       &  --         \\
20000 & 42.6296 &  43.0443  & 41.9921 & 41.9014 & 12.1434  & 17.3876  &  --       &  --         \\ \hline
\end{tabular}
\caption{Performance in Gflop/s of Cholesky Inversion
on Intel Tigerton computer, long real arithmetic
We use MKL-10.0.1.014 multithreaded LAPACK and BLAS.
Due to time limitation, 
the experiment was stopped for the packed storage format at $n=4000$.
}
\label{tab:zootinv_vendor}
}
\end{table}

\begin{table}
{\normalsize
 \center
\begin{tabular}{|c|c|cc|cc|cc|cc|} \hline
 r & n & \multicolumn{4}{c|}{RFPF} & \multicolumn{4}{c|}{LAPACK} \\ \cline{3-10}
 h & & \multicolumn{2}{c|}{NO TRANS}&\multicolumn{2}{c|}{TRANS}&\multicolumn{2}
{c|}{POTRS}&\multicolumn{2}{c|}{PPTRS} \\ \cline{3-10}
 s &    &   U  &   L  &   U  &   L  &   U  &   L  &  U  &  L  \\ \hline
 100 &  2000 &  0.1530 & 0.1439 & 0.1396& 0.1432&  0.1164 &   0.1093  &  0.1856 & 0.1482     \\
 100 &  4000 &  0.1516 & 0.1459 & 0.1486& 0.1503&  0.1077 &   0.1140  &  0.1545 & 0.1362     \\
 100 &  6000 &  0.1512 & 0.1451 & 0.1471& 0.1493&  0.1101 &   0.1065  &  0.1397 & 0.1223     \\
 100 &  8000 &  0.1490 & 0.1411 & 0.1429& 0.1458&  0.1085 &   0.1100  &  0.1192 & 0.1136     \\
 100 & 10000 &  0.1452 & 0.1408 & 0.1471& 0.1430&  0.1066 &   0.1088  &  0.1027 & 0.1019     \\
 100 & 12000 &  0.1407 & 0.1429 & 0.1452& 0.1404&  0.1079 &   0.1091  &  0.0958 & 0.0926     \\
 100 & 14000 &  0.1398 & 0.1406 & 0.1404& 0.1388&  0.1080 &   0.1100  &  0.0837 & 0.0843     \\
 100 & 16000 &  0.1374 & 0.1374 & 0.1411& 0.1405&  0.1075 &   0.1089  &  0.0786 & 0.0786     \\
 100 & 18000 &  0.1370 & 0.1366 & 0.1402& 0.1396&  0.1086 &   0.1087  &  0.0748 & 0.0745     \\
 100 & 20000 &  0.1362 & 0.1364 & 0.1394& 0.1425&  0.1065 &   0.1117  &  0.0699 & 0.0699     \\ \hline
\end{tabular}
\caption{Performance in Gflop/s of Cholesky Solution on 
Intel Tigerton computer, long real arithmetic.
We use MKL-10.0.1.014 multithreaded LAPACK and BLAS.}
\label{tab:zootsol_vendor}
}
\end{table}

\clearpage
\begin{table}[]
{\scriptsize \center
\setlength{\tabcolsep}{2.5mm}
\begin{tabular}{|c|c|cc|cccc|cc|} \hline
  n  & n  & Mflop/s  & \multicolumn{7}{|c|}{Times} \\ \cline{3-10}
     & pr & \multicolumn{2}{c|}{PFTRF} & \multicolumn{4}{c|}{in PFTRF}
     & \multicolumn{2}{c|}{LAPACK} \\ \cline{5-10}
     & oc &       &       & PO & TR & SY & PO & PO & PP \\
     &    &       &       & TRF & SM & RK & TRF & TRF & TRF \\ \hline
  1  & 2  &  3    &   4   & 5 & 6 & 7 & 8 & 9 & 10 \\ \hline
 1000 &  1 &  2695 &  0.12 & 0.02 & 0.05 & 0.04 & 0.02 &  0.12 &   0.94 \\
      &  5 &  7570 &  0.04 & 0.01 & 0.02 & 0.01 & 0.01 &  0.03 &   0.32 \\
      & 10 & 10699 &  0.03 & 0.01 & 0.01 & 0.01 & 0.00 &  0.02 &   0.16 \\
      & 15 & 18354 &  0.02 & 0.00 & 0.01 & 0.00 & 0.00 &  0.01 &   0.11 \\ \hline
 2000 &  1 &  2618 &  1.02 & 0.13 & 0.38 & 0.38 & 0.13 &  0.97 &   8.74 \\
      &  5 & 10127 &  0.26 & 0.04 & 0.10 & 0.09 & 0.04 &  0.24 &   3.42 \\
      & 10 & 17579 &  0.15 & 0.02 & 0.06 & 0.05 & 0.03 &  0.12 &   1.65 \\
      & 15 & 23798 &  0.11 & 0.02 & 0.04 & 0.04 & 0.01 &  0.13 &   1.11 \\ \hline
 3000 &  1 &  2577 &  3.49 & 0.45 & 1.33 & 1.28 & 0.44 &  3.40 &  30.42 \\
      &  5 & 11369 &  0.79 & 0.11 & 0.28 & 0.30 & 0.11 &  0.71 &  11.76 \\
      & 10 & 19706 &  0.46 & 0.06 & 0.19 & 0.16 & 0.05 &  0.38 &   6.16 \\
      & 15 & 29280 &  0.31 & 0.05 & 0.12 & 0.10 & 0.04 &  0.26 &   4.28 \\ \hline
 4000 &  1 &  2664 &  8.01 & 1.01 & 2.90 & 3.09 & 1.01 &  7.55 &  75.72 \\
      &  5 & 11221 &  1.90 & 0.26 & 0.68 & 0.72 & 0.24 &  1.65 &  25.73 \\
      & 10 & 21275 &  1.00 & 0.13 & 0.39 & 0.36 & 0.12 &  0.86 &  13.95 \\
      & 15 & 31024 &  0.69 & 0.09 & 0.28 & 0.24 & 0.08 &  0.59 &  10.46 \\ \hline
 5000 &  1 &  2551 & 16.34 & 2.04 & 6.16 & 6.10 & 2.04 & 15.79 & 154.74 \\
      &  5 & 11372 &  3.66 & 0.45 & 1.37 & 1.44 & 0.40 &  3.27 &  47.76 \\
      & 10 & 22326 &  1.87 & 0.25 & 0.78 & 0.62 & 0.22 &  1.73 &  28.13 \\
      & 15 & 32265 &  1.29 & 0.17 & 0.53 & 0.45 & 0.14 &  1.16 &  20.95 \\ \hline
\end{tabular}
\caption{Performance Times and Mflop/s of Cholesky 
Factorization on an IBM Power 4
computer, long real arithemic, using SMP parallelism on 1, 5, 10 and 15 processors.
Here vendor codes for Level 2 and 3 BLAS and POTRF are used, ESSL 
library version 3.3.
UPLO = 'L'.}\label{tab:ibmpar}
}
\end{table}

\begin{table}[t]
{\scriptsize \center
\begin{tabular}{|c|c|cc|cccc|cc|} \hline
  n  & n  & Mflop/s  & \multicolumn{7}{|c|}{Times} \\ \cline{3-10}
     & pr & \multicolumn{2}{c|}{PFTRF} & \multicolumn{4}{c|}{in PFTRF} 
     & \multicolumn{2}{c|}{LAPACK} \\ \cline{5-10}
     & oc &       &       & PO & TR & SY & PO & PO & PP \\
     &    &       &       & TRF & SM & RK & TRF & TRF & TRF \\ \hline
  1  & 2  &  3    &   4   & 5 & 6 & 7 & 8 & 9 & 10 \\ \hline
1000 &  1 &  1587 &  0.21 &  0.03 & 0.09 & 0.07 &  0.03 &  0.19 &   1.06 \\
     &  5 &  4762 &  0.07 &  0.02 & 0.02 & 0.02 &  0.02 &  0.07 &   1.13 \\
     & 10 &  5557 &  0.06 &  0.01 & 0.01 & 0.02 &  0.02 &  0.06 &   1.12 \\
     & 15 &  5557 &  0.06 &  0.02 & 0.01 & 0.01 &  0.02 &  0.06 &   1.11 \\ \hline
2000 &  1 &  1668 &  1.58 &  0.22 & 0.63 & 0.52 &  0.22 &  1.45 &  11.20 \\
     &  5 &  6667 &  0.40 &  0.07 & 0.13 & 0.13 &  0.07 &  0.38 &  11.95 \\
     & 10 &  8602 &  0.31 &  0.06 & 0.07 & 0.11 &  0.07 &  0.25 &  11.24 \\
     & 15 &  9524 &  0.28 &  0.06 & 0.06 & 0.08 &  0.08 &  0.23 &  11.66 \\ \hline
3000 &  1 &  1819 &  4.95 &  0.62 & 1.98 & 1.72 &  0.63 &  4.86 &  45.48 \\
     &  5 &  6872 &  1.31 &  0.20 & 0.42 & 0.48 &  0.20 &  1.38 &  55.77 \\
     & 10 & 12162 &  0.74 &  0.14 & 0.22 & 0.21 &  0.16 &  0.76 &  46.99 \\
     & 15 & 12676 &  0.71 &  0.14 & 0.16 & 0.30 &  0.16 &  0.61 &  45.71 \\ \hline
4000 &  1 &  1823 & 11.70 &  1.52 & 4.62 & 4.01 &  1.55 & 11.86 & 112.52 \\
     &  5 &  7960 &  2.68 &  0.40 & 0.94 & 0.92 &  0.42 &  2.74 & 112.77 \\
     & 10 & 14035 &  1.52 &  0.26 & 0.47 & 0.49 &  0.30 &  1.61 & 112.53 \\
     & 15 & 17067 &  1.25 &  0.24 & 0.37 & 0.35 &  0.29 &  1.29 & 111.67 \\ \hline
5000 &  1 &  1843 & 22.61 &  2.92 & 8.76 & 8.00 &  2.93 & 23.60 & 218.94 \\
     &  5 &  8139 &  5.12 &  0.77 & 1.81 & 1.80 &  0.74 &  5.45 & 221.58 \\
     & 10 & 14318 &  2.91 &  0.50 & 0.97 & 0.93 &  0.51 &  3.11 & 214.54 \\
     & 15 & 17960 &  2.32 &  0.45 & 0.72 & 0.68 &  0.47 &  2.40 & 225.08 \\ \hline
\end{tabular}
\caption{Performance in Times and Mflop/s of Cholesky Factorization 
on SUN UltraSPARC-IV computer, long real arithemic, with a different number of Processors,
testing the SMP Parallelism.
The implementation of PPTRF of sunperf does not show any SMP  
   parallelism.
 UPLO = 'L'.}\label{tab:sunpar}\label{tab:last}
}
\end{table}

\clearpage

\begin{center}\begin{table}\begin{center}
\begin{tabular}{|c|c|c|c|c|c|c|}
\hline
      & \multicolumn{2}{|c|}{factorization} &  \multicolumn{2}{|c|}{inversion}     &  \multicolumn{2}{|c|}{solution}      \\
\hline
      &PF/PO & PF/PP  & PF/PO & PF/PP  & PF/PO & PF/PP  \\
\hline
   50 & 0.99 &  1.47  &  1.23 &  1.29  &  1.10 &  3.43  \\
  100 & 1.20 &  1.87  &  1.03 &  1.66  &  1.04 &  2.98  \\
  200 & 0.97 &  2.55  &  0.97 &  2.11  &  1.12 &  3.04  \\
  400 & 1.02 &  2.88  &  1.05 &  2.50  &  1.08 &  3.00  \\
  500 & 0.99 &  2.91  &  0.95 &  2.53  &  1.09 &  3.02  \\
  800 & 0.99 &  3.25  &  0.98 &  2.95  &  1.15 &  3.95  \\
 1000 & 0.98 &  3.95  &  0.98 &  3.77  &  1.37 &  5.08  \\
 1600 & 0.99 &  4.46  &  1.07 &  4.43  &  1.19 &  6.46  \\
 2000 & 0.97 &  4.67  &  0.99 &  4.73  &  1.20 &  6.84  \\
 4000 & 0.91 &  5.82  &  1.04 &  6.69  &  1.77 & 16.05  \\
\hline
\end{tabular}
\caption{\label{tab:sun_rea_speedup}
Speedup of Cholesky Factorization/Inversion/Solution on
SUN UltraSPARC IV+ computer, long real  arithmetic.  The original data is presented in Appendix~\ref{appendix-table}  in Tables
\ref{tab:sunreafac}, \ref{tab:sunreainv} and \ref{tab:sunreasol}.
}
\end{center}\end{table}\end{center}

\begin{center}\begin{table}\begin{center}
\begin{tabular}{|c|c|c|c|c|c|c|}
\hline
      & \multicolumn{2}{|c|}{factorization} &  \multicolumn{2}{|c|}{inversion}     &  \multicolumn{2}{|c|}{solution}      \\
\hline
      & PF/PO & PF/PP & PF/PO & PF/PP  & PF/PO & PF/PP  \\
\hline
   50 &  1.16 &  1.31 &  1.26 &  1.23  &  0.92 &  1.92  \\
  100 &  1.01 &  1.32 &  1.09 &  1.53  &  1.02 &  1.84  \\
  200 &  1.01 &  1.41 &  0.98 &  1.60  &  1.08 &  1.76  \\
  400 &  0.97 &  1.45 &  1.01 &  1.56  &  1.06 &  1.68  \\
  500 &  0.97 &  1.47 &  1.00 &  1.62  &  1.06 &  1.76  \\
  800 &  0.98 &  2.19 &  1.01 &  2.20  &  1.09 &  3.17  \\
 1000 &  0.98 &  2.35 &  1.00 &  2.37  &  1.19 &  3.64  \\
 1600 &  0.97 &  2.41 &  0.94 &  2.50  &  1.12 &  4.13  \\
 2000 &  0.96 &  2.47 &  0.90 &  2.58  &  1.17 &  4.43  \\
 4000 &  0.99 &  4.11 &  0.95 &  5.31  &  1.21 & 13.39  \\
\hline
\end{tabular}
\caption{\label{tab:sun_com_speedup}
Speedup of Cholesky Factorization/Inversion/Solution on
SUN UltraSPARC IV+ computer, long complex  arithmetic.  The original
data is presented in Appendix~\ref{appendix-table}  in Tables
\ref{tab:suncomfac}, \ref{tab:suncominv} and \ref{tab:suncomsol}.
}
\end{center}\end{table}\end{center}

\begin{center}\begin{table}\begin{center}
\begin{tabular}{|c|c|c|c|c|c|c|}
\hline
      & \multicolumn{2}{|c|}{factorization} &  \multicolumn{2}{|c|}{inversion}     &  \multicolumn{2}{|c|}{solution}      \\
\hline
      & PF/PO&  PF/PP & PF/PO&  PF/PP  & PF/PO & PF/PP  \\
\hline
   50 &  1.34&   2.18 &  1.21&   1.97  &  0.78 &  5.15  \\
  100 &  1.05&   2.32 &  1.11&   2.30  &  0.87 &  4.48  \\
  200 &  1.05&   2.58 &  1.05&   2.56  &  0.97 &  3.61  \\
  400 &  1.00&   3.96 &  1.07&   2.65  &  1.02 &  3.13  \\
  500 &  0.97&   4.44 &  0.99&   2.61  &  0.97 &  2.90  \\
  800 &  0.95&   4.77 &  1.01&   2.42  &  1.18 &  2.69  \\
 1000 &  1.07&   4.96 &  1.05&   2.45  &  1.03 &  2.25  \\
 1600 &  1.14&  22.76 &  1.05&   3.36  &  1.13 &  8.71  \\
 2000 &  1.08&  38.20 &  1.28&  21.14  &  1.05 & 11.83  \\
 4000 &  0.93&  43.12 &  1.18&  23.77  &  1.01 & 23.92  \\
\hline
\end{tabular}
\caption{\label{tab:sgi_rea}
Speedup of Cholesky Factorization/Inversion/Solution on SGI
Altix 3700, Intel Itanium 2 computer, long real arithmetic.  The original 
data is presented in Appendix~\ref{appendix-table}  in Tables
\ref{tab:sgireafac}, \ref{tab:sgireainv} and \ref{tab:sgireasol}.
}
\end{center}\end{table}\end{center}

\begin{center}\begin{table}\begin{center}
\begin{tabular}{|c|c|c|c|c|c|c|}
\hline
      & \multicolumn{2}{|c|}{factorization} &  \multicolumn{2}{|c|}{inversion}     &  \multicolumn{2}{|c|}{solution}      \\
\hline
      &PF/PO & PF/PP  & PF/PO & PF/PP  & PF/PO & PF/PP  \\
\hline
   50 & 0.99 &  1.01  &  1.06 &  1.22  &  0.88 &  1.98  \\
  100 & 1.02 &  1.24  &  1.12 &  1.32  &  0.95 &  1.92  \\
  200 & 1.05 &  1.43  &  1.03 &  1.48  &  1.08 &  1.82  \\
  400 & 1.10 &  1.51  &  1.03 &  1.47  &  0.98 &  1.68  \\
  500 & 1.03 &  1.50  &  1.04 &  1.46  &  1.02 &  1.67  \\
  800 & 1.03 &  1.59  &  1.09 &  1.50  &  1.00 &  1.71  \\
 1000 & 1.04 &  3.28  &  1.25 &  5.35  &  1.02 &  2.52  \\
 1600 & 1.27 &  7.07  &  1.60 &  3.84  &  1.06 &  3.69  \\
 2000 & 1.02 & 12.39  &  1.03 & 12.79  &  0.98 &  6.04  \\
 4000 & 1.02 & 22.58  &  1.23 &  8.67  &  1.00 & 11.89  \\
\hline
\end{tabular}
\caption{\label{tab:sgi_com_speedup}
Speedup of Cholesky Factorization/Inversion/Solution on SGI
Altix 3700, Intel Itanium 2 computer, long complex  arithmetic.  The original
data is presented in Appendix~\ref{appendix-table}  in Tables
\ref{tab:sgicomfac}, \ref{tab:sgicominv} and \ref{tab:sgicomsol}.
}
\end{center}\end{table}\end{center}

\begin{center}\begin{table}\begin{center}
\begin{tabular}{|c|c|c|c|c|c|c|}
\hline
      & \multicolumn{2}{|c|}{factorization} & \multicolumn{2}{|c|}{inversion}      &  \multicolumn{2}{|c|}{solution}      \\
\hline
      & PF/PO & PF/PP & PF/PO & PF/PP  &  PF/PO & PF/PP \\
\hline
   50 &  0.71 &  1.47 &  0.76 &  1.30  &  0.80  & 3.34  \\ 
  100 &  0.99 &  2.10 &  0.75 &  1.62  &  0.85  & 2.94  \\ 
  200 &  1.07 &  2.40 &  0.93 &  2.02  &  1.00  & 2.72  \\ 
  400 &  1.02 &  3.18 &  0.93 &  2.35  &  0.99  & 4.05  \\ 
  500 &  0.99 &  3.19 &  0.92 &  2.41  &  1.05  & 4.17  \\ 
  800 &  1.05 &  3.72 &  0.99 &  2.95  &  1.01  & 5.63  \\ 
 1000 &  1.03 &  5.90 &  0.98 &  3.93  &  1.02  & 9.88  \\ 
 1600 &  1.00 &  9.09 &  1.02 &  6.24  &  1.00  &21.38  \\ 
 2000 &  1.01 & 10.89 &  1.00 &  6.40  &  0.96  &34.23  \\ 
 4000 &  0.99 & 10.08 &  1.05 &  6.44  &  0.99  &29.77  \\ 
\hline
\end{tabular}
\caption{\label{tab:ita_speedup}
Speedup of Cholesky Factorization/Inversion/Solution
on ia64 Itanium computer, long real arithmetic. The original data is presented  in Appendix~\ref{appendix-table}  in Tables
\ref{tab:itafac}, \ref{tab:itainv} and \ref{tab:itasol}.
}
\end{center}\end{table}\end{center}

\begin{center}\begin{table}\begin{center}
\begin{tabular}{|c|c|c|c|c|c|c|}
\hline
      & \multicolumn{2}{|c|}{factorization} &  \multicolumn{2}{|c|}{inversion}     &  \multicolumn{2}{|c|}{solution}      \\
\hline
      &PF/PO & PF/PP  & PF/PO & PF/PP  & PF/PO&  PF/PP  \\
\hline
   50 & 0.62 &  0.95  &  1.03 &  1.67  &  0.46&  10.10  \\
  100 & 0.75 &  1.34  &  1.38 &  2.23  &  0.68&  12.15  \\
  200 & 1.24 &  1.62  &  1.50 &  2.37  &  1.02&  12.08  \\
  400 & 1.18 &  1.98  &  1.07 &  2.57  &  1.02&   8.38  \\
  500 & 1.09 &  2.16  &  1.02 &  2.60  &  0.99&   7.11  \\
  800 & 1.16 &  2.15  &  1.11 &  3.18  &  1.15&   5.66  \\
 1000 & 1.01 &  2.31  &  1.02 &  3.24  &  1.00&   4.96  \\
 1600 & 1.06 &  2.42  &  0.98 &  3.14  &  1.00&   3.93  \\
 2000 & 1.01 &  2.61  &  1.00 &  3.24  &  1.02&   3.94  \\
 4000 & 1.00 &  2.77  &  0.99 &  3.28  &  1.03&   3.26  \\
\hline
\end{tabular}
\caption{\label{tab:nec_speedup}
Speedup of Cholesky Factorization/Inversion/Solution
on SX-6 NEC computer, long real arithmetic. The original data is presented  in Appendix~\ref{appendix-table}  in Tables
\ref{tab:necvfac}, \ref{tab:necvinv} and \ref{tab:necvsol}.
}
\end{center}\end{table}\end{center}

\begin{center}\begin{table}\begin{center}
\begin{tabular}{|c|c|c|c|c|c|c|}
\hline
      & \multicolumn{2}{|c|}{factorization} &  \multicolumn{2}{|c|}{inversion}     &  \multicolumn{2}{|c|}{solution}      \\
\hline
      &PF/PO & PF/PP  & PF/PO & PF/PP  & PF/PO & PF/PP  \\
\hline
 2000 & 1.10 & 29.11 &1.68 & 33.63 & 1.08 &  0.97 \\
 4000 & 1.09 & 56.58 &2.16 & 53.65 & 1.10 &  0.98 \\
 6000 & 1.14 & 63.45 &2.17 & 54.73 & 1.16 &  0.96 \\
 8000 & 1.25 &  --   &2.28 &  --   & 1.23 &   --  \\
10000 & 1.23 &  --   &2.37 &  --   & 1.24 &   --  \\
12000 & 1.17 &  --   &2.34 &  --   & 1.22 &   --  \\
14000 & 1.21 &  --   &2.42 &  --   & 1.18 &   --  \\
16000 & 1.22 &  --   &2.47 &  --   & 1.24 &   --  \\
18000 & 1.17 &  --   &2.45 &  --   & 1.17 &   --  \\
20000 & 1.19 &  --   &2.48 &  --   & 1.17 &   --  \\
\hline
\end{tabular}
\caption{\label{tab:zoot_speedup_reference}
Speedup of Cholesky Factorization/Inversion/Solution
on quad-socket quad-core Intel Tigerton computer, long real arithmetic.
We use reference LAPACK-3.2.0 (from netlib) and MKL-10.0.1.014 multithreaded BLAS.
The original data is in Appendix~\ref{appendix-table}  in Tables
\ref{tab:zootfac_vendor}, \ref{tab:zootinv_vendor} and \ref{tab:zootsol_vendor}.
For the solution phase, $nrhs$ is fixed to 100 for any $n$.
Due to time limitation, 
the experiment was stopped for the packed storage format at $n=6000$.
}
\end{center}\end{table}\end{center}

\begin{center}\begin{table}\begin{center}
\begin{tabular}{|c|c|c|c|c|c|c|}
\hline
      & \multicolumn{2}{|c|}{factorization} &  \multicolumn{2}{|c|}{inversion}     &  \multicolumn{2}{|c|}{solution}      \\
\hline
      &PF/PO & PF/PP  & PF/PO & PF/PP  & PF/PO & PF/PP  \\
\hline
 2000 & 0.92 &  1.52  &  1.71 & 34.92  &  1.31 &  0.82  \\
 4000 & 0.97 &  2.04  &  2.18 & 54.24  &  1.33 &  0.98  \\
 6000 & 0.99 &  1.83  &  2.21 &   --   &  1.37 &  1.08  \\
 8000 & 1.02 &  1.83  &  2.30 &   --   &  1.35 &  1.25  \\
10000 & 1.04 &  1.87  &  2.43 &   --   &  1.35 &  1.43  \\
12000 & 1.01 &  1.72  &  2.34 &   --   &  1.33 &  1.52  \\
14000 & 1.04 &  1.83  &  2.43 &   --   &  1.28 &  1.67  \\
16000 & 1.07 &  1.76  &  2.50 &   --   &  1.30 &  1.80  \\
18000 & 1.03 &  1.70  &  2.45 &   --   &  1.29 &  1.87  \\
20000 & 1.05 &  1.82  &  2.48 &   --   &  1.28 &  2.04  \\
\hline
\end{tabular}
\caption{\label{tab:zoot_speedup_vendor}
Speedup of Cholesky Factorization/Inversion/Solution
on quad-socket quad-core Intel Tigerton computer, long real arithmetic.
We use MKL-10.0.1.014 multithreaded LAPACK and BLAS.
The original data is presented  in Appendix~\ref{appendix-table}  in Tables
\ref{tab:zootfac_vendor}, \ref{tab:zootinv_vendor} and \ref{tab:zootsol_vendor}.
For the solution phase, $nrhs$ is fixed to 100 for any $n$.
Due to time limitation, 
the experiment was stopped for the packed storage format inversion at $n=4000$.
}
\end{center}\end{table}\end{center}

\end{appendix}

\end{document}